\documentclass[a4paper,11pt]{article}

\usepackage{jheppub} 

\usepackage[T1]{fontenc} 
\usepackage{amsmath,amssymb,mathrsfs,mathtools,graphicx}

\def\be{\begin{equation}}
\def\ee{\end{equation}}

\def\bea{\begin{eqnarray}}
\def\eea{\end{eqnarray}}

\def\vec[#1]{\boldsymbol{#1}}
\def\vecs[#1,#2]{\boldsymbol{{#1}_{#2}}}

\def\mes[#1]{d^{3}{#1}}

\def\del{\partial}

\newcommand{\half}{\frac{1}{2}}

\title{\boldmath Symmetry Constraints in Inflation, $\alpha$-vacua, and the Three Point Function}

\author{Ashish Shukla,}
\author{Sandip P. Trivedi,}
\author{and V. Vishal}
\affiliation{\it Department of Theoretical Physics, Tata Institute of Fundamental Research,\\  Colaba, Mumbai, 400005, India \\}

\emailAdd{ashish@theory.tifr.res.in}
\emailAdd{sandip@theory.tifr.res.in}
\emailAdd{vishal@theory.tifr.res.in}

\abstract{The Ward identities for conformal symmetries in single field models of inflation are studied in more detail in momentum space.
For a class of generalized single field models, where the inflaton action contains arbitrary powers of the scalar and its first derivative, we find that the Ward identities are valid. We also study a one-parameter family of vacua, called $\alpha$-vacua, which preserve conformal invariance in de Sitter space. We find that the Ward identities, upto contact terms,  are met for  the three point function of a scalar field in the probe approximation in these vacua. Interestingly, the corresponding non-Gaussian term in  the wave function does not satisfy the operator product expansion. For scalar perturbations in inflation, in the $\alpha$-vacua, we find that the Ward identities are not satisfied.
We argue that this is because the back-reaction on the metric of the full  quantum stress tensor has not been self-consistently incorporated. We also present a calculation, drawing on techniques from the AdS/CFT correspondence,  for the three point function of scalar perturbations in inflation in the Bunch-Davies vacuum.}

\begin{document}

\begin{flushright} 
\small{TIFR/TH/16-26}
\end{flushright}

\maketitle
\flushbottom

\section{Introduction}
\label{intro}
Inflation is a successful paradigm which explains the observed approximate homogeneity and isotropy of the universe. It also gives rise to perturbations due to quantum effects, which lead to the anisotropy of the CMB and seed the growth of large scale structure in the universe. During the inflationary epoch, the universe was approximately de Sitter space, which is a maximally symmetric solution to the Einstein's equations with a positive cosmological constant. The symmetry group of four dimensional de Sitter space is $O(1,4)$, with ten generators\footnote{For our analysis, we will be interested only in the connected subgroup of $O(1,4)$.}. It is interesting to ask what constraints are imposed by this big symmetry group, which is approximately preserved during inflation, on the quantum fluctuations generated during the inflationary epoch. Such an analysis has been carried out by a number of authors, see \cite{Antoniadis:1996dj, Larsen:2002et, Larsen:2003pf, McFadden:2010vh, Antoniadis:2011ib, McFadden:2011kk, Creminelli:2011mw, Bzowski:2011ab, Kehagias:2012pd, Kehagias:2012td, Schalm:2012pi, Bzowski:2012ih, McFadden:2014nta, Kehagias:2015jha}.

The symmetry algebra of $O(1,4)$ is the same as the symmetry algebra of a three dimensional Euclidean Conformal Field Theory. 
In \cite{Mata:2012bx, Ghosh:2014kba, Kundu:2014gxa}, following the seminal works \cite{Maldacena:2002vr} and \cite{Maldacena:2011nz}, single field slow-roll inflation was studied and it was shown that the symmetry constraints on the correlation functions of scalar and tensor perturbations can be expressed in terms of the Ward identities of conformal invariance.  These Ward identities give rise to the Maldacena consistency condition, and additional similar constraints arising due to the special conformal transformations. 

In fact, further study showed, \cite{Kundu:2015xta}, that these identities follow  from the constraints of  reparametrization invariance and should be  generally valid. This allows the breaking of conformal invariance during inflation, due to the evolution of inflaton, to be incorporated systematically even beyond leading order in the slow-roll parameters. Related references where the constraints are often thought of as following from non-linear realization of conformal symmetries include \cite{Weinberg:2003sw, Creminelli:2004yq, Cheung:2007sv, Weinberg:2008nf,  Creminelli:2011sq, Bartolo:2011wb, Creminelli:2012ed, Hinterbichler:2012nm, Senatore:2012wy, Assassi:2012zq, Creminelli:2012qr,  Goldberger:2013rsa, Hinterbichler:2013dpa, Creminelli:2013cga, Pimentel:2013gza, Berezhiani:2013ewa, Sreenath:2014nka, Mirbabayi:2014zpa, Joyce:2014aqa, Sreenath:2014nca, Binosi:2015obq, Chowdhury:2016yrh}.

The situation is analogous to what happens in a field theory which is not scale invariant. The correlations in such a theory still satisfy the Callan-Symanzik equation, which now involves contributions due to the non-vanishing of the beta functions. In the conformally invariant limit, the beta functions vanish and the Ward identities simplify and constrain the correlators in a more powerful way. In the near conformal limit, where the beta functions are small, there can still be significant constraints from the Ward identities. In the same way, for inflation it was found in \cite{Kundu:2015xta} that the Ward identities are generally valid since they arise from the constraints of spatial reparametrization invariance, which is a gauge symmetry of general relativity, and must hold very generally. In the slow-roll limit, where there is approximate conformal invariance, these conditions can impose significant constraints on the correlation functions for the scalar and tensor perturbations. 
 
An important aspect of this symmetry based analysis is that it is model independent. Constraints which arise, for example, in the approximately conformally invariant limit, probe basic features of the inflationary model in a model independent way. These constraints can have significant observational consequences, and can therefore give rise to model independent tests for the inflationary paradigm.

In this paper, we continue to explore these symmetry properties of the correlation functions for perturbations produced during inflation. In section \ref{sound}, we consider a class of models which are not of the standard slow-roll type. Instead, in these models, called generalized single field models, the inflaton  can roll quickly in units of the Hubble parameter, $H$, while the spacetime is still approximately de Sitter space. Using the earlier calculations of three and four point correlations for scalar perturbations in these models, \cite{Chen:2006nt} and \cite{Chen:2009bc}, we explicitly check that the Ward identities derived in \cite{Kundu:2015xta} are in fact valid. Some checks of these Ward identities were carried out earlier in \cite{Creminelli:2012ed}.
 
It is usually assumed in inflation that the initial state of the universe, when inflation commenced, was the Bunch-Davies vacuum. This assumption is well motivated. It corresponds to taking the modes with wavelength much smaller than $H^{-1}$  to be in their ``ground state'', i.e. in the state they would occupy in Minkowski space. The physical picture here is that at length scales much smaller than the Hubble scale the universe should be well approximated by Minkowski space. One positive feature of this choice is that the back-reaction due to the quantum stress tensor in this vacuum is small, and this makes the calculations, where only classical effects are included to leading order, self-consistent and well justified. 

However, since one of our purposes here is to examine various inflationary possibilities in a more model independent manner, we turn next to examining the Ward identities mentioned above in a class of vacua called $\alpha$-vacua, which are different from the Bunch-Davies vacuum. For a suitable choice of parameters these vacua preserve conformal invariance in de Sitter space, see \cite{PhysRevD.32.3136}. It is therefore quite interesting to ask if the Ward identities hold in these vacua as well, since the approximate conformal invariance of these vacua in inflation could then also lead to significant constraints on correlation functions. There is one drawback to these vacua, however. Modes of arbitrarily short wavelength in these vacua can be viewed as being highly excited as compared to their ground state in the Minkowski vacuum. As a result, the quantum stress tensor in these vacua is not small, and in fact is expected to diverge. This gives rise to a question as to whether such vacua can in fact arise in a self-consistent way in any inflationary model.
 
Since the thrust of our analysis is a model independent one, only tied to symmetry considerations, we lay aside this worry in the beginning. In section \ref{intml}, to begin with we study a probe scalar field, for which the quantum back-reaction can indeed be self-consistently neglected by taking the $M_{Pl}\rightarrow \infty$ limit (while keeping $H$ fixed). We verify that the $\alpha$-vacua do preserve the full conformal symmetry\footnote{This is true barring some subtleties involving zero modes for a massless field which we do not address, see \cite{PhysRevD.32.3136, PhysRevD.35.3771}.}. In an interacting theory obtained by adding a cubic term, we find that the resulting three point function does satisfy the Ward identities of conformal invariance, as expected. This analysis in the probe case serves as a test of some of the basic issues involving the $\alpha$-vacua. 
 
Next, we turn to the more complicated case of inflation. Working with the same vacua, we calculate the three point function for scalar perturbations and find that the Ward identities now do not hold. For example, we find that the Maldacena consistency condition is violated. We argue that this violation is because the back-reaction has not been included consistently for the analysis in these vacua. Unlike the probe case, we cannot set $M_{Pl}$ to be infinite here because the perturbations involved arise from gravity itself. It is therefore not consistent to neglect the back-reaction of the quantum stress tensor while incorporating quantum effects also suppressed in $M_{Pl}$ for the calculation of the three point function. This inconsistency, we argue, is why the Ward identities do not hold. 
 
To elaborate on this some more, the Ward identities, as we have mentioned above, arise because of spatial reparametrization invariance. The conditions ensuring this invariance are in fact part of the Einstein equations. By neglecting the back-reaction we do not meet the Einstein equations consistently. It is therefore not surprising that the Ward identities, which are consequences of these equations, are also not met. 
 
We end the paper in section \ref{bulk3pt} by discussing the scalar three point function in slow-roll inflation in some detail. This correlation function, which is observationally most significant in the study of non-Gaussianity, was first calculated in \cite{Maldacena:2002vr}. The Ward identities suggest a somewhat different way to calculate this correlation function. These  identities  relate the three point function to a scalar four point function in a particular limit, with the coefficient of the four point function being suppressed by a power of the slow-roll parameter ${\dot {\bar \phi}}/H$. This suggests that the leading slow-roll result for the three point function can be calculated from the four point function in the de Sitter approximation (where the slow-roll parameters can be set to vanish). We make this explicit in section \ref{bulk3pt} by carrying out the calculation along these lines. We show that replacing one of the legs in the four point calculation in de Sitter space with a factor of the slow-roll parameter ${\dot {\bar \phi}}/H$ does give the correct result for the three point function. This way of thinking about the three and higher point correlators, motivated by the AdS/CFT correspondence, and the resulting discussion of the Ward identities was implicit in some of the earlier literature, \cite{Ghosh:2014kba}, and has also played an important role in the recent discussions in \cite{Arkani-Hamed:2015bza}. 

The paper also includes appendices \ref{chenres}-\ref{bulkdet} which give additional important details. 

\textbf{Notation:} The Planck mass is given by $M_{Pl} = 1/\sqrt{8\pi G}$. We denote the conformal time coordinate by $\eta$. Spatial three vectors are denoted by boldface letters, e.g. $\vec[x], \vec[k]$ etc. $\vecs[k,a], a= 1,2, \ldots$ denotes the momentum vectors $\vecs[k,1], \vecs[k,2], \ldots$ etc, whereas $k^i, i = 1,2,3$ denotes the components of $\vec[k]$. The magnitude of a vector is denoted by the corresponding ordinary letter, e.g. $x \equiv |\vec[x]|$.  A dot above a quantity denotes ordinary time derivative, e.g. $\dot{f} \equiv df/dt$.

\section{Basic Properties of de Sitter Space}
\label{dsspace}
We start by presenting some key properties of de Sitter space, which we will refer to throughout the rest of the paper. Four dimensional de Sitter space in planar coordinates is given by the line element
\be
\label{dsp}
ds^2 = - \, dt^2 + e^{2Ht} dx^i dx_i ,
\ee
where $-\infty < t, x^i < \infty $. In our calculations, we will make use of the conformal time coordinate $\eta$, given by
\be
\label{defeta}
\eta = - \, \frac{1}{H} \, e^{- Ht} ,
\ee
where $-\infty < \eta \le 0$. The line element in eq.\eqref{dsp} then takes the form
\be
\label{meteta}
ds^2 = \frac{1}{H^2 \eta^2} (- \, d\eta^2 + dx^i dx_i).
\ee
Note that the coordinates $(t, \vec[x])$ or $(\eta, \vec[x])$ cover only half of  de Sitter space.

Four dimensional de Sitter space has the following isometries,
\begin{subequations}
\begin{align}
&\textit{(i) Translations:} ~~ x^i \rightarrow x^i + \epsilon^i;\label{iso1}\\
&\textit{(ii) Rotations:} ~~ x^i \rightarrow x^i + \omega^i_{\,j} x^j, \, \omega_{ij} = -\omega_{ji};\label{iso2}\\
&\textit{(iii) Dilatations:} ~~  \eta \rightarrow (1+\epsilon)\eta,\, x^i \rightarrow (1+\epsilon)x^i;\label{iso3}\\
\begin{split}
&\textit{(iv) Special Conformal Transformations:} ~~ \eta \rightarrow (1+2\vec[b]\cdot\vec[x])\eta,\label{iso4}\\
&\hspace{70mm} x^i \rightarrow x^i + 2\vec[b]\cdot\vec[x]\, x^i + b^i (\eta^2 - \vec[x]^2).
\end{split}
\end{align}
\end{subequations}
Here, the parameters $\epsilon^i, \omega_{ij}, \epsilon$ and $b^i$ are all infinitesimal. These isometries impose important constraints on the correlation functions of inflationary perturbations.

\section{Conformal Invariance and General Single Field Models of Inflation}
\label{sound}
The canonical single field slow-roll models of inflation are characterized by the action
\be
\label{sinf}
S = \frac{M_{Pl}^2}{2} \int d^4x \sqrt{-g} \, \Big( R - g^{\mu \nu} \del_\mu\phi \, \del_\nu\phi - 2 V(\phi) \Big),
\ee
where $V(\phi)$ is the potential for the inflaton $\phi$. For inflation to occur, the potential is assumed to be approximately flat over a range of values for $\phi$. The inflaton evolves slowly along this flat part during inflation, leading to exponential expansion. In the homogeneous limit, the inflaton is purely a function of time, $\phi \equiv \bar\phi(t)$, and the metric is the unperturbed FRW metric,
\be
\label{frw}
ds^2 = - \, dt^2 + a(t)^2 \delta_{ij} dx^i dx^j,
\ee
where $a(t)$ is the scale factor of the universe. The metric can equivalently be expressed in terms of the conformal time coordinate $\eta$ as\footnote{The general relation between the ordinary time $t$ and the conformal time $\eta$ is
\begin{equation*}
d\eta = \frac{dt}{a(t)}.
\end{equation*}}
\be
\label{homomet}
ds^2 = a(\eta)^2\, \big(-d\eta^2 +\delta_{ij} dx^i dx^j\big).
\ee
Also, the Hubble parameter is given by
\be
\label{hubble}
H = \frac{\dot{a}}{a}.
\ee
The homogeneous field $\bar{\phi}$ satisfies the equations
\begin{equation}
\begin{split}
\label{homoback}
3 H^2 &= \half \, \dot{\bar{\phi}}^2 + V(\bar{\phi}),\\
\dot{H} &= - \, \half\, \dot{\bar{\phi}}^2,\\
0 &= \ddot{\bar{\phi}} + 3 H \dot{\bar{\phi}} + \frac{dV(\bar{\phi})}{d\bar{\phi}}.
\end{split}
\end{equation}
The slow-roll conditions are imposed by setting the slow-roll parameters $\epsilon_1, \delta_1$ to be much less than unity, where
\be
\label{svp1}
\epsilon_1 = - \, \frac{\dot{H}}{H^2} ,
\ee
\be
\label{deldef}
\delta_1 = \frac{\ddot{H}}{2 H \dot{H}}.
\ee
The slow-roll criterion $\epsilon_1, \delta_1 \ll 1$ ensures that the universe remains approximately de Sitter during the inflationary phase. The slow-roll conditions can also be expressed in terms of the slow-roll parameters $\epsilon, \delta$, where
\be
\label{eps}
\epsilon = \half \frac{\dot{\bar{\phi}}^{\,2}}{H^2},
\ee
\be
\label{defdelp}
\delta = \frac{\ddot{\bar\phi}}{H\dot{\bar\phi}}.
\ee
Note that $\epsilon = \epsilon_1$ and $\delta = \delta_1$ in the canonical slow-roll model due to the background equations eq.\eqref{homoback}. Another set of slow-roll parameters are the potential slow-roll parameters, defined by
\begin{equation}
\begin{split}
\label{srdef}
&\epsilon_{\text{v}} =  \frac{1}{2} \left(\frac{V'}{V}\right)^2,\\
&\eta_{\text{v}} =  \frac{V''}{V}.
\end{split}
\end{equation}
In the slow-roll approximation, these are also related to the parameters $\epsilon_1, \delta$ due to the background equations eq.\eqref{homoback}, via $\epsilon_{\text{v}} = \epsilon_1$ and $\eta_{\text{v}} = \epsilon_1 - \delta$.

Perturbations to the homogeneous situation discussed above are introduced in the ADM formalism. The metric in the ADM formalism takes the form
\be
\label{admmet}
ds^2 = - N^2 dt^2 + h_{ij}(dx^i + N^i dt)(dx^j+ N^j dt),
\ee
where $h_{ij}$ is the induced metric on the spatial three surface labeled by time $t$, and $N, N^i$ are the lapse and shift functions, respectively. One needs to make a choice of gauge to fix the diffeomorphism invariance of the theory. A convenient choice is the \textit{synchronous gauge}, defined by imposing the conditions
\be
\label{syncgauge}
N = 1, N^i = 0.
\ee
The perturbed metric in this gauge has the form
\begin{equation}
\begin{split}
\label{defpm}
h_{ij} &= a^2\, [(1+2\zeta) \delta_{ij} + \widehat\gamma_{ij}], \\
\widehat\gamma_{ii} &= 0.
\end{split}
\end{equation}
where $\zeta, \widehat\gamma_{ij}$ are the scalar and tensor perturbations in the metric, respectively, with $\widehat\gamma_{ij}$ being traceless. The perturbed inflaton is given by
\be
\label{pinf}
\phi = \bar{\phi}(t) + \delta\phi(t,\vec[x]).
\ee

Note that in the ADM formalism $\phi, h_{ij}$ are the dynamical variables, whereas $N, N^i$ are Lagrange multipliers. One thus needs to impose the equations of motion of $N,N^i$ as constraints in the gauge eq.\eqref{syncgauge}. In the wave function of the universe approach, the equations of motion of $N,N^i$ correspond to time and spatial reparametrization invariance of the wave function. In \cite{Kundu:2015xta}, general Ward identities were derived for single field models of inflation as a consequence of these reparametrization invariance constraints. These Ward identities are satisfied by the correlation functions of the curvature perturbation $\zeta$, and the transverse and traceless tensor perturbations $\widehat{\gamma}_{ij}$.\footnote{In the gauge eq.\eqref{syncgauge}, the tensor perturbations $\widehat\gamma_{ij}$ can be made transverse, $\del_i \widehat\gamma_{ij} = 0$, at late times, using the spatial reparametrization $x^i \rightarrow x^i + v^i(\vec[x])$.} For instance, for a scaling transformation, we have
\begin{equation}
\begin{split}
\label{wardscg}
\bigg( 3(n-1) + \sum_{a=1}^n &k_a \, \frac{\del}{\del k_a} \bigg) \langle \zeta(\vecs[k,1]) \cdots \zeta(\vecs[k,n]) \rangle' = \\ &- \, \frac{1}{\langle \zeta(\vecs[k,n+1]) \zeta(-\vecs[k,n+1])\rangle'} \, \langle \zeta(\vecs[k,1]) \cdots \zeta(\vecs[k,n+1]) \rangle'\bigg|_{\vecs[k,n+1] \rightarrow 0},
\end{split}
\end{equation} 
where a $'$ on a correlation function denotes the suppression of the overall momentum conserving $\delta$-function; for e.g.
\be
\label{defprs}
 \langle \zeta(\vecs[k,1]) \cdots \zeta(\vecs[k,n]) \rangle =  (2\pi)^3 \delta^3\Big(\sum _{a=1}^n \vecs[k,a] \Big) \, \langle \zeta(\vecs[k,1]) \cdots \zeta(\vecs[k,n]) \rangle'.
\ee
Similarly, for special conformal transformations, we have the Ward identity
\begin{equation}
\begin{split}
\label{wardsctg}
\big\langle \delta(\zeta(\vecs[k,1])) \cdots \zeta(\vecs[k,n]) \big\rangle &+ \cdots + \big\langle \zeta(\vecs[k,1]) \cdots \delta(\zeta(\vecs[k,n])) \big\rangle = \\ &- 2 \bigg( \vec[b]\cdot\frac{\del}{\del \vecs[k,n+1]}\bigg) \frac{\langle \zeta(\vecs[k,1]) \cdots \zeta(\vecs[k,n+1]) \rangle}{\langle \zeta(\vecs[k,n+1]) \zeta(-\vecs[k,n+1])\rangle'} \, \bigg|_{\vecs[k,n+1] \rightarrow 0},
\end{split}
\end{equation}
where $\delta(\zeta(\vec[k]))$ is given by
\begin{equation}
\begin{split}
\label{comchz}
\delta(\zeta(\vec[k])) = \widehat{\mathcal{L}}^{\,\vec[b]}_{\vec[k]} \, \zeta(\vec[k]) &+ 6\, b^m k^i \int \frac{d^3\tilde{k}}{(2\pi)^3}\, \frac{1}{\tilde{k}^2}\, \zeta(\vec[k] - \tilde{\vec[k]}) \, \widehat\gamma_{im}(\tilde{\vec[k]}) \\ &+ 2 \, b^m k^i \int \frac{d^3\tilde{k}}{(2\pi)^3}\, \frac{1}{\tilde{k}^2} \, \widehat\gamma_{ij}(\vec[k] - \tilde{\vec[k]}) \, \widehat\gamma_{jm}(\tilde{\vec[k]}),
\end{split}
\end{equation}
and the operator $\widehat{\mathcal{L}}^{\,\vec[b]}_{\vec[k]}$ is given by
\be
\label{deflh}
\widehat{\mathcal{L}}^{\,\vec[b]}_{\vec[k]} = 2 \, \Big( \vec[k] \cdot \frac{\del}{\del \vec[k]}\Big) \Big( \vec[b] \cdot \frac{\del}{\del \vec[k]} \Big) - (\vec[b]\cdot\vec[k]) \Big( \frac{\del}{\del \vec[k]} \cdot \frac{\del}{\del \vec[k]} \Big) + 6 \bigg( \vec[b] \cdot \frac{\del}{\del \vec[k]} \bigg).
\ee
Here, we have reproduced the Ward identities satisfied by the correlation functions of $\zeta$. Similar Ward identities are satisfied by the correlation functions of the tensor perturbation $\widehat\gamma_{ij}$ as well. For details, see \cite{Kundu:2015xta}.

We would now like to check the validity of these Ward identities for more general single field models of inflation, where the matter part of the Lagrangian density is an arbitrary function of the scalar field $\phi$ and its first derivatives. These models of inflation follow from the action (see \cite{ArmendarizPicon:1999rj, Garriga:1999vw})
\be
\label{genmact}
S = \half \int d^4x \sqrt{-g} \, \big[ M_{Pl}^2 R + 2 P(X,\phi)\big],
\ee
where
\be
\label{defx}
X = - \, 
\half \, g^{\mu \nu} \del_\mu\phi \del_\nu\phi,
\ee
and $P(X,\phi)$ is an arbitrary function of $X, \phi$. The speed of sound parameter $c_s$ which characterizes these general single field models of inflation is defined as
\be
\label{defsound}
c_s^{\,2} \equiv \frac{P_{,X}}{P_{,X} + 2 X P_{,XX}}.
\ee
Clearly, for the canonical slow-roll model of inflation, where $P(X,\phi) = X - V(\phi)$, the speed of sound is $c_s = 1$. For models with more general form of the matter Lagrangian $P(X,\phi)$ than the canonical slow-roll model, we have $c_s \neq 1$.

In these general models of single field inflation, one defines three ``slow-variation parameters,'' given by $\epsilon_1$, eq.\eqref{svp1}, and
\be
\label{svp2}
\eta_1 = \frac{\dot{\epsilon}_1}{\epsilon_1 H} ,
\ee
\be
\label{svp3}
s = \frac{\dot{c}_s}{c_s H}.
\ee
For inflation to occur, the three slow-variation parameters must be small,
\be
\label{svc}
\epsilon_1, \eta_1, s \ll 1.
\ee
However, the parameter $\epsilon$, eq.\eqref{eps}, which is small in the canonical slow-roll model of inflation, need not be small for the more general models.\footnote{Consider, for instance, the DBI model of inflation \cite{Silverstein:2003hf, Alishahiha:2004eh}. In this model, one has
\begin{equation*}
P(X,\phi) = - \frac{1}{f(\phi)} \sqrt{1 - 2 X f(\phi)} + \frac{1}{f(\phi)} - V(\phi),
\end{equation*}
where the inflaton $\phi$ is the position of a D3 brane moving in a warped throat, and $f(\phi)$ is the warping factor. The energy density and pressure for this model are given by
\begin{equation*}
\rho = \frac{1}{f(\phi)\sqrt{1-2Xf(\phi)}} - \frac{1}{f(\phi)} + V(\phi),
\end{equation*}
and
\begin{equation*}
p = - \frac{1}{f(\phi)} \sqrt{1 - 2 X f(\phi)} + \frac{1}{f(\phi)} - V(\phi).
\end{equation*}
The speed of sound can be calculated using eq.\eqref{defsound}, and is given by
\begin{equation*}
c_s = \sqrt{1 - 2Xf(\phi)}.
\end{equation*}
Working in the homogeneous limit, the inflaton becomes purely a function of time, $\phi \equiv \phi(t)$. We then have $X = \dot{\phi}^{\,2}/2$. The speed of sound then becomes
\begin{equation*}
c_s = \sqrt{1 -\dot{\phi}^{\,2} f(\phi)}.
\end{equation*}
In this model, the inflaton evolves relativistically, and the parameter $\epsilon$ defined in eq.\eqref{eps} is not small. Consequently, the speed of sound is very small, $c_s \ll 1$. This gives an approximate expression for $\dot\phi$,
\begin{equation*}
|\dot\phi| \approx \frac{1}{\sqrt{f(\phi)}}.
\end{equation*}
Also, the Friedmann and continuity equations have the form
\begin{equation*}
\begin{split}
&3 M_{Pl}^{2} H^2 = \rho,\\
&\dot{\rho} = -3H(\rho + p).
\end{split}
\end{equation*}
Using these equations, one finds that the expression for the slow-variation parameter $\epsilon_1$, defined in eq.\eqref{svp1}, is given by
\begin{equation*}
\begin{split}
\epsilon_1 &= \frac{3 {\dot\phi}^2}{2\left( \frac{1}{f}(1-c_s) + c_s V \right)} \\
&\approx \frac{3}{2} \frac{1}{1+c_s f V},
\end{split}
\end{equation*}
where the approximate expression follows from the condition that $c_s \ll 1$. To get a de Sitter like phase of exponential expansion, one must have $\epsilon_1 \ll 1$. Thus, the potential must satisfy the condition
\begin{equation*}
2 c_s f V \gg 1.
\end{equation*}
}

The two, three and four point functions for the curvature perturbation $\zeta$ in these models of inflation have been computed explicitly. For our purpose, we follow the references \cite{Chen:2006nt, Chen:2009bc}. We have reproduced their results in appendix \ref{chenres} for completeness.

For the case of $n=2$, the scaling Ward identity eq.\eqref{wardscg} becomes,
\begin{equation}
\begin{split}
\label{mcc1}
\bigg( 3 + \sum_{a=1}^2 k_a \, \frac{\del}{\del k_a} \bigg)& \langle \zeta(\vecs[k,1]) \zeta(\vecs[k,2]) \rangle' = - \, \frac{\langle \zeta(\vecs[k,1]) \zeta(\vecs[k,2]) \zeta(\vecs[k,3]) \rangle'}{\langle \zeta(\vecs[k,3]) \zeta(-\vecs[k,3])\rangle'} \bigg|_{\vecs[k,3] \rightarrow 0},
\end{split}
\end{equation} 
which is also known as the Maldacena consistency condition. An explicit check for its validity in the $P(X,\phi)$ models of inflation was performed in \cite{Chen:2006nt}, and it was found to hold true. 

We now check the Ward identities eqs.\eqref{wardscg} and \eqref{wardsctg} for the case of $n = 3$. Their explicit form is
\begin{equation}
\begin{split}
\label{wscg}
\bigg( 6 + \sum_{a=1}^3 k_a \, \frac{\del}{\del k_a} \bigg)& \langle \zeta(\vecs[k,1]) \zeta(\vecs[k,2]) \zeta(\vecs[k,3]) \rangle' = - \, \frac{\langle \zeta(\vecs[k,1]) \zeta(\vecs[k,2]) \zeta(\vecs[k,3]) \zeta(\vecs[k,4]) \rangle'}{\langle \zeta(\vecs[k,4]) \zeta(-\vecs[k,4])\rangle'} \bigg|_{\vecs[k,4] \rightarrow 0},
\end{split}
\end{equation} 
for the scaling transformation, and for the special conformal transformation we have
\begin{equation}
\begin{split}
\label{wsctg}
\bigg(\sum_{a=1}^3 \widehat{\mathcal{L}}_{\, \vecs[k,a]}^{\, \vec[b]}\bigg) \langle\zeta(\vecs[k,1]) \zeta(\vecs[k,2]) \zeta(\vecs[k,3])\rangle = - 2 \bigg( \vec[b]\cdot\frac{\del}{\del \vecs[k,4]}\bigg) \frac{\langle \zeta(\vecs[k,1]) \zeta(\vecs[k,2])  \zeta(\vecs[k,3]) \zeta(\vecs[k,4]) \rangle}{\langle \zeta(\vecs[k,4]) \zeta(-\vecs[k,4])\rangle'} \, \bigg|_{\vecs[k,4] \rightarrow 0}.
\end{split}
\end{equation}
Using the expressions for the three and four point functions given in appendix \ref{chenres}, we performed a check of these Ward identities on Mathematica. We find that the scaling Ward identity  is met, since the LHS and RHS of eq.\eqref{wscg} vanish individually. Similarly, we find that the special conformal Ward identity, eq.\eqref{wsctg}, is also met. As discussed in appendix \ref{chenres}, the four point function has two parts, one coming from a contact interaction term and the other from an intermediate scalar exchange. An interesting point to note is that the dominant contribution to the RHS of eq.\eqref{wsctg} in the limit $\vecs[k,4] \rightarrow 0$ comes from the four point contact interaction term, whereas the intermediate scalar exchange contribution is subleading.
Let us note before we conclude this section that the special conformal Ward identity eq.\eqref{wsctg} above is a generalization of the identity given in eq.(37) of \cite{Creminelli:2012ed}. Eq.\eqref{wsctg} reduces to the identity in \cite{Creminelli:2012ed} if one substitutes $\vec[b] \propto \vecs[k,4]$.

\section{The Probe Approximation: A Massless Interacting Scalar Field in de Sitter Space}
\label{intml}
We now proceed to check the validity of the Ward identities for a family of vacuum states which respect conformal invariance, called the ``$\alpha$-vacua'' states, \cite{PhysRevD.32.3136}. The properties of these vacuum states were also investigated in \cite{PhysRevD.31.754, Bousso:2001mw, Kaloper:2002cs, Goldstein:2003ut, Kundu:2011sg, Kundu:2013gha}. Our eventual aim is to investigate the possibility of having the $\alpha$-vacua as initial states in the inflationary scenario. However, in the present section, we consider the simpler case of a massless scalar field in de Sitter space with a cubic interaction term. This helps us in defining the interacting $\alpha$-vacua states. We then compute the three point function for the probe scalar field in an interacting $\alpha$-vacua state, and check for the conformal invariance of the resulting answer. The more interesting case of $\alpha$-vacua states in inflation is considered in the next section.

\subsection{Defining the $\alpha$-vacua}
\label{defalps}
We would like to first introduce the $\alpha$-vacua states in a free theory. These vacuum states can be most conveniently defined in terms of mode functions which are a generalization of the mode functions for a field in the Bunch-Davies vacuum state. We start by considering a massless free scalar field in de Sitter space. The action is given by
\begin{equation}
\begin{split}
\label{mlact}
S &= - \, \half \int \mes[x]\, d\eta \, \sqrt{-g} \, g^{\mu\nu} \del_{\mu}\varphi \, \del_{\nu}\varphi \\
&= \half \int \mes[x]\, d\eta \, \frac{1}{\eta^2 H^2} \left((\del_\eta \varphi)^2 - (\del_i \varphi)^2\right),
\end{split}
\end{equation}
where we have made use of the de Sitter metric given in eq.\eqref{meteta}. The equation of motion for $\varphi$ can be obtained by varying the action eq.\eqref{mlact}. It is given by
\be
\label{eqmvarp}
\varphi'' -\frac{2}{\eta} \, \varphi' - \del^2\varphi = 0,
\ee
where a $'$ denotes a derivative with respect to the conformal time $\eta$, $' \equiv \del/\del\eta$. On quantization, the mode expansion for the scalar field $\varphi$ in the Bunch-Davies vacuum is given by
\be
\label{bdmode1}
\varphi(\eta, \vec[x]) = \int \frac{d^3k}{(2\pi)^3} \Big[ a_{\vec[k]} \, u_{k}(\eta) + a^{\dagger}_{-\vec[k]} \, u_{k}^*(\eta) \Big] \text{e}^{i\vec[k]\cdot\vec[x]},
\ee
where the mode functions $u_k(\eta)$ are given by
\be
\label{bdmode2}
u_k(\eta) = \frac{H}{\sqrt{2 k^3}} (1+ik\eta) \, \text{e}^{-ik\eta}.
\ee
The free Bunch-Davies vacuum state $|0\rangle$ itself is defined as the state that gets annihilated by the operator $a_{\vec[k]}$,
\be
\label{}
a_{\vec[k]} |0\rangle = 0 \, \forall \, \vec[k].
\ee
The operators $a_{\vec[k]}, a^\dagger_{\vec[k]}$ satisfy the canonical commutation relations
\be
\label{cancombd}
\big[a_{\vec[k]}, a_{\vec[p]}^\dagger\big] = (2\pi)^3 \delta^3(\vec[k] - \vec[p]) , \, \, \big[a_{\vec[k]}, a_{\vec[p]}\big] = 0, \, \,  \big[a_{\vec[k]}^\dagger, a_{\vec[p]}^\dagger\big] = 0.
\ee

In \cite{PhysRevD.32.3136}, a two real parameter family of vacuum states for the free field $\varphi$ invariant under the connected de Sitter group was constructed. We denote such two parameter vacuum states by  $|\alpha,\beta\rangle$. The mode expansion for the massless free scalar field in an $|\alpha,\beta\rangle$ vacuum state is given by
\be
\label{newmodep}
\varphi(\eta, \vec[x]) = \int \frac{d^3k}{(2\pi)^3} \Big[ b_{\vec[k]} \, \tilde{u}_{k}(\eta) + b^{\dagger}_{-\vec[k]} \, \tilde{u}_{k}^*(\eta) \Big] \text{e}^{i\vec[k]\cdot\vec[x]},
\ee
where the mode functions $\tilde{u}_k(\eta)$ are given by
\be
\label{model}
\tilde{u}_k(\eta) = \frac{H}{\sqrt{2 k^3}} \left\{A (1-ik\eta) \, \text{e}^{ik\eta} + B (1+ik\eta) \, \text{e}^{-ik\eta} \right\},
\ee
with the normalization condition\footnote{The condition eq.\eqref{ipcond}, when combined with eq.\eqref{cancom2}, leads to the standard commutation relation between the field $\varphi$ and its conjugate momentum $\pi$, given by $[\varphi(\eta, \vec[x]), \pi(\eta,\vec[x]')] = i \delta^3(\vec[x] - \vec[x]').$}
\be
\label{ipcond}
|B|^2 - |A|^2 = 1.
\ee
The choice of parametrization that satisfies eq.\eqref{ipcond} and respects de Sitter invariance is (see \cite{PhysRevD.32.3136})
\begin{equation}
\label{paramet}
\begin{split}
A &= - i e^{i \beta} \text{sinh}(\alpha), \\
B &= \text{cosh}(\alpha).
\end{split}
\end{equation}
Thus, for every choice of values for the two real parameters $(\alpha, \beta)$, we get a unique set of mode functions $\tilde{u}_k(\eta)$, eq.\eqref{model}, with the coefficients $A,B$ determined from eq.\eqref{paramet}. The operators $b_{\vec[k]}, b^\dagger_{\vec[k]}$ appearing in the mode expansion act as the annihilation and creation operators, respectively, for the corresponding $|\alpha,\beta\rangle$ vacuum state. In particular,
\be
\label{defb2}
b_{\vec[k]} |\alpha,\beta\rangle = 0 \, \forall \, \vec[k],
\ee
with $b_{\vec[k]}, b_{\vec[k]}^\dagger$ satisfying the canonical commutation relations
\be
\label{cancom2}
\big[b_{\vec[k]}, b_{\vec[p]}^\dagger\big] = (2\pi)^3 \delta^3(\vec[k] - \vec[p]) , \, \, \big[b_{\vec[k]}, b_{\vec[p]}\big] = 0, \, \,  \big[b_{\vec[k]}^\dagger, b_{\vec[p]}^\dagger\big] = 0.
\ee

From eq.\eqref{paramet}, we see that the Bunch-Davies vacuum is a special case of the $|\alpha,\beta\rangle$ vacuum states, with $\alpha = 0$. More explicitly, the vacua $|\alpha, \beta \rangle$ are related to the Bunch-Davies vacuum $|0\rangle$ by\footnote{See \cite{Mukhanov:2007zz} for a pedagogical discussion.}
\be
\label{vacrel}
|\alpha, \beta \rangle = \prod_{\vec[k]}\,\frac{1}{\sqrt{|B|}} \, \text{exp}\left( \, \frac{A^*}{2B^*} \, a^\dagger_{\vec[k]} a^\dagger_{-\vec[k]}\right)\,|0\rangle.
\ee
We can see that the $|\alpha, \beta \rangle$ vacuum state contains excitations of the Bunch-Davies vacuum in pairs of particles which carry equal and opposite momenta. One can explicitly check that the state $|\alpha, \beta \rangle$ satisfies
\be
\label{check1}
\hat{\vec[P]} |\alpha, \beta \rangle \equiv \int \frac{d^3p}{(2\pi)^3} \, \vec[p] \, a^\dagger_{\vec[p]} a_{\vec[p]} \, |\alpha, \beta \rangle = 0.
\ee

The two sets of creation and annihilation operators $a_{\vec[k]}, a^\dagger_{\vec[k]}$ and $b_{\vec[k]}, b^\dagger_{\vec[k]}$ are related to each other by the Bogolyubov transformations
\begin{equation}
\label{bogol}
\begin{split}
&b_{\vec[k]} = B^* a_{\vec[k]} - A^* a^\dagger_{-\vec[k]}, \\
&a_{\vec[k]} = B b_{\vec[k]} + A^* b^\dagger_{-\vec[k]}.
\end{split}
\end{equation}

In this paper, for simplicity and clarity, we will restrict ourselves to the case where $\beta=0$.
As argued in \cite{PhysRevD.32.3136}, to get de Sitter invariance in global de Sitter space,  we should set the parameter $\beta=0$. Strictly speaking, this is true only for a massive scalar. The massless case being considered here is different due to subtleties coming from zero modes. We will ignore these subtleties in this paper, since our primary interest is  modes with non-zero spatial momentum, $\vec[k] \neq 0$. Also, in  our discussion, we will  study  invariance only under the connected subgroup of the de Sitter symmetry group $O(1,4)$.

Setting $\beta=0$  gives a one real parameter family of full de Sitter invariant vacuum states denoted by $|\alpha\rangle$; these states are called the ``$\alpha$-vacua'' states. From eq.\eqref{model} and eq.\eqref{paramet}, the mode functions for the $\alpha$-vacua states are given by
\be
\label{modealp}
\tilde{u}_k(\eta) = \frac{H}{\sqrt{2 k^3}} \left\{\text{cosh}(\alpha) (1+ik\eta)\, \text{e}^{-ik\eta} - i\, \text{sinh}(\alpha) (1-ik\eta) \, \text{e}^{ik\eta} \right\}.
\ee

Following eq.\eqref{vacrel}, the $\alpha$-vacua states can be expressed in terms of the free Bunch-Davies vacuum $|0\rangle$ as
\be
\label{alphad}
|\alpha\rangle = \frac{1}{\mathcal{N}} \, \text{exp}\Big( \frac{i}{2} \, \text{tanh}(\alpha) \int \frac{\mes[k]}{(2\pi)^3} \, a^\dagger_{\vec[k]} a^\dagger_{-\vec[k]} \Big) |0\rangle,
\ee
where we have substituted the values of $A, B$ from eq.\eqref{paramet} into eq.\eqref{vacrel}, and set $\beta=0$. In eq.\eqref{alphad}, $\mathcal{N}$ is an overall normalization constant.

After defining the free $\alpha$-vacua states, we now move on to define the interacting $\alpha$-vacua states. Suppose that there is an interaction term for the field $\varphi$ in the action eq.\eqref{mlact}, and the theory is not free. We propose that for such an interacting theory, the corresponding interacting $\alpha$-vacua, denoted by $|\alpha\rangle_{\mathrm{I}}$, are given by an expression of the form similar to the one in eq.\eqref{alphad},
\be
\label{alphai}
|\alpha\rangle_{\mathrm{I}} = \frac{1}{\mathcal{N}} \, \text{exp}\Big( \frac{i}{2} \, \text{tanh}(\alpha) \int \frac{\mes[k]}{(2\pi)^3} \, a^\dagger_{\vec[k]} a^\dagger_{-\vec[k]} \Big) |\Omega\rangle,
\ee
where $|\Omega\rangle$ is the Bunch-Davies vacuum of the interacting theory. As we show in appendix \ref{isogen}, these vacua preserve  conformal invariance.\footnote{There are some subtleties having to do with surface terms which appear in commutators for the special conformal transformations; see eq.\eqref{kbcomm3}. Also, we are once again ignoring subtleties having to do with the zero modes for the massless case, \cite{PhysRevD.32.3136, PhysRevD.35.3771}.}
 
Notice that the creation operators $a_{\vec[k]}^\dagger$ which appear in eq.\eqref{alphai} are defined in terms of the free field $\varphi$ through the mode expansion eq.\eqref{bdmode1}. 
This expansion is valid in the interaction picture as well, as the interaction picture field $\varphi_{I}$ behaves like a free field. Thus eq.\eqref{alphai} can be taken to be the definition of the $|\alpha\rangle_{\mathrm{I}}$ vacuum in the interaction picture.

\subsection{The Three Point Function in the $|\alpha\rangle_{\mathrm{I}}$ Vacua}
\label{ininrev}
In this subsection, we consider the three point function which arises in an $\alpha$-vacuum for the interacting theory of a massless scalar field in de Sitter space, with a cubic self interaction term. The action is given by
\be
\label{intact}
S = - \int d^4x \, \sqrt{-g} \left( \half g^{\mu\nu} \del_{\mu}\varphi \, \del_{\nu}\varphi + \frac{\lambda}{3!} \, \varphi^3 \right),
\ee
with the metric given in eq.\eqref{dsp}. From the action eq.\eqref{intact}, we can read off the interaction Hamiltonian, which is given by
\begin{equation}
\label{intham}
H_{int} = \int \mes[x] ~ \text{e}^{3Ht} \, \frac{\lambda}{3!} \, \varphi^3.
\end{equation}

We will compute the three point function using the \textit{in-in formalism}. The general expression for the vacuum expectation value of an operator $\mathcal{O}$ at time $t$ in the in-in formalism is given by
\be
\label{inindef}
\langle \text{vac}| \mathcal{O} |\text{vac} \rangle(t) = \langle \text{vac}|\tilde{\mathcal{T}} \bigg( \text{e}^{i \int\limits_{t_0}^t H_{int}^I(t') \, dt'} \bigg) \mathcal{O}_I(t) \, \mathcal{T} \bigg(\text{e}^{-\, i \int\limits_{t_0}^t H_{int}^I(t') \, dt'} \bigg)|\text{vac} \rangle,
\ee
where the vacuum state $|\text{vac}\rangle$ is defined on an initial time slice located at $t = t_0$, $\mathcal{O}_I$ is the operator $\mathcal{O}$ written in the interaction picture, and $H_{int}^I$ is the interaction Hamiltonian in the interaction picture. $\mathcal{T}$ and $\tilde{\mathcal{T}}$ are the time and anti-time ordering operators, respectively. Working to the leading order in the coupling constant, eq.\eqref{inindef} reduces to
\be
\label{iired}
\langle \text{vac}| \mathcal{O} |\text{vac} \rangle(t) = \langle \text{vac}| \mathcal{O}_I(t) |\text{vac} \rangle - i \int\limits_{t_0}^t dt' \, \langle \text{vac}| \left[\mathcal{O}_I(t), H_{int}^I(t')\right] |\text{vac} \rangle.
\ee
For the purpose of calculations, it will be beneficial if we express the vacuum expectation value in terms of conformal time. The expression then becomes
\begin{equation}
\label{iicon}
\langle \text{vac}| \mathcal{O} |\text{vac} \rangle(\eta) = \langle \text{vac}| \mathcal{O}_I(\eta) |\text{vac} \rangle - i \int\limits_{\eta_0}^\eta d\eta'\, \langle \text{vac}| \left[\mathcal{O}_I(\eta), H_{int}^I(\eta')\right] |\text{vac} \rangle.
\end{equation}

We now come back to the case of our interest, eq.\eqref{intact}. Going to the interaction picture and converting to conformal time, we have
\be
\label{hintcorr}
H_{int}^I(\eta) = \int d^3x \, \frac{1}{\eta^4 H^4} \, \frac{\lambda}{3!} \, \varphi_I^3(\eta,\vec[x]).
\ee
Thus, following eq.\eqref{iicon}, the three point function of $\varphi$ is given by
\begin{equation}
\begin{split}
\label{tpfv}
\langle \text{vac}| \varphi(\vecs[k,1]) \varphi(\vecs[k,2]) \varphi&(\vecs[k,3]) |\text{vac} \rangle(\eta) = \langle \text{vac}| \varphi_I(\eta, \vecs[k,1]) \varphi_I(\eta, \vecs[k,2]) \varphi_I(\eta, \vecs[k,3]) |\text{vac} \rangle\\
&-i \, \frac{\lambda}{3!} \int\limits_{\eta_0}^\eta d\eta' \frac{1}{(\eta' H)^4} \int \bigg[ \prod_{a=1}^3 \frac{d^3p_a}{(2\pi)^3}\bigg] \, (2\pi)^3 \delta^3\Big(\sum_{a=1}^3 \vecs[p,a]\Big) \times\\ 
\langle \text{vac}| \Big[\varphi_I(\eta, \vecs[k,1]) &\varphi_I(\eta, \vecs[k,2]) \varphi_I(\eta, \vecs[k,3]), \varphi_I(\eta', \vecs[p,1]) \varphi_I(\eta', \vecs[p,2]) \varphi_I(\eta', \vecs[p,3])\Big] |\text{vac} \rangle.
\end{split}
\end{equation}
This is the expression which we will use for our computations.

\subsubsection{Computing the Three Point Function}
\label{allordc}
We now compute the three point function for the probe field $\varphi$ in the vacuum state $|\alpha\rangle_{\mathrm{I}}$, eq.\eqref{alphai}, using eq.\eqref{tpfv}. The mode expansion for the interaction picture free field $\varphi_I$ is given by eq.\eqref{newmodep}, with the modes $\tilde{u}_k(\eta)$ given by eq.\eqref{modealp}. We will replace  the interacting Bunch-Davies vacuum state $|\Omega\rangle$ on the RHS of eq.\eqref{alphai}  with the  free  vacuum $|0\rangle$, annihilated by $a_{\vec[k]}$, accompanied by a change in the contour for the $\eta$-integral in eq.\eqref{tpfv}, so that it is now carried out over complex values of $\eta$ given by
\be
\label{changet}
\eta\rightarrow \eta(1 \pm i \theta),
\ee
where $\theta$ is a positive infinitesimal parameter. The sign of the imaginary term in eq.\eqref{changet} is determined by the requirement that the resulting time evolution projects on to the ground state of the interacting theory, $|\Omega \rangle$. We see that after making this replacement, and the choice eq.\eqref{changet} of contour for the $\eta$-integral, the calculation in the $|\alpha\rangle_{\mathrm{I}}$ vacua reduces to one in free field theory. 

In effect, in making these changes, we are assuming that the  system can be prepared in the $|\alpha\rangle_{\mathrm{I}}$ vacuum in the far past in two steps. First, we start in the free vacuum $|0\rangle$  at $\eta=-\infty$ and evolve it to obtain the state $|\Omega\rangle$ (this is the standard procedure used for computations in the interacting Bunch-Davies vacuum); then we act with the operator $\text{exp}\Big( \frac{i}{2} \, \text{tanh}(\alpha) \int \frac{\mes[k]}{(2\pi)^3} \, a^\dagger_{\vec[k]} a^\dagger_{-\vec[k]} \Big)$ on this state to obtain $|\alpha\rangle_{\mathrm{I}}$, eq.\eqref{alphai}. We are not being very careful about separating these two steps in the procedure discussed above, and there could be a worry about this.  
The final result we get will turn out to be conformally invariant (up to contact terms),  and this is one reason to believe  that the procedure above works. 
In fact, the whole calculation turns out to be quite insensitive to this issue.  
In appendix \ref{3ptbd}, we show that for the $\varphi^3$ interaction being considered here, eq.\eqref{intham}, the free vacuum $|0\rangle$ and the interacting vacuum $|\Omega \rangle$ agree in the far past. 
We also find, in the calculations below, that the rotated contour of eq.\eqref{changet} does not turn out to be required  to ensure convergence of the integrals. We could have therefore, for purposes of this calculation, as well not have rotated the contour and simply replaced $|\Omega\rangle$ by $|0\rangle$ in the far past $\eta\rightarrow -\infty$. 

From eq.\eqref{tpfv}, with the vacuum $|\text{vac}\rangle$ now being $|\alpha\rangle$, along with the accompanying choice of the complex contour for the $\eta$-integral as discussed above, we get  after a straightforward calculation that at late time,
\begin{equation}
\begin{split}
\label{3ptapn}
{}_{\mathrm{I}}\langle \alpha |\varphi&(\vecs[k,1]) \varphi(\vecs[k,2]) \varphi(\vecs[k,3]) |\alpha\rangle_{\mathrm{I}} = \lambda H^2 \, (2\pi)^3 \delta^3\Big(\sum_{a=1}^3 \vecs[k,a]\Big) \Big[\prod_{a=1}^3 \frac{1}{(2k_a^{\,3})}\Big] \times \\
&\Bigg[ \bigg( \frac{\pi}{6} \,\text{sinh}(4\alpha) - \frac{8}{9} \, \text{cosh}^2(2\alpha)\bigg) \sum_{a=1}^3 k_a^{\,3} + \bigg( \frac{5}{3} - \text{cosh}(4\alpha)\bigg) k_1k_2k_3 \\
&\hspace{1mm} -\frac{2}{3} \, \text{cosh}^2(2\alpha) \sum_{a \neq b} k_a k_b^{\,2} + \frac{1}{3} \, \big[\tilde{S} + \tilde{S}^*\big] \sum_{a=1}^3 k_a^{\,3}\\
&\hspace{1mm}+\frac{1}{3} \, \text{sinh}^2(2\alpha) \bigg(  \big[V_1 + V_1^*\big] \big(-k_1^{\,3}+k_2^{\,3}+k_3^{\,3}\big)\\
&\hspace{25mm} + \big[V_2 + V_2^*\big] \big(k_1^{\,3}-k_2^{\,3}+k_3^{\,3}\big) + \big[V_3 + V_3^*\big] \big(k_1^{\,3}+k_2^{\,3}-k_3^{\,3}\big) \bigg) \Bigg],
\end{split}
\end{equation}
where $\tilde{S}$ is given by
\be
\label{stformc}
\tilde{S} = \int\limits_{-\infty(1-i\theta)}^0\, \frac{d\eta}{\eta} ~\text{e}^{i K\eta},
\ee
with $K = k_1 + k_2 + k_3$, and $V_a, a=(1,2,3) $ are given by
\be
\label{v1}
V_a = \int\limits_{-\infty(1-i \,\theta)}^0 \frac{d\eta}{\eta} ~ \text{e}^{i(K - 2k_a)\eta} .
\ee
Notice that if we set $\alpha = 0$, eq.\eqref{3ptapn} reproduces the result for the three point function in the Bunch-Davies vacuum. This is discussed  in ref \cite{Kundu:2014gxa}; see also appendix \ref{3ptbd}.

The integrals eqs.\eqref{stformc} and \eqref{v1} are divergent as $\eta\rightarrow 0$. Introducing a cut-off $\varepsilon$, where $\varepsilon \rightarrow 0^-$, gives 
\be
\label{valS}
\tilde{S} \approx \int\limits_{-\infty(1-i\theta)}^\varepsilon \, \frac{d\eta}{\eta} ~ \text{e}^{i K\eta} = - \, \Gamma[0,-iK\varepsilon] = \gamma + ln(-iK\varepsilon) + O(\varepsilon) ,
\ee
where $\gamma$ is the Euler-Mascheroni constant. Similarly for  $V_a, a=(1,2,3)$ we get
\be
\label{defVa}
V_a \approx \int\limits_{-\infty(1-i \,\theta)}^\varepsilon \frac{d\eta}{\eta} ~ \text{e}^{i(K - 2k_a)\eta} = - \, \Gamma[0,-i(K-2k_a)\varepsilon] = \gamma + ln[-i(K-2k_a)\varepsilon] + O(\varepsilon).
\ee
Note that $\vecs[k,1], \vecs[k,2], \vecs[k,3]$ form the three sides of a triangle, with lengths $k_1, k_2, k_3,$ respectively. The triangle inequality ensures that $k_1+k_2>k_3$ etc, so the quantity $(K-2k_a)$ appearing in the expression for $V_a$ is always positive. Using eqs.\eqref{valS} and \eqref{defVa}, the result eq.\eqref{3ptapn} can now be written as
\begin{equation}
\begin{split}
\label{3pta}
&{}_{\mathrm{I}}\langle \alpha |\varphi(\vecs[k,1]) \varphi(\vecs[k,2]) \varphi(\vecs[k,3]) |\alpha\rangle_{\mathrm{I}} = \lambda H^2 \, (2\pi)^3 \delta^3\Big(\sum_{a=1}^3 \vecs[k,a]\Big) \Big[\prod_{a=1}^3 \frac{1}{(2k_a^{\,3})}\Big] \times \\
&\Bigg[ \bigg( \frac{\pi}{6} \,\text{sinh}(4\alpha) - \frac{8}{9} \, \text{cosh}^2(2\alpha)\bigg) \sum_{a=1}^3 k_a^{\,3} + \bigg( \frac{5}{3} - \text{cosh}(4\alpha)\bigg) k_1k_2k_3 \\
&\hspace{1mm} -\frac{2}{3} \, \text{cosh}^2(2\alpha) \sum_{a \neq b} k_a k_b^{\,2} + \frac{2}{3} \big(\gamma + ln|\varepsilon|\big)\, \text{cosh}^2(2\alpha) \sum_{a=1}^3 k_a^{\,3} + \frac{2}{3} \Big(\sum_{a=1}^3 k_a^{\,3}\Big) ln(K)\\
&\hspace{1mm}+\frac{2}{3} \, \text{sinh}^2(2\alpha) \Big[ \big(-k_1^{\,3} + k_2^{\,3} + k_3^{\,3}\big) \, ln(-k_1 + k_2 +k_3)\\
&\qquad+  \big(k_1^{\,3} - k_2^{\,3} + k_3^{\,3}\big) \, ln(k_1 - k_2 +k_3) + \big(k_1^{\,3} + k_2^{\,3} - k_3^{\,3}\big) \, ln(k_1 + k_2 -k_3)\Big]\Bigg],
\end{split}
\end{equation}
Eq.\eqref{3pta}  is the final expression for the three point function of the probe field. We next discuss the symmetry properties of this result.  

\subsubsection{Checking for Conformal Invariance}
It is clear that the presence  of the cut-off $\varepsilon$ in eq.\eqref{3pta} violates scale invariance in the theory. This feature can be seen in the Bunch-Davies vacuum itself, which corresponds to setting $\alpha=0$ in eq.\eqref{3pta}, giving  
\begin{equation}
\begin{split}
\label{ibdmt}
\langle \Omega| \varphi(\vecs[k,1]) &\varphi(\vecs[k,2]) \varphi(\vecs[k,3]) |\Omega \rangle = \frac{2}{3} \, \lambda H^2 \, (2\pi)^3 \delta^3\Big(\sum_{a=1}^3 \vecs[k,a]\Big) \Big[\prod_{a=1}^3 \frac{1}{2k_a^{\,3}}\Big] \times\\
&\Bigg[ -\frac{4}{3} \sum_{a=1}^3 k_a^{\,3} -\sum_{a \neq b} k_a k_b^{\,2} + k_1k_2k_3 + \Big(\gamma+ln|\varepsilon|+ln(K)\Big) \Big(\sum_{a=1}^3 k_a^{\,3}\Big) \Bigg].
\end{split}
\end{equation}
In the more general case of $\alpha$-vacua, there are  additional terms proportional to $V_a$ in eq.\eqref{3ptapn} which  give rise to additional logarithmic pieces, as can be easily seen by comparing eq.\eqref{3pta} with  eq.\eqref{ibdmt}. These again violate  scale invariance.

The symmetry properties of the three point function in eq.\eqref{3pta} can be conveniently studied by examining the coefficient function for the cubic term in the late time wave function. The wave function $\Psi[\varphi]$ has the schematic form
\be
\label{wfunsch1}
\Psi[\varphi] = \text{exp}\bigg(-\half \int \varphi \varphi \, \langle OO \rangle + \frac{1}{3!} \int \varphi \varphi \varphi \, \langle OOO \rangle + \cdots\bigg),
\ee
where
\be
\label{oo1}
\langle O(\vecs[k,1]) O(\vecs[k,2])\rangle = (2\pi)^3 \delta^3(\vecs[k,1]+\vecs[k,2]) \, \frac{k_1^{\,3}}{H^2}\, \frac{1}{ \text{cosh}(2\alpha)},
\ee
and $\langle OOO \rangle, \ldots$ are the coefficient functions.
Note that the operators $O$ behave as marginal scalar operators in a three dimensional Euclidean CFT.
From eqs.\eqref{3pta} and \eqref{wfunsch1}, we can deduce that the coefficient function $\langle OOO \rangle$ has the form
\begin{equation}
\begin{split}
\label{oooav}
&\langle O(\vecs[k,1]) O(\vecs[k,2]) O(\vecs[k,3]) \rangle' = \frac{\lambda}{2H^4}\, \frac{1}{\text{cosh}^3(2\alpha)} \,\Bigg[ \bigg( \frac{\pi}{6} \,\text{sinh}(4\alpha) - \frac{8}{9} \, \text{cosh}^2(2\alpha)\bigg) \sum_{a=1}^3 k_a^{\,3} \\
&+ \bigg( \frac{5}{3} - \text{cosh}(4\alpha)\bigg) k_1k_2k_3 - \frac{2}{3} \, \text{cosh}^2(2\alpha) \sum_{a \neq b} k_a k_b^{\,2} + \frac{2}{3} \big(\gamma + ln|\varepsilon|\big)\, \text{cosh}^2(2\alpha) \sum_{a=1}^3 k_a^{\,3} \\
&+\frac{2}{3} \Big(\sum_{a=1}^3 k_a^{\,3}\Big) ln(K)+\frac{2}{3} \, \text{sinh}^2(2\alpha) \Big[ \big(-k_1^{\,3} + k_2^{\,3} + k_3^{\,3}\big) \, ln(-k_1 + k_2 +k_3)\\
&\qquad\qquad+  \big(k_1^{\,3} - k_2^{\,3} + k_3^{\,3}\big) \, ln(k_1 - k_2 +k_3) + \big(k_1^{\,3} + k_2^{\,3} - k_3^{\,3}\big) \, ln(k_1 + k_2 -k_3)\Big] \Bigg].
\end{split}
\end{equation}
From the RHS above, we see that the terms which go like $ln|\varepsilon|$ and break scale invariance are contact terms, since they are proportional to $\sum_a k_a^{\,3}$. 
Up to these contact terms, the result preserves the Ward identities for conformal invariance. 

In fact, the Ward identity for special conformal transformations is met identically, 
\be
\label{sctwa}
\Big( \sum_{a=1}^3 \mathcal{L}^{\,\vec[b]}_{\vecs[k,a]} \Big) \langle O(\vecs[k,1]) O(\vecs[k,2]) O(\vecs[k,3]) \rangle' = 0,
\ee
where the operator $\mathcal{L}^{\vec[b]}_{\vec[k]}$ is given by
\begin{equation}
\begin{split}
\label{defl}
\mathcal{L}^{\,\vec[b]}_{\vec[k]} &= 2 \Big( \vec[k]\cdot\frac{\del}{\del\vec[k]} \Big) \Big( \vec[b]\cdot\frac{\del}{\del\vec[k]} \Big) - (\vec[b]\cdot\vec[k]) \Big(\frac{\del}{\del\vec[k]} \cdot\frac{\del}{\del\vec[k]} \Big) \\
&= (\vec[b]\cdot\vec[k]) \Big( - \frac{2}{k} \frac{\del}{\del k} + \frac{\del^2}{\del k^2} \Big).
\end{split}
\end{equation}
On the other hand, the Ward identity for scale invariance is 
\begin{equation}
\begin{split}
\label{scc2}
\bigg( \sum_{a=1}^3 \vecs[k,a]\cdot\frac{\del}{\del\vecs[k,a]}\bigg) &\langle O(\vecs[k,1]) O(\vecs[k,2]) O(\vecs[k,3]) \rangle' \\
&= 3 \,\langle O(\vecs[k,1]) O(\vecs[k,2]) O(\vecs[k,3]) \rangle' + \frac{\lambda}{3H^4 \, \text{cosh}(2\alpha)} \, \Big(\sum_{a=1}^3 k_a^{\,3}\Big),
\end{split}
\end{equation}
with the extra term on the RHS arising from the violation of scale invariance mentioned above. 

In the Bunch-Davies case, eq.\eqref{oooav} with $\alpha = 0$, we have
\begin{equation}
\begin{split}
\label{ooobdv}
\langle O(\vecs[k,1]) O(\vecs[k,2]) O(\vecs[k,3]) \rangle' =\frac{\lambda}{3H^4} \bigg[- \frac{4}{3}\sum_{a=1}^3 k_a^{\,3} + k_1 k_2 k_3 -\sum_{a \neq b} k_a k_b^{\,2} + \big(\gamma + ln\big(|\varepsilon|K\big)\big) \sum_{a=1}^3 k_a^{\,3} \bigg].
\end{split}
\end{equation}
The result in eq.\eqref{ooobdv} is consistent with the Operator Product Expansion (OPE), and this in fact explains why the $\log$ violation of scale invariance occurs in momentum space. 
This is discussed in appendix B of \cite{Kundu:2014gxa}. Briefly, the OPE for two marginal operators gives 
\be
\label{OPE1}
O(0) O(\vec[x]) = \frac{A}{x^3} \, O(0) + \cdots \, .
\ee
Now, the coefficient function in momentum space is related to the position space expression via a Fourier transform,
\begin{equation}
\begin{split}
\label{arge}
\langle O(\vecs[k,1]) O(\vecs[k,2]) &O(\vecs[k,3])\rangle = \int d^3x_1 \, d^3x_2 \, d^3x_3 \, e^{-\,i (\sum_{a} \vecs[k,a]\cdot\vecs[x,a])} \langle O(\vecs[x,1]) O(\vecs[x,2]) O(\vecs[x,3])\rangle \\
&= (2\pi)^3 \, \delta^3\big(\sum_{a=1}^3 \vecs[k,a]\big) \int d^3x \, d^3y \, e^{-\,i(\vecs[k,2]\cdot\vec[x] + \vecs[k,3]\cdot\vec[y])} \langle O(0) O(\vec[x]) O(\vec[y])\rangle\\
&\approx (2\pi)^3 \, \delta^3\big(\sum_{a=1}^3 \vecs[k,a]\big) \bigg(\int d^3x\, e^{-\,i\vecs[k,2]\cdot\vec[x]} \frac{A}{x^3} \bigg) \bigg( \int d^3y \, e^{-\,i \vecs[k,3]\cdot\vec[y]} \, \frac{1}{H^2}\frac{1}{y^6}\bigg),
\end{split}
\end{equation}
where in the second step we have made use of the notation $\vecs[x,2]-\vecs[x,1] = \vec[x]$ and $\vecs[x,3]-\vecs[x,1] = \vec[y]$, while in the third step we have used the OPE eq.\eqref{OPE1}, and the position space analog of eq.\eqref{oo1} with $\alpha = 0$, given by
\be
\label{oopa}
\langle O(\vec[x]) O(\vec[y]) \rangle = \frac{1}{H^2} \, \frac{1}{|\vec[x] - \vec[y]|^6}.
\ee
From eq.\eqref{arge}, we see that the Fourier transform is not well defined and a cut-off needs to be introduced to carry it out,
\be
\int d^3x \, e^{-i\vecs[k,2]\cdot\vec[x]} \frac{1}{x^3} \sim ln(\varepsilon k_2), \, \text{where} ~\varepsilon \rightarrow 0.
\label{ldk2}
\ee
This gives rise to the violation of scale invariance in momentum space.
Making use of eq.\eqref{ldk2} and the fact that $O(\vec[k])$ has mass dimension $-1$, eq.\eqref{arge} gives
\be
\langle O(\vecs[k,1]) O(\vecs[k,2]) O(\vecs[k,3])\rangle' \sim \frac{A}{H^2} \, k_3^{\, 3} \, ln(\varepsilon k_2).
\label{tpl1}
\ee
The result eq.\eqref{tpl1} agrees with the limiting behaviour obtained from eq.\eqref{ooobdv}, upto additional contact terms, in the limit when $k_1 \approx k_2 \gg k_3$.

The general solutions to the Ward identities of conformal invariance were studied in \cite{Coriano:2013jba, Bzowski:2013sza, Kundu:2014gxa}. It was found that there are four independent solutions, all of which violated scale invariance upto contact terms. These are of the form given in eq.(B.22) of \cite{Kundu:2014gxa}. Requiring that the OPE in eq.\eqref{OPE1} is valid gives rise to a unique answer, which is the Bunch-Davies result, eq.\eqref{ooobdv}. The result we have obtained for the three point function for a general $\alpha$-vacuum can be shown to be a linear combination of the four solutions mentioned above. 
However, it does not agree with the OPE in eq.\eqref{OPE1}. In fact, one can show that the additional terms present for $\alpha$-vacua cannot be reproduced by assuming that there is an operator product expansion for the operators $O$. For instance,  in the limit $k_1 \approx k_2 \gg k_3$, the leading non-contact term for the $\alpha$-vacua result in eq.\eqref{oooav}  is
\be
\label{anc}
\frac{\lambda}{H^4} \, \frac{\sinh^2(2\alpha)}{\cosh^3(2\alpha)} \, k_2 (\vecs[k,2]\cdot\vecs[k,3]) \, ln \bigg(\frac{k_2 k_3 + \vecs[k,2]\cdot\vecs[k,3]}{k_2 k_3 - \vecs[k,2]\cdot\vecs[k,3]}\bigg). 
\ee
The  OPE  cannot  give rise to such a term.  Instead, if the OPE is valid, the  leading non-contact term in this limit  must be the product of two functions, one only dependent on $\vecs[k,2]$ and the other only on  $\vecs[k,3]$, analogous  to what happens in eq.\eqref{tpl1}.

Let us end this section with a few more comments. 
First, from the point of view of a possible dS/CFT duality, it would be interesting to ask what kind of a field theory would be dual to the $\alpha$-vacua, in view of the fact that the three point function has the more general form found above, which is not consistent with the existence of an operator product expansion. 
Second, it might seem surprising at first that the form of the three point function is not uniquely determined by conformal invariance, and the Ward identities allow for four linearly independent solutions. In position space it is certainly the case, that upto possible contact terms, there is a unique solution to the Ward identities.
However, it does not follow automatically that the Ward identities in momentum space give the same number of linearly independent solutions.
This is true only if the correct boundary conditions are imposed. Requiring the OPE to be valid  is one way to impose these boundary conditions. More generally, however, the Ward identities are first order equations in position space but second order in momentum space, and thus the number of independent solutions in the two cases can be different. It would be interesting to try and work out the form of the four independent solutions in position space, but the Fourier transforms are not easy to do and we will not attempt this here. 

Finally, in the probe approximation we have discussed above, we work in the limit $M_{Pl}\rightarrow \infty$, keeping the Hubble scale $H$ fixed. The quantum stress tensor in the $\alpha$-vacua actually diverges, as discussed in appendix \ref{back-reaction}. 
However, since $M_{Pl}\rightarrow \infty$, the back reaction on the geometry, in a formal sense, is suppressed in this approximation.

\section{Correlation Functions with $\alpha$-vacua in Inflation}
\label{infla}
Having dealt with the properties of correlation functions of scalar fields in de Sitter space, we now consider the more general scenario of inflation  with $\alpha$-vacua as initial states. We compute the two and three point functions for the scalar perturbation $\zeta$, and discuss the Maldacena consistency condition.

\subsection{Basic Setup}
\label{basicinf}
 Following the approach in \cite{Maldacena:2002vr}, we consider the single field slow-roll model of inflation, which is described by the action eq.\eqref{sinf}. With the ADM form of the metric, eq.\eqref{admmet}, the action eq.\eqref{sinf} takes the form
\begin{equation}
\begin{split}
\label{admsct}
S = \half \int d^4x \sqrt{h} \, \Big( N R^{(3)} - 2 NV &+ \frac{1}{N} (E_{ij}E^{ij} - E^2) \\&+ \frac{1}{N} (\dot{\phi} - N^i \del_i\phi)^2 - N h^{ij} \del_i\phi \, \del_j\phi \Big),
\end{split}
\end{equation}
where $R^{(3)}$ is the Ricci scalar curvature of the spatial slice, and\footnote{Note that the covariant derivative in eq.\eqref{extrinsic}, $\nabla_i$, is to be taken with respect to the 3-metric $h_{ij}$.}
\begin{equation}
\begin{split}
\label{extrinsic}
E_{ij} := N K_{ij} &= \half (\dot{h}_{ij} - \nabla_i N_j - \nabla_j N_i), \\
E &= h_{ij} E^{ij},
\end{split}
\end{equation}
with $K_{ij}$ being the extrinsic curvature. 

We need to make a gauge choice to fix the diffeomorphism invariance of the theory. For the derivation of the Ward identities, it is more convenient to work in the synchronous gauge, eq.\eqref{syncgauge}; see \cite{Kundu:2015xta} for a detailed discussion. However, for calculating the inflationary correlation functions, it is more convenient to choose one of the following two gauges, 
\begin{equation}
\begin{split}
\label{gaugedef}
&\text{Gauge 1}: \delta\phi = 0, \, \zeta \neq 0, \, \del_i\gamma_{ij} = 0, \, \text{or} \\
&\text{Gauge 2}: \delta\phi \neq 0,\, \zeta = 0, \, \del_i\gamma_{ij} = 0.
\end{split}
\end{equation}
For our present calculations, we will work in Gauge 1. In this gauge, the inflaton is homogeneous and the scalar perturbations are present in the metric; also the tensor perturbation $\gamma_{ij}$ is both transverse and traceless. Our focus will be on computing the correlation functions for the scalar field $\zeta$ at late times, when the modes of interest have exited the horizon.

Once we have chosen gauge 1, we can expand the action in eq.\eqref{admsct} to the desired order in the scalar and tensor perturbations $\zeta, \gamma_{ij}$. This requires us to solve the equations of motion for the Lagrange multipliers $N, N^i$ in terms of the perturbations, and their solutions to be substituted back into the action eq.\eqref{admsct}; see \cite{Maldacena:2002vr} for details. From the expression for the action upto the desired order, one can calculate the correlation functions of interest.

\subsection{The Two Point Function}
\label{genord}
Expanding the action eq.\eqref{admsct} upto the quadratic order for the scalar perturbation $\zeta$, we get
\be
\label{quadord}
S^{(2)} = \int d^4x \, \epsilon \, \Big( a^3 \dot{\zeta}^2 - a (\del\zeta)^2 \Big),
\ee
where $\epsilon$ is the slow-roll parameter defined in eq.\eqref{eps}. Working to the order where $\epsilon$ can be taken to be time independent, the action in eq.\eqref{quadord} can be mapped to the action of a massless free scalar field in exact de Sitter space, eq.\eqref{mlact}, with the identification
\be
\label{iden}
\zeta = \frac{1}{\sqrt{2\epsilon}} \, \varphi.
\ee
The free field $\zeta$ can thus be mode-expanded as follows,
\be
\label{newmodez}
\zeta(\eta, \vec[x]) = \int \frac{d^3k}{(2\pi)^3} \Big[ b_{\vec[k]} \, v_{k}(\eta) + b^{\dagger}_{-\vec[k]} \, v_{k}^*(\eta) \Big] \text{e}^{i\vec[k]\cdot\vec[x]},
\ee
where the mode functions $v_k(\eta)$ are given by
\be
\label{modeu}
v_k(\eta) = \frac{H}{\sqrt{4\epsilon k^3}} \left\{ \text{cosh}(\alpha) \, (1+ik\eta) \, \text{e}^{-ik\eta} - i \, \text{sinh}(\alpha) \, (1-ik\eta) \, \text{e}^{ik\eta}\right\}.
\ee
As before, the operators $b_{\vec[k]}, b^\dagger_{\vec[k]}$ act as the annihilation and creation operators for the free $\alpha$-vacua,
\be
\label{defb}
b_{\vec[k]} |\alpha\rangle = 0 \, \forall \vec[k].
\ee

The two point function can now be calculated in a straightforward manner,
\begin{equation}
\begin{split}
\label{pt2}
{}_{\mathrm{I}}\langle \alpha | \zeta(\vecs[k,1]) \zeta(\vecs[k,2]) |\alpha\rangle_{\mathrm{I}} (\eta)  &= \langle \alpha | \zeta_{\,\text{I}}(\eta, \vecs[k,1])  \zeta_{\,\text{I}}(\eta, \vecs[k,2]) |\alpha\rangle \\
&= (2\pi)^3 \delta^3(\vecs[k,1] + \vecs[k,2]) \, |v_{k_1}(\eta)|^2 ,
\end{split}
\end{equation}
which on substituting the expression for $v_{k_1}(\eta)$ from eq.\eqref{modeu} and taking the late time limit gives
\be
\label{pt2s}
{}_{\mathrm{I}}\langle \alpha | \zeta(\vecs[k,1]) \zeta(\vecs[k,2]) |\alpha\rangle_{\mathrm{I}} = (2\pi)^3 \delta^3(\vecs[k,1] + \vecs[k,2]) \, \frac{H^2}{4 \epsilon k_1^{\,3}}\, \text{cosh}(2\alpha).
\ee
From eq.\eqref{pt2s}, it is clear that for small $\alpha$,  the leading order term is independent of $\alpha$, and the subleading term is proportional to $\alpha^2$; there is no term linear in $\alpha$.

\subsection{The Three Point Function}
\label{3pgen}
We now proceed to calculate the three point function for the scalar field $\zeta$ in the interacting $\alpha$-vacuum. From \cite{Maldacena:2002vr}, we find that the third order term in the action eq.\eqref{admsct} is given by
\be
\label{thirdord}
S^{(3)} = \int d^4x \left[ \epsilon^2 \left\{ a^3 \tilde{\zeta} \, \dot{\tilde{\zeta}}^2   + a \tilde{\zeta} \big(\del\tilde{\zeta}\big)^2 \right\} - 2 \epsilon a^3 \dot{\tilde{\zeta}}\, \del_i\tilde{\zeta} \, \del_i\chi\right],
\ee
where eq.\eqref{thirdord} follows from the eq.(3.9) of \cite{Maldacena:2002vr} after dropping the terms subleading in the slow-roll parameters, and incorporating the changes due to the field redefinition
\begin{equation}
\begin{split}
\label{redef}
\zeta &= \tilde{\zeta} + \lambda \, \tilde{\zeta}^{\,2}\,,\\
\lambda &= \half \frac{\ddot{\bar{\phi}}}{H \dot{\bar{\phi}}} + \frac{1}{4} \frac{\dot{\bar{\phi}}^2}{H^2}.
\end{split}
\end{equation}
Also,
\be
\label{defchi}
\chi = \epsilon \, \del^{-2}\dot{\tilde{\zeta}}.
\ee
We need the three point function for the variable $\zeta$ and not $\tilde{\zeta}$. The two are related via
\begin{equation}
\begin{split}
\label{rel2p}
\langle \zeta(\vecs[x,1]) \zeta(\vecs[x,2]) \zeta(\vecs[x,3]) \rangle = & \langle \tilde{\zeta}(\vecs[x,1]) \tilde{\zeta}(\vecs[x,2]) \tilde{\zeta}(\vecs[x,3]) \rangle \\ &+ 2 \lambda \Big[ \langle \zeta(\vecs[x,1]) \zeta(\vecs[x,2]) \rangle \langle \zeta(\vecs[x,1]) \zeta(\vecs[x,3]) \rangle + \text{Perms} \Big].
\end{split}
\end{equation}
The strategy is to first calculate the three point function for the variable $\tilde{\zeta}$, and then add to it the contribution due to the field redefinition, eq.\eqref{rel2p}, to get the three point function for $\zeta$. The interaction Hamiltonian for the field $\tilde{\zeta}$ can be easily read off from the action eq.\eqref{thirdord},
\begin{equation}
\label{ihamtz}
H_{int} =  - \int \mes[x] \Big[ \epsilon^2 \big\{ a^2 \tilde{\zeta} \, \tilde{\zeta}\,'^{\,2}   + a^2 \tilde{\zeta} \big(\del\tilde{\zeta}\big)^2 \big\} - 2 \epsilon a^3 \tilde{\zeta}\,' \del_i\tilde{\zeta} \, \del_i\chi \Big].
\end{equation}
Knowing the interaction Hamiltonian eq.\eqref{ihamtz}, we can now perform the calculation for the three point function of $\tilde{\zeta}$ using the in-in formalism, eq.\eqref{iicon},
\begin{equation}
\begin{split}
\label{def3ptz}
{}_{\mathrm{I}}\langle \alpha | \tilde{\zeta}(\vecs[k,1])  \tilde{\zeta}(\vecs[k,2]) \tilde{\zeta}(\vecs[k,3])|\alpha\rangle_{\mathrm{I}} (\eta)  =\langle \alpha | \tilde{\zeta}_{\,\text{I}}(\eta, \vecs[k,1])  \tilde{\zeta}_{\,\text{I}}(\eta, \vecs[k,2]) \tilde{\zeta}_{\,\text{I}}(\eta, \vecs[k,3])|\alpha\rangle \\
- \, i \int\limits_{\eta_0}^\eta d\eta' \,\langle \alpha | \left[\tilde{\zeta}_{\,\text{I}}(\eta, \vecs[k,1])  \tilde{\zeta}_{\,\text{I}}(\eta, \vecs[k,2]) \tilde{\zeta}_{\,\text{I}}(\eta, \vecs[k,3]), H_{int}^I(\eta')\right] |\alpha\rangle\,,
\end{split}
\end{equation}
where $\tilde{\zeta}_{\,\text{I}}$ is the interaction picture free field with the mode expansion eq.\eqref{newmodez}.  Also, $H_{int}^I$ is the interaction Hamiltonian in the interaction picture, which can be obtained from eq.\eqref{ihamtz} by replacing the field $\tilde{\zeta}$ with the free field $\tilde{\zeta}_{\,\text{I}}$. The state  $|\alpha\rangle_{\mathrm{I}}$ is the interacting $\alpha$-vacuum eq.\eqref{alphai}; this can be related to the free vacuum $|\alpha\rangle$ as in the case of the probe $\varphi^3$ theory discussed in section \ref{allordc}, by rotating the time integration contour with the  addition of  a small imaginary part, eq.\eqref{changet}. Some details of the calculation are given in appendix \ref{calci}. The final late time result is given by\footnote{See \cite{Kundu:2013gha} for a similar and related discussion.}
\begin{equation}
\begin{split}
\label{tpall2}
&{}_{\mathrm{I}}\langle \alpha | \zeta(\vecs[k,1]) \zeta(\vecs[k,2]) \zeta(\vecs[k,3])|\alpha\rangle_{\mathrm{I}} = (2\pi)^3 \delta^3\Big(\sum_{a=1}^3 \vecs[k,a]\Big) \Big[\prod_{a=1}^3 \frac{1}{(2k_a^{\,3})}\Big] \frac{H^4}{4\epsilon} \times\\
&\hspace{2mm}\Bigg[- \sum_{a=1}^3 k_a^{\,3} + \sum_{a\neq b}k_a k_b^{\,2} + \frac{8}{K} \sum_{a<b}k_a^{\,2} k_b^{\,2}  + \frac{2(\epsilon + \delta)}{\epsilon} \, \text{cosh}^2(2\alpha) \Big(\sum_{a=1}^3 k_a^{\,3} \Big)\\
&\hspace{4mm}+ \text{sinh}^2(2\alpha)\Bigg\{ - \sum_{a=1}^3 k_a^{\,3} + \sum_{a\neq b}k_a k_b^{\,2}
+ 8 \bigg(\sum_{a<b}k_a^{\,2} k_b^{\,2}\bigg) \times \\
&\hspace{26mm} \bigg(\frac{1}{k_1+k_2-k_3} + \frac{1}{k_1-k_2+k_3} + \frac{1}{-k_1+k_2+k_3}\bigg) \Bigg\}\Bigg],
\end{split}
\end{equation}
where $\epsilon, \delta$ are defined in eqs.\eqref{eps}, \eqref{defdelp}. From eq.\eqref{tpall2}, we immediately see that by setting $\alpha = 0$, we recover the  three point function for the field $\zeta$ in the interacting Bunch-Davies vacuum computed in \cite{Maldacena:2002vr}.
We also see that the non-Gaussianity can be significantly enhanced in the $\alpha$-vacua, growing as $\text{e}^{2\alpha}$ for $\alpha\gg 1$. This is due to the last two terms on the  RHS of eq.\eqref{tpall2}, going like $\cosh^2(\alpha)$ and $\sinh^2\alpha$, respectively. The second of these, with coefficient $\sinh^2\alpha$, is proportional to a non-contact term. Some references where estimates of non-Gaussianity in the $\alpha$-vacua states are discussed in more detail are \cite{Holman:2007na, Ganc:2011dy, Aravind:2013lra}; see also \cite{Ashoorioon:2010xg, Ashoorioon:2013eia}.

It is worth noting that the result obtained for inflation does not have any infrared divergences, and does not require an infrared cutoff as $\eta\rightarrow 0$  to obtain a finite answer.  
As a result, there is no violation of scale invariance due to the presence of $ln(\varepsilon)$ terms, unlike what was seen for the probe $\varphi^3$ theory. 
However, compared to the probe $\varphi^3$ case, it is important here to work with the contour for $\eta$ with a small imaginary part so that convergence is ensured at early times $\eta\rightarrow -\infty$. 

\subsubsection{Squeezed Limit and the Consistency Condition}
\label{mcccheck}
The Maldacena consistency condition is given by
\be
\label{maldacc}
\lim_{\vecs[k,3] \rightarrow 0} \langle \zeta(\vecs[k,1]) \zeta(\vecs[k,2]) \zeta(\vecs[k,3])\rangle' = -\, n_s \, \langle \zeta(\vecs[k,1]) \zeta(-\vecs[k,1]) \rangle' \langle \zeta(\vecs[k,3]) \zeta(-\vecs[k,3]) \rangle' ,
\ee
where $n_s$ is the scalar spectral tilt, given by
\be
\label{stilt}
n_s = -2(2\epsilon + \delta).
\ee
From eq.\eqref{pt2s}, we can easily compute the RHS of the consistency condition,
\be
\label{ccrhs}
-\, n_s \, \langle \zeta(\vecs[k,1]) \zeta(-\vecs[k,1]) \rangle' \langle \zeta(\vecs[k,3]) \zeta(-\vecs[k,3]) \rangle' = \frac{2\epsilon + \delta}{\epsilon^2} \, \frac{H^4}{8} \frac{1}{k_1^{\,3} k_3^{\,3}} \, \text{cosh}^2(2\alpha).
\ee
Now, the LHS of the consistency condition, computed using the result eq.\eqref{tpall2}, is given by\footnote{$\theta_{\vecs[k,1],\vecs[k,3]}$ denotes the angle between $\vecs[k,1]$ and $\vecs[k,3]$.}
\begin{equation}
\label{cclhs}
\lim_{\vecs[k,3] \rightarrow 0} \langle \zeta(\vecs[k,1]) \zeta(\vecs[k,2]) \zeta(\vecs[k,3])\rangle' = \frac{1}{k_1^{\,3} k_3^{\,3}} \frac{H^4}{8} \bigg(\frac{2\epsilon + \delta}{\epsilon^2} \, \text{cosh}^2(2\alpha) + \frac{4}{\epsilon} \,\frac{k_1}{k_3}\, \frac{\text{sinh}^2(2\alpha)}{\text{sin}^2\theta_{\vecs[k,1],\vecs[k,3]}}\bigg).
\end{equation}
We can clearly see that due to the presence of the second term in the parentheses of eq.\eqref{cclhs}, the consistency condition is not met for $\alpha \neq 0$. In fact, the ratio $k_1/k_3 \rightarrow \infty$ in the limit $\vecs[k,3] \rightarrow 0$, so the second term in eq.\eqref{cclhs} dominates . 

As was emphasized in section \ref{sound}, the consistency conditions are a consequence of the underlying spatial  reparametrization invariance, \cite{Kundu:2015xta}. This would suggest that they should  remain valid for $\alpha$-vacua also, at odds with what we have found above. 

The underlying explanation is simply that the calculation for the three point function, and also the two point function, are not reliable for non-zero $\alpha$. 
In fact, the zeroth order solution itself is not reliable, since for $\alpha$-vacua, the quantum stress tensor  makes a non-trivial and in fact diverging contribution to the total stress-energy tensor. 
This contribution has been neglected in the zeroth order solution, which only included the classical stress-energy due to the inflaton.
As a result of this, the zeroth order solution itself does not  actually solve Einstein's equations.
Now, the requirement of spatial reparametrization invariance arises from the Einstein's equations (obtained by varying the shift and lapse functions in the action), and is also therefore not met by the zeroth order solution.
It is thus not surprising that the consequences of spatial reparametrization  invariance are also not satisfied by the perturbations about this inconsistent zeroth order solution.

An estimate of the quantum stress-energy tensor in $\alpha$-vacua is given in appendix \ref{back-reaction}.

\section{Bulk Calculation of the Scalar Three Point Function in Inflation}
\label{bulk3pt}
The three point function for the scalar perturbation $\zeta$, eq.\eqref{defpm}, in inflation was computed in \cite{Maldacena:2002vr} using the in-in formalism. In this section, we present an alternate approach for computing the same. This approach arises as a consequence of the scaling and special conformal Ward identities relating the three and four point functions at the leading order in the slow-roll approximation, see \cite{Kundu:2014gxa}. The Ward identities suggest that the three point function must follow from a computation of the four point function, with one of the external legs replaced by the time derivative of the homogeneous background, $\dot{\bar\phi}$. Working in the Bunch-Davies vacuum, we will show that this is indeed the case, providing another check for the validity of the Ward identities. Our method follows the approach utilized in \cite{Ghosh:2014kba} for computing the inflationary four point function of $\zeta$. The discussion here is also related to \cite{Arkani-Hamed:2015bza} and \cite{Lee:2016vti}, where related ideas are used to examine the effect of higher spin fields on non-Gaussianity. 

The present technique for computing the inflationary three point function relies on an important analogy between calculations in dS and AdS spaces. We first calculate the wave function, in terms of the late time values for the perturbations, and then the correlation functions can be computed from the wave function. 
To compute the inflationary three point function for $\zeta$, we need to evaluate the wave function $\Psi[\delta\phi]$, where $\delta\phi$ is the perturbation to the inflaton, eq.\eqref{pinf}, and is related to $\zeta$ by a change of gauge. The wave function $\Psi[\delta\phi]$ has the schematic form
\be
\label{wfunsch}
\Psi[\delta\phi] = \text{exp}\bigg[\frac{M_{Pl}^2}{H^2} \bigg(-\half \int \delta\phi \delta\phi \, \langle OO \rangle + \frac{1}{3!} \int \delta\phi \delta\phi \delta\phi \, \langle OOO \rangle + \cdots\bigg) \bigg],
\ee
where
\be
\label{oo}
\langle O(\vecs[k,1]) O(\vecs[k,2])\rangle = (2\pi)^3 \delta^3(\vecs[k,1]+\vecs[k,2]) \, k_1^{\,3},
\ee
and $\langle OOO \rangle, \ldots$ are the coefficient functions. Once we have the wave function $\Psi[\delta\phi]$, in particular the cubic coefficient $\langle OOO\rangle$, we can get the three point function for $\delta\phi$ by using
\be
\label{qmform}
\langle \delta\phi \delta\phi \delta\phi \rangle = \frac{\int [\mathcal{D}\delta\phi]  \, \delta\phi \delta\phi \delta\phi \, \big|\Psi[\delta\phi]\big|^2 }{\int [\mathcal{D}\delta\phi]  \,\big|\Psi[\delta\phi]\big|^2},
\ee
which is the standard quantum mechanical prescription to compute expectation values. Knowing the three point function for $\delta\phi$, we can get the three point function for $\zeta$ by simply performing a change of gauge. Thus, the whole computation boils down to computing the cubic term $\langle OOO \rangle $ in the wave function $\Psi[\delta\phi]$. 

\begin{figure}
\begin{center}
\includegraphics[scale=0.70]{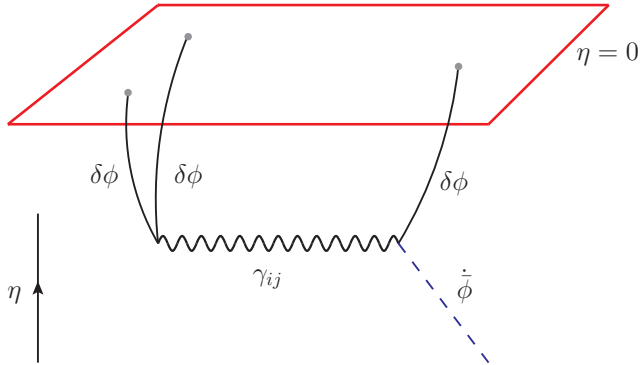}
\caption{The three point function in de Sitter space, with one leg being the time-derivative of the background $\bar\phi$.}
\label{dsdia}
\end{center}
\end{figure}

The analogy with AdS space also suggests that this cubic term can be calculated using the analogue of Feynman-Witten bulk-to-boundary propagators. In fact, as already mentioned, the three point function gets related to the four point function with one leg replaced by the background value of the inflaton, ${\dot{\bar\phi}}$, as shown in figure \ref{dsdia}, see \cite{Kundu:2014gxa, Arkani-Hamed:2015bza}. The propagators in figure \ref{dsdia} are bulk-to-boundary propagators in de Sitter space, and the interaction vertex is given in figure \ref{vertex}, which arises by expanding the action about the inflationary background as will be explained shortly. 
In fact, since de Sitter space can be analytically continued to Euclidean AdS (EAdS) space, the whole calculation can be done conveniently by first working in EAdS space and then continuing the result back to de Sitter space.  Note that the Bunch-Davies vacuum in de Sitter space  corresponds to choosing the boundary condition that deep in the interior of EAdS space all perturbations become regular. 
This is the procedure we follow in the computation below. 
 
To be more specific, consider  the metric of  four dimensional $\text{EAdS}$ space in  Poincare coordinates,
\be
\label{eads}
ds^2 = \frac{\text{R}_{\text{AdS}}^{\,2}}{z^2} \, \bigg(dz^2 + \sum_{i=1}^3 dx_i \, dx^i \bigg),
\ee
where $\text{R}_{\text{AdS}}$ is the $\text{EAdS}$ radius, and the coordinate $z \in [0,\infty)$. Under the analytic continuation
\be
\label{anaz}
z = -\,i\eta ,
\ee
and
\be
\label{anar}
\text{R}_{\text{AdS}} = \frac{i}{H},
\ee
the metric eq.\eqref{eads} goes to the metric of four dimensional de Sitter space, eq.\eqref{meteta}. Also, the partition function in $\text{EAdS}$ space is related to the wave function in de Sitter space via the same analytic continuation. In the semiclassical approximation, where one can replace the path integral involved in the calculation of the partition function or the wave function by its saddle point value, one can write
\be
\label{dsads}
Z_{\text{EAdS}}[\Phi(x)] \equiv \text{e}^{-S^{\text{EAdS}}_{on-shell}[\Phi(x)]} \xleftrightarrow[\text{R}_{\text{AdS}} = \frac{i}{H}]{z = -\,i\eta} \Psi[\Phi(x)] \equiv \text{e}^{iS^{\text{dS}}_{on-shell}[\Phi(x)]} ,
\ee
where the EAdS partition function $Z_{\text{EAdS}}$ is a functional of the boundary value of the field $\Phi(x)$ as $z \rightarrow 0$, whereas the wave function $\Psi$ in de Sitter space is a functional of the late time value of the field $\Phi(x)$ as $\eta \rightarrow 0$. The other boundary condition imposed while computing the on-shell action is to demand regularity of the solution deep in the interior, $z \rightarrow \infty$, of $\text{EAdS}$; as mentioned above, this corresponds to the choice of Bunch-Davies vacuum in the far past, $\eta \rightarrow -\infty$, in de Sitter space.

\subsection{Computing the Coefficient Function $\langle OOO \rangle$}
\label{ooocalc}
\begin{figure}
\begin{center}
\includegraphics[scale=0.65]{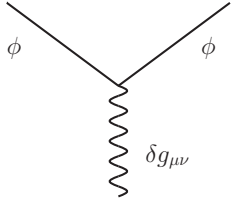}
\caption{The qualitative three point bulk interaction vertex between two scalars and a graviton.}
\label{vertex}
\end{center}
\end{figure}

In the $\text{EAdS}$ space, with the metric given by eq.\eqref{eads}, we start with the action
\be
\label{eadsact}
S = \frac{M_{Pl}^2}{2} \int d^4x \sqrt{g} \, \Big(R - 2\Lambda - (\nabla \phi)^2 - 2V(\phi)\Big),
\ee
where $\phi(z,\vec[x]) = \bar{\phi}(z) + \delta\phi(z,\vec[x])$ is the inflaton written in AdS coordinates, and $\Lambda$ is the cosmological constant, which is related to $\text{R}_{\text{AdS}}$ by
\be
\label{ccon}
\Lambda = - \, \frac{3}{\text{R}_{\text{AdS}}^2} .
\ee
We expand the metric perturbatively as
\be
\label{permet}
g_{\mu\nu} = \bar{g}_{\mu\nu} + \delta g_{\mu\nu},
\ee
where $\bar{g}_{\mu\nu}$ is the unperturbed background metric given in eq.\eqref{eads}, and $\delta g_{\mu\nu}$ is the perturbation. Substituting eq.\eqref{permet} into the action eq.\eqref{eadsact} and expanding, we get
\be
\label{actexp}
S = S_0 + S^{(2)}_{\text{grav}} - \half \, M_{Pl}^2 \int d^4x \sqrt{\bar{g}} \, \bar{g}^{\mu\nu} \del_{\mu}(\delta\phi) \del_{\nu}(\delta\phi) + S_{int} ,
\ee
where $S_0$ is the action for the unperturbed background, $S^{(2)}_{\text{grav}}$ is the part of the action which is quadratic in the metric perturbation $\delta g_{\mu\nu}$, and $S_{int}$ is the interaction term, given by
\be
\label{adsint}
S_{int} = \half \, M_{Pl}^2 \int d^4x \sqrt{\bar{g}} \, \delta g_{\mu\nu} T^{\mu\nu},
\ee
where the energy-momentum tensor $T_{\mu\nu}$ for the scalar field $\phi$ is given by
\be
\label{emads}
T_{\mu\nu} = \del_{\mu}\phi \, \del_{\nu}\phi - \half\,\bar{g}_{\mu\nu} \Big(\bar{g}^{\alpha \beta} \del_{\alpha}\phi \, \del_{\beta}\phi + 2 V(\phi)\Big).
\ee
The interaction term eq.\eqref{adsint} gives rise to an interaction vertex between two scalars and a graviton, depicted qualitatively in figure \ref{vertex}.

From $S^{(2)}_{\text{grav}}$, we can compute the propagator for the graviton. For doing so, we choose the gauge\footnote{This gauge naturally goes over to the gauge $N=1, N^i =0$ under analytic continuation, which is used in our inflationary calculations.}
\be
\label{gravgauge}
\delta g_{zz} = 0, \, \delta g_{zi} = 0,
\ee
with $i = 1,2,3$. The graviton propagator in this gauge is given by \cite{Raju:2010by, Raju:2011mp}
\begin{equation}
\begin{split}
\label{gravprop}
&\mathcal{G}_{ij,kl}(z_1,\vecs[x,1];z_2,\vecs[x,2]) \\
&= \int \frac{d^3\vec[k]}{(2\pi)^3} \, \text{e}^{i\vec[k]\cdot(\vecs[x,1]-\vecs[x,2])} \int_{0}^{\infty} \frac{dp^2}{2} \Bigg[ \frac{J_{\frac{3}{2}}(pz_1) J_{\frac{3}{2}}(pz_2)}{\sqrt{z_1 z_2} (\vec[k]^2 + p^2)} \half (\mathcal{T}_{ik} \mathcal{T}_{jl} + \mathcal{T}_{il} \mathcal{T}_{jk} - \mathcal{T}_{ij} \mathcal{T}_{kl})\Bigg],
\end{split}
\end{equation}
where
\be
\label{tij}
\mathcal{T}_{ij} = \delta_{ij} + \frac{k_i k_j}{p^2}.
\ee
Note that the graviton propagator in eq.\eqref{gravprop} is not transverse. We can however decompose it into a transverse part and a longitudinal part. The transverse graviton propagator is given by
\begin{equation}
\begin{split}
\label{gproptt}
&\tilde{\mathcal{G}}_{ij,kl}(z_1,\vecs[x,1];z_2,\vecs[x,2]) \\
&= \int \frac{d^3\vec[k]}{(2\pi)^3} \, \text{e}^{i\vec[k]\cdot(\vecs[x,1]-\vecs[x,2])} \int_{0}^{\infty} \frac{dp^2}{2} \Bigg[ \frac{J_{\frac{3}{2}}(pz_1) J_{\frac{3}{2}}(pz_2)}{\sqrt{z_1 z_2} (\vec[k]^2 + p^2)} \half (\tilde{\mathcal{T}}_{ik} \tilde{\mathcal{T}}_{jl} + \tilde{\mathcal{T}}_{il} \tilde{\mathcal{T}}_{jk} - \tilde{\mathcal{T}}_{ij} \tilde{\mathcal{T}}_{kl})\Bigg],
\end{split}
\end{equation}
where
\be
\label{ttij}
\tilde{\mathcal{T}}_{ij} = \delta_{ij} - \frac{k_i k_j}{k ^2}.
\ee
The longitudinal part is then essentially the difference between the full propagator, eq.\eqref{gravprop}, and the transverse piece, eq.\eqref{gproptt}.

From eq.\eqref{actexp}, we see that the scalar field $\delta\phi$ behaves essentially like a free scalar field in EAdS space, with only gravitational interactions. Thus for a particular momentum mode carrying momentum $\vec[k]$, we have
\be
\label{modedp}
\delta\phi_{\vec[k]}(x,\vec[z]) = \phi_0(\vec[k]) (1+kz)\,\text{e}^{-kz}\, \text{e}^{i\vec[k]\cdot\vec[x]},
\ee
where we have chosen the solution for $\delta\phi$ which is regular as $z\rightarrow\infty$.

\begin{figure}[h]
\begin{center}
\includegraphics[scale=0.65]{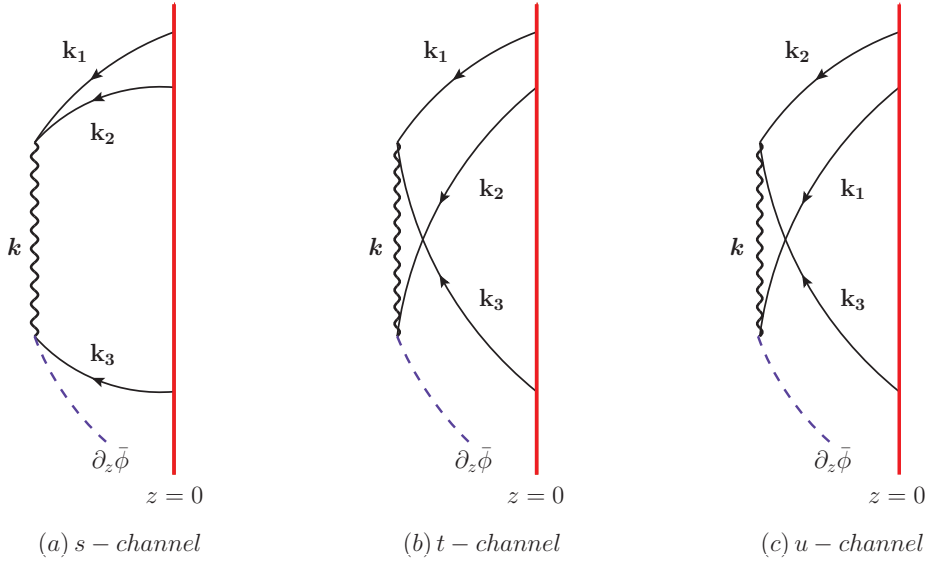}
\caption{Feynman-Witten diagrams for the \textit{s, t}, and \textit{u} channel processes contributing to the calculation of the EAdS partition function. $z=0$ is the $\text{EAdS}$ boundary. The wavy line denotes the bulk-to-bulk graviton propagator, the solid lines represent the bulk-to-boundary propagators for the scalar field $\delta\phi$, and the dashed line represents the background. The exchanged graviton carries momentum $\vec[k]$.}
\label{stuch}
\end{center}
\end{figure}

To calculate the EAdS partition function, we need to compute the on-shell action. In particular, to extract the unknown coefficient $\langle OOO\rangle$, we need to evaluate the contribution from the Feynman-Witten diagrams of figure \ref{stuch}.
This contribution is given by
\begin{equation}
\begin{split}
\label{3pons}
S^{EAdS}_{on-shell} = \half\, M_{\text{Pl}}^2 \, \text{R}_{\text{AdS}}^2 &\int \frac{dz_1}{z_1^4} \frac{dz_2}{z_2^4} \, d^3x_1 d^3x_2 \, \bar{g}^{i_1 i_2} \bar{g}^{j_1 j_2} T_{i_1 j_1}(z_1,\vecs[x,1])\times\\
&\mathcal{G}_{i_2 j_2,k_2 l_2}(z_1,\vecs[x,1];z_2,\vecs[x,2]) \bar{g}^{k_1 k_2} \bar{g}^{l_1 l_2} T_{k_1 l_1}(z_2,\vecs[x,2]).
\end{split}
\end{equation}
As already discussed above, we can write the graviton propagator as a sum of a transverse and a longitudinal part. This gives us
\be
\label{sadswr}
S^{EAdS}_{on-shell} = \half\, M_{\text{Pl}}^2 \, \text{R}_{\text{AdS}}^2 \,(\mathcal{W} + 2 \mathcal{R}) ,
\ee
where $\mathcal{W}$ is the transverse graviton contribution,
\begin{equation}
\label{tgcont}
\mathcal{W} = \int dz_1 dz_2\, d^3x_1 d^3x_2 \,  T_{ij}(z_1,\vecs[x,1]) \tilde{\mathcal{G}}_{ij,kl}(z_1,\vecs[x,1];z_2,\vecs[x,2]) T_{kl}(z_2,\vecs[x,2]),
\end{equation}
where we have used $\bar{g}^{ij} = z^2 \delta^{ij}$. The contribution to the on-shell action from the longitudinal part of the exchanged graviton is written in the form of a ``remainder'' term $\mathcal{R}$, which has the form
\be
\label{remainder}
\mathcal{R} = \mathcal{R}_1 + \mathcal{R}_2 + \mathcal{R}_3 ,
\ee
with $\mathcal{R}_1, \mathcal{R}_2, \mathcal{R}_3$ given by
\begin{equation}
\begin{split}
\label{r123pos}
&\mathcal{R}_1 = - \int \frac{dz}{z^2} \, d^3x\,  T_{zj}(z,\vec[x]) \, \frac{1}{\del^{\,2}} \, T_{zj}(z,\vec[x]),\\
&\mathcal{R}_2 = - \half \int \frac{dz}{z} \, d^3x\,  \del_j T_{zj}(z,\vec[x]) \, \frac{1}{\del^{\,2}} \, T_{zz}(z,\vec[x]),\\
&\mathcal{R}_3 = - \frac{1}{4} \int \frac{dz}{z^2} \, d^3x\,  \del_j T_{zj}(z,\vec[x]) \, \bigg(\frac{1}{\del^{\,2}}\bigg)^2 \, \del_i T_{zi}(z,\vec[x]).
\end{split}
\end{equation}

We can now perform the computation of the on-shell action. Some details of the calculation are given in appendix \ref{bulkdet}. The contribution from the transverse part of the graviton exchanged vanishes, see appendix \ref{transverse}. The contribution from the longitudinal part is calculated in appendix \ref{longitudinal}, and is given by
\begin{equation}
\begin{split}
\label{rfin}
\mathcal{R} = \half \, \frac{\dot{\bar\phi}}{H} \, (2\pi)^3\delta^3\Big(\sum_{a=1}^3& \vecs[k,a]\Big) \Big( \prod_{a=1}^3 \phi_0(\vecs[k,a])\Big) \times \\
&\Bigg[- \half \sum_{a=1}^3 k_a^{\,3} + \half \sum_{a \neq b} k_a k_b^{\,2} + \frac{4}{K}\sum_{a<b} k_a^{\,2} k_b^{\,2} \Bigg].
\end{split}
\end{equation}
Using eq.\eqref{sadswr}, we see that the EAdS on-shell action is
\begin{equation}
\begin{split}
\label{eadssf}
S^{EAdS}_{on-shell} = \half \, M_{\text{Pl}}^2 \, \text{R}_{\text{AdS}}^2& \, \frac{\dot{\bar\phi}}{H} \, (2\pi)^3\delta^3\Big(\sum_{a=1}^3 \vecs[k,a]\Big) \Big( \prod_{a=1}^3 \phi_0(\vecs[k,a])\Big) \times \\
&\Bigg[- \half \sum_{a=1}^3 k_a^{\,3} + \half \sum_{a \neq b} k_a k_b^{\,2} + \frac{4}{K}\sum_{a<b} k_a^{\,2} k_b^{\,2} \Bigg].
\end{split}
\end{equation}
From eq.\eqref{eadssf}, by taking derivatives with respect to the boundary value $\phi_0$ of the field $\delta\phi$, we obtain the three point coefficient function $\langle OOO \rangle$,
\begin{equation}
\begin{split}
\label{dsoc}
\langle O(\vecs[k,1]) O(\vecs[k,2]) O(\vecs[k,3]) \rangle = -\,\half \, &\frac{\dot{\bar\phi}}{H} \, (2\pi)^3\delta^3\Big(\sum_{a=1}^3 \vecs[k,a]\Big) \times \\
&\Bigg[- \half \sum_{a=1}^3 k_a^{\,3} + \half \sum_{a \neq b} k_a k_b^{\,2} + \frac{4}{K}\sum_{a<b} k_a^{\,2} k_b^{\,2} \Bigg],
\end{split}
\end{equation}
where we have made use of the analytic continuation eq.\eqref{anar}.

\subsection{The Three Point Function $\langle \zeta \zeta \zeta \rangle$}
\label{detzeta}
We now proceed to compute the inflationary three point function $\langle \zeta \zeta \zeta \rangle$. From the wave function eq.\eqref{wfunsch}, one finds that
\be
\label{dporel}
\langle \delta\phi(\vecs[k,1])  \delta\phi(\vecs[k,2])  \delta\phi(\vecs[k,3]) \rangle = \frac{1}{4} \, \frac{H^4}{M_{\text{Pl}}^4} \, \frac{\langle O(\vecs[k,1]) O(\vecs[k,2]) O(\vecs[k,3]) \rangle}{\prod_{a=1}^3 \langle O(\vecs[k,a]) O(-\vecs[k,a]) \rangle'}.
\ee
Substituting the result eq.\eqref{dsoc} in eq.\eqref{dporel}, and using eq.\eqref{oo}, we get
\begin{equation}
\begin{split}
\label{dp3}
\langle \delta\phi(\vecs[k,1])  \delta\phi(\vecs[k,2])  \delta\phi(\vecs[k,3]) \rangle = -\, \frac{1}{8} \, &\frac{H^4}{M_{\text{Pl}}^4} \, \frac{\dot{\bar\phi}}{H}\, (2\pi)^3\delta^3\Big(\sum_{a=1}^3 \vecs[k,a]\Big) \bigg(\prod_{a=1}^3 \frac{1}{k_a^{\,3}} \bigg)\times \\
&\Bigg[- \half \sum_{a=1}^3 k_a^{\,3} + \half \sum_{a \neq b} k_a k_b^{\,2} + \frac{4}{K}\sum_{a<b} k_a^{\,2} k_b^{\,2} \Bigg].
\end{split}
\end{equation}
Now, to obtain the three point function for the perturbation $\zeta$, we need to perform a change of gauge from $\delta\phi$ to $\zeta$ in eq.\eqref{dp3}. The second order change of gauge relating $\delta\phi$ and $\zeta$ is given by (see \cite{Maldacena:2002vr} for details)
\be
\label{change}
\zeta = - \, \frac{H}{\dot{\bar\phi}} \, \delta\phi + \half \bigg( \half + \frac{\ddot{\bar\phi} H}{\dot{\bar\phi}^{\,3}}\bigg) \delta\phi^2 .
\ee
Performing the change of gauge eq.\eqref{change} in eq.\eqref{dp3}, we get the final result
\begin{equation}
\begin{split}
\label{3pzf}
\langle \zeta(\vecs[k,1])  \zeta(\vecs[k,2])  \zeta(\vecs[k,3]) \rangle = (2\pi)^3&\delta^3\Big(\sum_{a=1}^3 \vecs[k,a]\Big) \bigg(\prod_{a=1}^3 \frac{1}{(2k_a^{\,3})} \bigg) \frac{H^4}{M_{\text{Pl}}^4} \, \frac{H^2}{\dot{\bar\phi}^2} \times \\
&\Bigg[\bigg( \half + \frac{2 H \ddot{\bar\phi}}{\dot{\bar\phi}^{\,3}}\bigg) \sum_{a=1}^3 k_a^{\,3} + \half \sum_{a \neq b} k_a k_b^{\,2} + \frac{4}{K}\sum_{a<b} k_a^{\,2} k_b^{\,2} \Bigg],
\end{split}
\end{equation}
which is indeed the expression for the three point function of $\zeta$ computed in \cite{Maldacena:2002vr}.

The present method of calculation also provides us with an understanding of which region in the bulk makes a significant contribution to the 
late time $\eta\rightarrow 0$ correlation functions. E.g. for the three point function, when $k_3\rightarrow 0$ keeping $k_1, k_2$ fixed, the corresponding bulk-to-boundary propagators for the scalars, shown in figure \ref{dsdia}, go deep inside the bulk. Thus the behaviour in the far past becomes important. This is a version of the UV-IR connection in de Sitter or inflationary space times, and has also been discussed in \cite{Arkani-Hamed:2015bza}.

\section{Conclusions}
\label{conclusion}
We have studied the Ward identities for scale and special conformal transformations in the context of inflation and de Sitter space in  this paper. It was argued earlier, \cite{Kundu:2015xta}, that these Ward identities follow from the coordinate reparametrization symmetries of the system. The coordinate reparametrization invariance can be used to set the perturbation in the inflaton to vanish, $\delta \phi=0$, at late times. The resulting perturbations in single field models then correspond to scalar perturbations $\zeta$, and tensor perturbations $\widehat\gamma_{ij}$ in the metric. The residual spatial reparametrization symmetries present give rise to Ward identities for the correlation functions of these perturbations. See \cite{Kundu:2015xta} for details.

For generalized models of single field inflation, it was  shown here  that the Ward identities are indeed valid, as would be expected from the general nature of the arguments leading to these  identities. We should mention that some of these Ward identities were checked in an earlier work \cite{Creminelli:2012ed}. 

We also explored a class of vacua, called $\alpha$-vacua, which preserve conformal invariance. For these vacua, we found that the scalar three point function $\langle\zeta\zeta\zeta\rangle$ did not meet the Maldacena consistency condition, which is the Ward identity for scale invariance. We argued that this is because the background inflationary solution, about which the perturbations have been computed, is itself not self-consistent. In the $\alpha$-vacua, the quantum stress tensor diverges, and thus the back-reaction of the quantum stress tensor cannot be neglected. The background solution though neglects this effect and only incorporates the classical potential and small corrections due to the rolling of the inflaton. 

We also explored the nature of the $\alpha$-vacua in some detail for a probe scalar field. We showed directly, by constructing the conserved charges, that  these vacua preserve conformal invariance in the interacting theory, upto subtleties having to do with zero modes and possible surface terms, see appendix \ref{isogen}. We also calculated the late time three point function in the $\alpha$-vacua for a probe massless scalar field. We found that the result is conformally invariant, upto contact terms. However, interestingly, the corresponding non-Gaussian term in the wave function does not satisfy the operator product expansion. The implications for a possible dS/CFT correspondence are left for the future.

Finally, we described an alternate calculation for the three point function for scalar perturbations in standard slow-roll inflation in the Bunch-Davies vacuum. This calculation is motivated by techniques drawn from the AdS/CFT correspondence and is related to other recent papers,
including \cite{Ghosh:2014kba, Arkani-Hamed:2015bza, Lee:2016vti}, and could be useful in thinking about the implications of additional fields during inflation, including those with higher spin. 

\acknowledgments
We especially thank Nilay Kundu for collaboration in the initial stages of this work. We also thank Paolo Creminelli, Atish Dabholkar, Thomas Hartman, Norihiro Iizuka,  Lavneet Janagal, Bhawik Jani, Jorge Norena, Shiraz Minwalla and Marko Simonovic for insightful discussions and comments. AS would like to thank the organizers of the $10^{th}$ Asian Winter School on Strings, Particles and Cosmology at Okinawa Institute of Science and Technology, Japan, for their hospitality during the course of this work. SPT acknowledges support from the J. C. Bose fellowship of the DST, Government of India. We thank the DAE, Government of India for support. Most of all, we thank the people of India for generously supporting research in String Theory. 

\appendix

\section{Correlation Functions of $\zeta$ for the $P(X,\phi)$ Models of Inflation}
\label{chenres}
In this appendix, to make the paper self-contained, we present the results for the two, three and four point functions of the curvature perturbation $\zeta$ for the general single field models of inflation introduced in section \ref{sound}. These results are taken from \cite{Chen:2006nt, Chen:2009bc}.

The two point function is given by
\be
\label{gen2p}
\langle \zeta(\vecs[k,1]) \zeta(\vecs[k,2])\rangle = (2\pi)^3 \delta^{\,3}(\vecs[k,1] + \vecs[k,2]) \, \frac{H^2}{M_{Pl}^2} \, \frac{1}{c_s} \, \frac{1}{4\epsilon k_1^3}.
\ee

The three point function to the leading order in the ``slow-variation'' parameter $\epsilon$ is given by
\begin{equation}
\label{gen3p}
\langle \zeta(\vecs[k,1]) \zeta(\vecs[k,2]) \zeta(\vecs[k,3])\rangle = (2\pi)^3 \delta^{\,3}\Big(\sum_{a=1}^3 \vecs[k,a]\Big) \, \frac{H^4}{M_{Pl}^4} \, \frac{2}{c_s^{\,2} \epsilon^2} \prod_{a=1}^3 \bigg(\frac{1}{2k_a^{\,3}}\bigg) \, \mathcal{A} ,
\end{equation}
where\footnote{The term $\mathcal{A}$ can have subleading corrections which are $O(\epsilon)$. For details, see \cite{Chen:2006nt}.}
\begin{equation}
\begin{split}
\label{cala}
\mathcal{A} = &\bigg( \frac{1}{c_s^{\,2}} -1 - \frac{2 \lambda}{\Sigma} \bigg) \frac{3 k_1^2 k_2^2 k_3^2}{2 K^3} \\
&+ \bigg( \frac{1}{c_s^{\,2}} -1\bigg) \bigg(-\frac{1}{K} \sum_{a<b} k_a^2 k_b^2 + \frac{1}{2K^2} \sum_{a \neq b} k_a^2 k_b^3 + \frac{1}{8} \sum_{a=1}^3 k_a^{\,3} \bigg),
\end{split}
\end{equation}
with $\lambda$ and $\Sigma$ being defined as
\be
\label{deflam}
\lambda = X^2 P_{,XX} + \frac{2}{3} X^3 P_{,XXX} \, ,
\ee
and
\be
\label{defsig}
\Sigma = X P_{,X} + 2 X^2 P_{,XX} .
\ee
Note that $K = k_1 + k_2 + k_3$. Another important quantity which we will need later is
\be
\label{defmu}
\mu = \half X^2 P_{,XX} + 2 X^3 P_{,XXX} + \frac{2}{3} X^4 P_{,XXXX}.
\ee

The four point function receives contributions from a contact interaction term, as well as from an intermediate scalar exchange. The complete expression for the four point function to the leading order is given by
\begin{equation}
\label{gen4p}
\langle \zeta(\vecs[k,1]) \zeta(\vecs[k,2]) \zeta(\vecs[k,3]) \zeta(\vecs[k,4])\rangle = (2\pi)^3 \delta^{\,3}\Big(\sum_{a=1}^4 \vecs[k,a]\Big) \, \frac{H^6}{M_{Pl}^6} \, \frac{2}{c_s^{\,3} \epsilon^3} \prod_{a=1}^4 \bigg(\frac{1}{2k_a^{\,3}}\bigg) \, \mathcal{T} ,
\end{equation}
where $\mathcal{T}$ is given by
\begin{equation}
\begin{split}
\label{calt}
\mathcal{T} = &\bigg( \frac{\lambda}{\Sigma} \bigg)^2 \mathcal{T}_{s1} + \frac{\lambda}{\Sigma} \bigg( \frac{1}{c_s^{\,2}} -1\bigg) \mathcal{T}_{s2} + \bigg(\frac{1}{c_s^{\,2}} -1\bigg)^2 \mathcal{T}_{s3} \\
&+ \bigg( \frac{\mu}{\Sigma} - \frac{9\lambda^2}{\Sigma^2} \bigg) \mathcal{T}_{c1} + \bigg( \frac{3\lambda}{\Sigma} - \frac{1}{c_s^{\,2}} + 1\bigg) \mathcal{T}_{c2} + \bigg( \frac{1}{c_s^{\,2}} - 1\bigg)\mathcal{T}_{c3}.
\end{split}
\end{equation}
Here, $\mathcal{T}_{s1}, \mathcal{T}_{s2}, \mathcal{T}_{s3}$ are contributions coming from an intermediate scalar exchange, and are given by
\begin{equation}
\begin{split}
\label{calts1}
\mathcal{T}_{s1} = &\bigg[\frac{9}{8}\, k_1^2 k_2^2 k_3^2 k_4^2 k_{12} \, \frac{1}{(k_1 + k_2 + k_{12})^3 M^3} \\
&+ \frac{9}{4}\, k_1^2 k_2^2 k_3^2 k_4^2 k_{12} \, \frac{1}{M^3} \bigg( \frac{6M^2}{\tilde{K}^5} + \frac{3M}{\tilde{K}^4} + \frac{1}{\tilde{K}^3} \bigg)\bigg] + 23\: \text{Permutations},
\end{split}
\end{equation}
\begin{equation}
\begin{split}
\label{calts2}
\mathcal{T}_{s2} = \bigg[&-\, \frac{3}{32} \, (\vecs[k,3]\cdot\vecs[k,4])\, k_{12} k_1^2 k_2^2 \, \frac{1}{(k_1 + k_2 + k_{12})^3} \, F(k_3, k_4, M)\\
&-\,\frac{3}{16} \, (\vecs[k,12]\cdot\vecs[k,4])\,\frac{k_1^2 k_2^2 k_3^2}{k_{12}} \, \frac{1}{(k_1 + k_2 + k_{12})^3} \, F(k_{12}, k_4, M)\\
&-\, \frac{3}{16} \, (\vecs[k,3]\cdot\vecs[k,4])\, k_{12} k_1^2 k_2^2 \,G_{ab}(k_3,k_4) -\frac{3}{8} \, (\vecs[k,12]\cdot\vecs[k,4])\,\frac{k_1^2 k_2^2 k_3^2}{k_{12}} \,G_{ab}(k_{12},k_4)\\
&-\, \frac{3}{16} \, (\vecs[k,1]\cdot\vecs[k,2])\, k_{12} k_3^2 k_4^2 \,G_{ba}(k_1,k_2) +\frac{3}{8} \, (\vecs[k,12]\cdot\vecs[k,2])\,\frac{k_1^2 k_3^2 k_4^2}{k_{12}} \,G_{ba}(-k_{12},k_2)\bigg]\\
&+ 23\: \text{Permutations} ,
\end{split}
\end{equation}
\begin{equation}
\begin{split}
\label{calts3}
\mathcal{T}_{s3} = &\bigg[\frac{1}{128} \,(\vecs[k,1]\cdot\vecs[k,2]) (\vecs[k,3]\cdot\vecs[k,4]) k_{12}\, F(k_1,k_2,k_1+k_2+k_{12}) \, F(k_3,k_4,M)\\
&+ \frac{1}{32} \,(\vecs[k,1]\cdot\vecs[k,2]) (\vecs[k,12]\cdot\vecs[k,4]) \frac{k_3^2}{k_{12}} \, F(k_1,k_2,k_1+k_2+k_{12}) \, F(k_{12},k_4,M)\\
&- \frac{1}{32} \,(\vecs[k,12]\cdot\vecs[k,2]) (\vecs[k,12]\cdot\vecs[k,4]) \frac{k_1^2 k_3^2}{k_{12}^3} \, F(k_{12},k_2,k_1+k_2+k_{12}) \, F(k_{12},k_4,M)\\
&+ \frac{1}{64} \,(\vecs[k,1]\cdot\vecs[k,2]) (\vecs[k,3]\cdot\vecs[k,4]) k_{12}\,G_{bb}(k_1,k_2,k_3,k_4)\\
&+ \frac{1}{32} \,(\vecs[k,1]\cdot\vecs[k,2]) (\vecs[k,12]\cdot\vecs[k,4]) \frac{k_3^2}{k_{12}} \,G_{bb}(k_1,k_2,k_{12},k_4)\\
&- \frac{1}{32} \,(\vecs[k,12]\cdot\vecs[k,2]) (\vecs[k,3]\cdot\vecs[k,4]) \frac{k_1^2}{k_{12}} \,G_{bb}(-k_{12},k_2,k_3,k_4)\\
&- \frac{1}{16} \, (\vecs[k,12]\cdot\vecs[k,2]) (\vecs[k,12]\cdot\vecs[k,4]) \frac{k_1^2 k_3^2}{k_{12}^3} \,G_{bb}(-k_{12},k_2,k_{12},k_4) \bigg] + 23\: \text{Permutations},
\end{split}
\end{equation}
where we have used the notation
\begin{equation}
\begin{split}
\label{notation1}
&\tilde{K} = k_1 + k_2 + k_3 + k_4 , \\
&\vecs[k,12] = \vecs[k,1] + \vecs[k,2] ,\\
&k_{12} = |\vecs[k,12]| ,\\
&M = k_3 + k_4 + k_{12}.
\end{split}
\end{equation}
Also, the functions $F, G_{ab}, G_{ba}$ and $G_{bb}$ used in eqs.\eqref{calts2} and \eqref{calts3} above are defined as
\be
\label{deff}
F(u,v,m) \equiv \frac{1}{m^3} \big[ 2 u v + (u+v)m + m^2 \big] ,
\ee
\begin{equation}
\begin{split}
\label{defgab}
G_{ab}(u,v) &\equiv \frac{1}{M^3 \tilde{K}^3} \big[ 2 u v + (u+v)M + M^2 \big] \\
&+ \frac{3}{M^2 \tilde{K}^4} \big[ 2 u v + (u+v)M \big] + \frac{12}{M \tilde{K}^5}\, uv ,
\end{split}
\end{equation}
\begin{equation}
\begin{split}
\label{defgba}
G_{ba}(u,v) &\equiv \frac{1}{M^3 \tilde{K}} + \frac{1}{M^3 \tilde{K}^2} (u+v+M) \\
&+ \frac{1}{M^3 \tilde{K}^3} \big[ 2 u v + 2(u+v)M + M^2 \big] + \frac{3}{M^2 \tilde{K}^4} \big[ 2 u v + (u+v)M \big]\\
&+ \frac{12}{M \tilde{K}^5}\, uv ,
\end{split}
\end{equation}
and
\begin{equation}
\begin{split}
\label{defgbb}
&G_{bb}(u,v,x,y) \equiv \frac{1}{M^3 \tilde{K}} \big[ 2 x y + (x+y)M + M^2 \big]\\
&\:+ \frac{1}{M^3 \tilde{K}^2} \big[ 2xy(u+v) + \big( 2xy +(u+v)(x+y)\big) M + (u+v+x+y)M^2\big]\\
&\:+ \frac{2}{M^3 \tilde{K}^3} \big[ 2uvxy + \big( 2xy(u+v) +uv(x+y)\big) M \\ 
&\hspace{30mm}+ (uv+ux+uy+vx+vy+xy)M^2\big]\\
&\:+ \frac{6}{M^2 \tilde{K}^4} \, uvxy \, \bigg( 2 + M\bigg(\frac{1}{u} + \frac{1}{v} + \frac{1}{x} + \frac{1}{y}\bigg)\bigg) + \frac{24}{M \tilde{K}^5} \, uvxy .
\end{split}
\end{equation}

Also, $\mathcal{T}_{c1}, \mathcal{T}_{c2}, \mathcal{T}_{c3}$ in eq.\eqref{calt} are contributions coming from the four point contact interaction, and are given by
\be
\label{caltc1}
\mathcal{T}_{c1} = 36 \, \frac{k_1^2 k_2^2 k_3^2 k_4^2}{\tilde{K}^5},
\ee
\begin{equation}
\label{caltc2}
\mathcal{T}_{c2} = - \frac{1}{8} \frac{k_1^2 k_2^2 (\vecs[k,3]\cdot\vecs[k,4])}{\tilde{K}^3} \bigg( 1 + \frac{3(k_3 + k_4)}{\tilde{K}} + \frac{12 k_3 k_4}{\tilde{K}^2} \bigg) + 23 \, \text{Permutations},
\end{equation}
and
\begin{equation}
\begin{split}
\label{caltc3}
\mathcal{T}_{c3} = \frac{1}{32} \frac{(\vecs[k,1]\cdot\vecs[k,2])(\vecs[k,3]\cdot\vecs[k,4])}{\tilde{K}}\bigg( 1 &+ \frac{\sum_{a<b} k_a k_b}{\tilde{K}^2} + \frac{3k_1k_2k_3k_4}{\tilde{K}^3} \sum_{a=1}^4 \frac{1}{k_a} \\
&+ \frac{12k_1k_2k_3k_4}{\tilde{K}^4}\bigg) + 23 \: \text{Permutations}.
\end{split}
\end{equation}

\section{Isometries, Conserved Charges, and the Symmetries of $\alpha$-vacua}
\label{isogen}
The isometries of de Sitter space have been discussed in section \ref{dsspace}, see eqs.\eqref{iso1}-\eqref{iso4}. The presence of isometries gives rise to conserved currents. Consider the matter part of the action given schematically as
\be
\label{genact}
S_m = \int d^4x \sqrt{-g}~ L(g^{\mu\nu}, \varphi, \del_\mu \varphi) ,
\ee
where $L$ is the Lagrangian of the system. If the isometry is given by
\be
\label{kill}
x^\mu \rightarrow x^\mu + \xi^\mu(x) ,
\ee
where $\xi^\mu$ is infinitesimal, then the current
\be
\label{curr1}
J^\mu = \sqrt{-g} \, \xi_\nu T^{\mu\nu} 
\ee
is conserved.\footnote{The conservation of the current, $\del_\mu J^\mu = 0$, follows from the fact that $\xi^\mu$ satisfies the Killing equation
\begin{equation*}
\nabla_{\mu} \xi_\nu +  \nabla_{\nu} \xi_\mu = 0,
\end{equation*}
and the energy-momentum tensor is covariantly conserved,$\nabla_{\mu} T^{\mu\nu} = 0$.} 
Here, $T^{\mu\nu}$ is the energy-momentum tensor for matter, given by
\begin{equation}
\begin{split}
\label{tmndef}
T^{\mu\nu} &= \frac{2}{\sqrt{-g}} \frac{\delta S_m}{\delta g_{\mu\nu}} \\
&= g^{\mu\nu} L - \frac{\del L}{\del({\del_\mu \phi})} \del^\nu\varphi .
\end{split}
\end{equation}
Associated with the conserved current $J^\mu$, we have the conserved charge
\be
\label{gench}
\mathcal{Q} = \int_{\Sigma_t} \mes[x] \, J^0(x),
\ee
where the integration is on a spacelike surface $\Sigma_t$ labeled by the time coordinate $t$. The conserved charges act as generators of the respective transformations on the fields in the system. For instance, under the transformation eq.\eqref{kill}, we have for a scalar,
\begin{equation}
\label{chp}
\varphi(x) \rightarrow \varphi'(x) = U \varphi(x) U^{-1} = \varphi(x^\mu - \xi^\mu),
\end{equation}
where $U$ is the operator obtained by exponentiating the conserved charge $\mathcal{Q}$, 
\be
\label{defu}
U  = \text{exp}({i \theta \mathcal{Q}}),
\ee
with $\theta$ being the parameter of the transformation.

\subsection{Conserved Charges for a Massless Free Scalar Field}
\label{conschg}
Following the discussion above, we can construct the conserved charges corresponding to the de Sitter isometries for the massless free field theory. The action is given by eq.\eqref{mlact}. The energy-momentum tensor for the field $\varphi$ can be calculated using eq.\eqref{tmndef},
\begin{equation}
\label{mlem}
T^{\mu\nu} = g^{\mu\alpha} g^{\nu\beta} \del_{\alpha}\varphi \, \del_{\beta}\varphi - \half \, g^{\mu\nu} g^{\alpha\beta} \del_{\alpha}\varphi \, \del_{\beta}\varphi.
\end{equation}
Then, from eq.\eqref{gench}, the conserved charge for spatial translations, eq.\eqref{iso1}, is given by
\begin{equation}
\begin{split}
\label{conc1}
\mathbf{P}_i &= \frac{1}{\eta^4 H^4} \int d^3x \,  :T^0_{\,i}:\\
&= \int \frac{\mes[k]}{(2\pi)^3} \, k_i \, a^\dagger_{\vec[k]} a_{\vec[k]} \, ,
\end{split}
\end{equation}
where in deriving the last step we have used the mode expansion in eq.\eqref{bdmode1}, with the modes $u_k(\eta)$ given by eq.\eqref{bdmode2}. Similarly, for rotational symmetry eq.\eqref{iso2}, we obtain the conserved charge
\begin{equation}
\begin{split}
\label{conc2a}
\mathbf{L}_{ij} &= \frac{1}{\eta^4 H^4} \int d^3x \, :\Big(x_j T^0_{\,i} - x_i T^0_{\,j}\Big):\\
&= i \int \frac{\mes[k]}{(2\pi)^3} \, a^\dagger_{\vec[k]} \left( k_i \frac{\del}{\del k^j} - k_j \frac{\del}{\del k^i} \right) a_{\vec[k]}.
\end{split}
\end{equation}
Consider next the case of dilatations, eq.\eqref{iso3}. The conserved charge is
\begin{equation}
\begin{split}
\label{conc3b}
\mathbf{D} &= \frac{1}{\eta^4 H^4} \int d^3x :x^\mu \, T^0_{\,\mu}:\\
&= i \int \frac{\mes[k]}{(2\pi)^3} \, a^\dagger_{\vec[k]} \left( k_i \frac{\del}{\del k^i} + \frac{3}{2} \right) a_{\vec[k]}.
\end{split} 
\end{equation}
Finally, we consider the special conformal transformation, eq.\eqref{iso4}. The conserved charge is given by
\begin{equation}
\begin{split}
\label{conc4a}
\mathbf{K}_i &= \frac{1}{\eta^4 H^4} \int d^3x :\Big(2 \eta x_i T^0_{\,0} + 2 x_i x^j T^0_{\,j} + (\eta^2 - \vec[x]^2) T^0_{\,i}\Big):\,,\\
&= \int \frac{\mes[k]}{(2\pi)^3} \bigg( \frac{9}{4} \frac{k_i}{k^2} a^\dagger_{\vec[k]} a_{\vec[k]} - 3 a^\dagger_{\vec[k]} \frac{\del a_{\vec[k]}}{\del k^i} + a^\dagger_{\vec[k]} k_i \frac{\del^2 a_{\vec[k]}}{\del k^j \del k_j} - 2 a^\dagger_{\vec[k]} k_j \frac{\del^2 a_{\vec[k]}}{\del k_j \del k^i} \bigg).
\end{split}
\end{equation}

From equations \eqref{conc1}, \eqref{conc2a}, \eqref{conc3b} and \eqref{conc4a}, due to the normal ordering, it is clear that the Bunch-Davies vacuum, defined by the condition that $a_{\vec[k]} |0\rangle = 0 ~\forall~ \vec[k]$, is annihilated by the generators of the de Sitter isometries, and is thus conformally invariant.

\subsection{Symmetries of the  $\alpha$-vacua in the Massless Free Field Theory}
\label{symfree}
We now consider $\alpha$-vacua in the massless free field theory. 
The expression for the free $\alpha$-vacua in terms of the Bunch-Davies vacuum is given in eq.\eqref{alphad}. For compactness, we rewrite the expression as
\begin{equation}
\label{alphadef}
|\alpha \rangle = c_1 \,\text{exp}\left(c_2\int\frac{d^3 k}{(2\pi)^3} \, a_{\vec [k]}^{\dagger}a_{-\vec [k]}^{\dagger}\right)|0\rangle ,
\end{equation}
where $c_1 = 1/\mathcal{N}$ and $c_2 = \frac{i}{2} \, \text{tanh}(\alpha)$. 

The late time two point function for the probe in this vacuum is given by
\be
\label{}
\langle \alpha |\varphi(\vecs[k,1]) \varphi(\vecs[k,2]) |\alpha \rangle = (2\pi)^3 \delta^3(\vecs[k,1] + \vecs[k,2]) \, \frac{H^2}{2k^3} \, \text{cosh}(2\alpha),
\ee
and can be shown to be conformally invariant. This shows that the $\alpha$-vacua are conformally invariant. More correctly, this is true upto subtleties involving zero modes, which we will ignore here.  

In what follows, we will study  the symmetry properties more directly in terms of the action of the conserved charges on the state eq.\eqref{alphadef}. Let $\mathbf{U}$ denote an element of the de Sitter isometry group. We then have
\begin{equation}
\begin{split}
\label{alptrans}
\mathbf{U}|\alpha\rangle&=\mathbf{U}\left[c_1\,\text{exp}\left(c_2\int\frac{d^3 k}{(2\pi)^3} \, a_{\vec [k]}^{\dagger}a_{-\vec [k]}^{\dagger}\right)|0\rangle\right]\\
&=c_1\,\text{exp}\left(c_2\int\frac{d^3 k}{(2\pi)^3} \,\mathbf{U} a_{\vec [k]}^{\dagger}\mathbf{U}^{-1}\,\mathbf{U}a_{-\vec [k]}^{\dagger}\mathbf{U}^{-1}\right)|0\rangle,
\end{split}
\end{equation}
where we have used the invariance of the Bunch-Davies vacuum under the action of the de Sitter isometry group, $\mathbf{U}|0\rangle=|0\rangle$. The operator $a_{\vec [k]}^{\dagger}$ can be expressed in terms of the field $\varphi_{\vec[k]}$ and its conjugate momentum $\pi_{\vec[k]}$ as
\begin{equation}
\label{invmode}
a_{\vec[k]}^{\dagger}=\left(i\sqrt{\frac{k}{2}}\frac{1}{\eta H^2} \, \varphi_{-\vec[k]}(\eta) -\frac{i}{\sqrt{2k^3}}(1+ik\eta) \,\pi_{-\vec[k]}(\eta)\right)e^{-ik\eta},
\end{equation}
where
\begin{equation}
\label{invmodeh}
\varphi_{-\vec[k]}(\eta) =\int d^3 x\, e^{i \vec[k]\cdot \vec[x]}\,\varphi({\eta,\vec[x]}), \quad \pi_{-\vec[k]}(\eta)=\int d^3 x \,e^{i \vec[k]\cdot\vec[x]} \,\pi({\eta,\vec[x]}).
\end{equation}
The transformation of $a_{\vec[k]}^\dagger$ under the action of $\mathbf{U}$ is thus given by
\begin{equation}
\begin{split}
 \label{adtrans}
 \mathbf{U}a_{\vec [k]}^{\dagger}\mathbf{U}^{-1}&=i\sqrt{\frac{k}{2}}\frac{1}{\eta H^2}\left(\int d^3 x\, e^{i \vec [k]\cdot \vec [x]}\,\mathbf{U}\varphi({\eta,\vec [x]})\mathbf{U}^{-1}\right)e^{-ik\eta}
 \\&\qquad-\frac{i}{\sqrt{2k^3}}(1+ik\eta)\left(\int d^3 x\, e^{i \vec [k]\cdot \vec [x]}\,\mathbf{U}\pi({\eta,\vec [x]})\mathbf{U}^{-1}\right) e^{-ik\eta}.
\end{split}
\end{equation}
Using the expression eq.\eqref{adtrans}, we can now check for the invariance of the $\alpha$-vacua under the de Sitter isometries on a case by case basis. For translations $\vec[x] \rightarrow \vec[x]+\vec[r]$, it is easy to see that $a_{\vec [k]}^{\dagger}$ picks up a phase,
\be
\label{adagmom13}
\mathbf{U}_{\textsc{t}}a_{\vec [k]}^{\dagger}\mathbf{U}_{\textsc{t}}^{-1} =e^{i \vec [k]\cdot \vec [r]}a_{\vec [k]}^{\dagger}\, ,
\ee
which on substituting in eq.\eqref{alptrans} gives 
\be
\label{traninv}
\mathbf{U}_{\textsc{t}}|\alpha\rangle=|\alpha\rangle.
\ee
We see that $\alpha$-vacua are translationally invariant. Similarly, for spatial rotations $\vec[x] \rightarrow \Lambda\vec[x]$, we have
\begin{equation}
\label{adagrot2}
\mathbf{U}_{\textsc{r}}a_{\vec [k]}^{\dagger}\mathbf{U}_{\textsc{r}}^{-1} = a_{\Lambda^{\text{T}} \vec [k]}^{\dagger},
\end{equation}
where $\Lambda^{\text{T}}$ denotes the transpose of $\Lambda$. Using the result of eq.\eqref{adagrot2} in eq.\eqref{alptrans}, we see that
\be
\label{rotinva}
\mathbf{U}_{\textsc{r}}|\alpha\rangle=|\alpha\rangle,
\ee
where in getting eq.\eqref{rotinva}, we have performed a change of variable $\vec[k]\rightarrow\Lambda\vec[k]$ and used $\Lambda^{\text{T}} \Lambda =1$. Eq.\eqref{rotinva} proves that the $\alpha$-vacua are rotationally invariant as well.

Now consider a scaling transformation that scales the coordinates $(\eta,\vec[x])$ to $(\lambda \eta,\lambda \vec[x])$. Then the action of the scaling operator on $\varphi({\eta,\vec[x]})$ and $\pi({\eta,\vec [x]})$ is
\begin{equation}
\begin{split}
\label{phisc}
&\mathbf{U}_{\textsc{s}}\varphi({\eta,\vec[x]})\mathbf{U}_{\textsc{s}}^{-1}=\varphi(\lambda^{-1}\eta,\lambda^{-1}\vec[x]),\\
&\mathbf{U}_{\textsc{s}}\pi({\eta,\vec[x]}) \mathbf{U}_{\textsc{s}}^{-1} = \frac{1}{\lambda^3}\, \pi(\lambda^{-1}\eta,\lambda^{-1}\vec[x]),
\end{split}
\end{equation}
where the RHS of the second transformation has a factor of $\lambda^{-3}$ because the scaling dimension of $\pi({\eta,\vec [x]})$ is 3. Using eq.\eqref{phisc} in eq.\eqref{adtrans}, and performing manipulations similar to the previous cases, we find that
\begin{equation}
\label{adagsc2}
\mathbf{U}_\textsc{s}a_{\vec [k]}^{\dagger}\mathbf{U}_{\textsc{s}}^{-1} = \lambda^{3/2}a_{\lambda \vec[k]}^{\dagger},
\end{equation}
which when used in eq.\eqref{alptrans} gives
\be
\label{sclinv}
\mathbf{U}_{\textsc{s}}|\alpha\rangle=|\alpha\rangle,
\ee
proving that scaling is also a symmetry of the $\alpha$-vacua states.

Finally, to argue for the invariance of $\alpha$-vacua for the case of special conformal transformations, we follow a different approach. This is because if we try to calculate $\mathbf{U}_{\textsc{k}} a_{\vec [k]}^{\dagger}\mathbf{U}_{\textsc{k}}^{-1}$ just as we did for the other isometries, there is no obvious way to express the result as a function of $a_{\vec [k]}^{\dagger}$ because of the nonlinear nature of the SCT. We start by considering eq.\eqref{alphadef}, which can also be written as
\begin{equation}
\label{redalp}
|\alpha\rangle=c_1\,\text{exp}\left( {c_2 \hat{\mathbf{B}}}  \right)|0\rangle; \quad \hat{\mathbf{B}}=\int\frac{d^3 k}{(2\pi)^3} \, a_{\vec [k]}^{\dagger}a_{-\vec [k]}^{\dagger}.
\end{equation}
Now, for the $\alpha$-vacua to be invariant under an SCT, we must have
\begin{equation}
  \label{alpsct}
  \mathbf{U}_{\textsc{k}}|\alpha\rangle=|\alpha\rangle ,
\end{equation}
where $\mathbf{U}_{\textsc{k}} = e^{-i\vec[b]\cdot\mathbf{K}}$, with $\vec[b]$ being the transformation parameter. Thus, we must have 
\begin{equation}
\label{condscti}
\big[e^{-i\vec[b]\cdot\mathbf{K}},e^{\hat{\mathbf{B}}}\big]=0,
\end{equation}
or equivalently,
\be
\label{condsimp}
\big[\mathbf{K}_i,\hat{\mathbf{B}}\big]=0.
\ee

Using the expression for $\mathbf{K}_i$ given in eq.\eqref{conc4a}, we have
\begin{equation}
\label{kbcomm}
\big[\mathbf{K}_i,\hat{\mathbf{B}}\big]=2 \int \frac{\mes[k]}{(2\pi)^3}\left(-k_i a^\dagger_{\vec[k]}\frac{\del^2 a_{-\vec[k]}^{\dagger}}{\del k^j \del k_j}+2k_j a^\dagger_{\vec[k]}\frac{\del^2 a_{-\vec[k]}^{\dagger}}{\del k_j \del k^i}+3 a^\dagger_{\vec[k]} \frac{\del a_{-\vec[k]}^{\dagger}}{\del k^i}\right).
\end{equation}
Now, the RHS of the above expression is actually a surface term, and can be dropped. To see this, we change the integration variable $\vec[k]\rightarrow-\vec[k]$ in eq.\eqref{kbcomm}, and perform several integrations by parts to get
\begin{equation}
\label{kbcomm3}
\big[\mathbf{K}_i,\hat{\mathbf{B}}\big]=\int \frac{\mes[k]}{(2\pi)^3}\left[\frac{\del}{\del k^i}\left(a^\dagger_{\vec[k]}k^j \frac{\del a_{-\vec[k]}^{\dagger}}{\del k^j}\right)\,-\,\frac{\del}{\del k^j}\left\{a^{\dagger}_{\vec[k]}\left(k_i \frac{\del}{\del k_j}-k^j\frac{\del}{\del k^i}\right)a_{-\vec[k]}^{\dagger}\right\}\right].
\end{equation}
Therefore, we see that upto surface terms in momentum space, the condition eq.\eqref{condsimp} is indeed met. The surface terms will not contribute in any state where only finite momenta are excited. However, there is a subtlety in concluding from the above argument that the $\alpha$-vacua we are considering are invariant under $\mathbf{K}_i$, since arbitrarily high $\vec[k]$ modes are excited in these states above the Bunch-Davies vacuum. However, we saw at the beginning of the section that for the free theory, the two point function directly shows that the $\alpha$-vacua are indeed conformally invariant. 

\subsection{Massless Interacting Scalar Field and Symmetries of the $|\alpha\rangle_{\text{I}}$ Vacua}
\label{ialpsym}
We now briefly discuss symmetry properties of the interacting $\alpha$-vacua, which we have denoted by $|\alpha\rangle_{\text{I}}$. Consider the action eq.\eqref{intact} for a massless interacting scalar field in de Sitter space, which we discuss below. This example will in fact illustrate a more general issue regarding the form conserved charges take in the Heisenberg and Interaction pictures. 

For the interacting theory, eq.\eqref{intact}, the energy-momentum tensor has the form
\begin{equation}
\label{emfi}
T^{\mu\nu} = g^{\mu\alpha} g^{\nu\beta} \del_{\alpha}\varphi \, \del_{\beta}\varphi - \half \, g^{\mu\nu} g^{\alpha\beta} \del_{\alpha}\varphi \, \del_{\beta}\varphi - g^{\mu\nu} \frac{\lambda}{3!} \, \varphi^3,
\end{equation}
and depends on the $\lambda \varphi^3$ interaction. 
As a result, the  expressions for the conserved charges also change from the non-interacting case, since they depend on the form of $T_{\mu\nu}$. In particular,  the generators of dilatation and special conformal transformations take the form, 
\begin{equation}
\begin{split}
\label{dilint}
\mathbf{D^{\text{int}}} = \mathbf{D} - \frac{1}{\eta^3 H^4} \, \frac{\lambda}{3!} \int d^3x :\varphi^3: \, ,
\end{split}
\end{equation}
and 
\begin{equation}
\begin{split}
\label{sctcci}
\mathbf{K}_i^{\text{int}} = \mathbf{K}_i - \frac{2}{\eta^3 H^4} \, \frac{\lambda}{3!}  \int d^3x :x_i \, \varphi^3:.
\end{split}
\end{equation}
However, the generators for translations and rotations stay the same, eqs.\eqref{conc1} and \eqref{conc2a}, for the interacting theory as well.

Note that in writing these expressions down we are  working in the Heisenberg picture. The expressions for the charges, eqs.\eqref{conc1}, \eqref{conc2a}, \eqref{dilint} and \eqref{sctcci}, have been obtained from the action eq.\eqref{intact}  using the Noether procedure. Varying the same action gives the equation of motion for the field $\varphi$ in the Heisenberg picture. 

In contrast, in the interaction picture the form of the charges is different, and in fact the same as in the free field theory without interactions. This can be understood as follows. Consider dilatations and special conformal transformations. The charges must generate the symmetry transformations 
\begin{equation}
\begin{split}
\label{shs1}
&\varphi(\eta, \vec[x]) \xrightarrow{Dilatation} \varphi\big((1-\epsilon)\eta, (1-\epsilon)\vec[x]\big),\\ &\varphi(\eta,\vec[x]) \xrightarrow{SCT} \varphi\big((1-2\vec[b]\cdot\vec[x])\eta, \vec[x]- 2(\vec[b]\cdot\vec[x])\vec[x]-\vec[b](\eta^2-\vec[x]^2)\big),
\end{split}
\end{equation}
where the dilatations and special conformal transformations are given by eqs.\eqref{iso3} and \eqref{iso4} respectively. In the interaction picture, $\varphi$ continues to have the free field mode expansion, eq.\eqref{bdmode1}.
Thus the form of the charges in terms of $a_{\vec[k]}, a^\dagger_{\vec[k]}$, continues to be the same as given in eq.\eqref{conc3b} and eq.\eqref{conc4a} respectively, so that the field $\varphi$ transforms correctly as given by eq.\eqref{shs1}. 

Now, as the charges in the interaction picture continue to have the same form as for the free theory, they commute past the exponential in the definition of the interacting $\alpha$-vacuum in eq.\eqref{alphai}, as discussed in appendix \ref{symfree}. Assuming that the interacting Bunch-Davies vacuum $|\Omega\rangle$ is invariant under the $O(1,4)$ symmetry algebra then implies that the interacting $\alpha$-vacua defined in eq.\eqref{alphai} also share the same symmetry algebra.\footnote{Once again, there are subtleties due to boundary terms for $\mathbf{K}_i$, see discussion below eq.\eqref{kbcomm3}, and also due to the presence of zero modes.}

\section{Three Point Function in the Bunch-Davies Vacuum}
\label{3ptbd}
The late time three point function for the probe field in the Bunch-Davies vacuum can be obtained from eq.\eqref{3pta} by setting the parameter $\alpha = 0$,
\begin{equation}
\begin{split}
\label{ibd4}
\langle \Omega|& \varphi(\vecs[k,1]) \varphi(\vecs[k,2]) \varphi(\vecs[k,3]) |\Omega \rangle = \frac{2}{3} \, \lambda H^2 \, (2\pi)^3 \delta^3\Big(\sum_{a=1}^3 \vecs[k,a]\Big) \Big[\prod_{a=1}^3 \frac{1}{2k_a^{\,3}}\Big] \times\\
&\bigg( -\frac{4}{3} \sum_{a=1}^3 k_a^{\,3} - \sum_{a \neq b} k_a k_b^{\,2} + k_1k_2k_3 + \big(\gamma + ln|\varepsilon|\big) \Big(\sum_{a=1}^3 k_a^{\,3}\Big) + ln(K) \Big(\sum_{a=1}^3 k_a^{\,3}\Big)\bigg).
\end{split}
\end{equation}

One may have tried to compute the three point function in the interacting Bunch-Davies vacuum not by following the standard in-in procedure, but by explicitly computing the state $|\Omega\rangle$ perturbatively as a series of corrections to the free vacuum $|0\rangle$, to the desired order in the parameter $\lambda$ on the initial time slice, and then evolving the resulting state in time. For the present case, one has to the leading order in $\lambda$,\footnote{Though the system is inherently time dependent, we are justified to use time independent perturbation theory because we are working on a fixed time slice at $\eta = \eta_0$.}
\begin{equation}
\label{omg1}
|\Omega\rangle = |0\rangle + \sum_{m}\mbox{}^{'} \frac{\langle m| H_{int}^I(\eta_0)|0\rangle}{E_0 - E_m} \, |m\rangle,
\end{equation}
where $|m\rangle$ is the $m^{th}$ energy eigenstate of the free Hamiltonian at $\eta = \eta_0$, and $H_{int}^I(\eta_0)$ is the interaction Hamiltonian in the interaction picture at $\eta = \eta_0$, eq.\eqref{hintcorr}, where $\eta_0 \rightarrow -\infty$. Note that the $'$ symbol on the summation in eq.\eqref{omg1} implies the exclusion of the $m=0$ term. Eq.\eqref{omg1} gives
\begin{equation}
\label{omg2}
|\Omega\rangle = |0\rangle - \frac{\lambda}{3!} \, \frac{1}{\eta_0^4 H^4} \, \sum_{m}\mbox{}^{'} \frac{1}{E_m} \bigg( \langle m| \int d^3x \, \varphi_I^3(\eta_0,\vec[x]) |0\rangle\bigg)|m\rangle.
\end{equation}
Now, the mode expansion of $\varphi_I(\eta_0,\vec[x])$ considering $\eta_0 \rightarrow -\infty$ is
\be
\label{apmodex}
\varphi_I(\eta_0,\vec[x]) \approx i \eta_0 H \int \frac{d^3k}{(2\pi)^3} \frac{1}{\sqrt{2k}} \big( a_{\vec[k]} \text{e}^{-ik\eta_0} - a^\dagger_{-\vec[k]} \text{e}^{ik\eta_0} \big) \text{e}^{i\vec[k]\cdot \vec[x]}.
\ee
Substituting eq.\eqref{apmodex} into eq.\eqref{omg2}, we get
\begin{equation}
\begin{split}
\label{omg3}
|\Omega\rangle = |0\rangle &+ \frac{i\lambda}{3!} \frac{1}{\eta_0 H} \sum_{m}\mbox{}^{'} \frac{1}{E_m} \, |m\rangle \times\\
& \int d^3x \, \langle m| \left[\prod_{a=1}^3 \left( \int \frac{d^3k_a}{(2\pi)^3} \frac{1}{\sqrt{2k_a}} \big( a_{\vecs[k,a]} \text{e}^{-ik_a\eta_0} - a^\dagger_{-\vecs[k,a]} \text{e}^{ik_a\eta_0} \big)\, \text{e}^{i\vecs[k,a]\cdot\vec[x]}\right)\right] |0\rangle.
\end{split}
\end{equation}
One can get contributions only from single and three particle states in eq.\eqref{omg3}. On explicit computation, we find that the single particle contribution vanishes, and we are left with only the three particle contribution, giving
\begin{equation}
\begin{split}
\label{omg4}
|\Omega\rangle = |0\rangle -\frac{i\lambda}{3!} \frac{1}{\eta_0 H} \int \prod_{a=1}^3 \bigg(\frac{d^3k_a}{(2\pi)^3} \, \frac{1}{\sqrt{2k_a}}\bigg) (2\pi)^3 \delta^3\Big(\sum_{a=1}^3 \vecs[k,a]\Big) \frac{\text{e}^{iK\eta_0}}{K} \,a_{\vecs[k,1]}^\dagger  a_{\vecs[k,2]}^\dagger a_{\vecs[k,3]}^\dagger|0\rangle.
\end{split}
\end{equation}
We see that in the limit $\eta_0 \rightarrow -\infty$, $|\Omega\rangle$ becomes $|0\rangle$. This would then give rise to the same expression for the three point function as in eq.\eqref{ibd4}. Note that the vanishing of the correction term due to $\eta_0$ sitting in the denominator happens only for the case of cubic self interaction that we have considered; if one considers quartic or higher powers in the interaction term, then there will be genuine corrections to $|0\rangle$ and the three point function.

\section{Evaluating the Three Point Function in the $|\alpha\rangle_{\text{I}}$ Vacuum}
\label{calci}
In this appendix, we provide some details for the evaluation of the inflationary three point function in the $|\alpha\rangle_{\mathrm{I}}$ vacuum, section \ref{3pgen}. We follow the standard in-in calculation procedure which is outlined in section \ref{intml}. We start by performing the field redefinition eq.\eqref{redef}. The interaction Hamiltonian for the redefined field is given by eq.\eqref{ihamtz}. The complete three point function of $\zeta$ is then calculated by combining the contributions coming from the three point function of $\tilde{\zeta}$, eq.\eqref{def3ptz}, and the field redefinition terms; see eq.\eqref{rel2p}. We make use of the prescription eq.\eqref{changet}, and the free field mode expansion given by eq.\eqref{newmodez}. A straightforward but lengthy calculation then gives
\begin{equation}
\begin{split}
\label{tpall1}
{}_{\mathrm{I}}\langle \alpha | \zeta(\vecs[k,1]) &\zeta(\vecs[k,2]) \zeta(\vecs[k,3])|\alpha\rangle_{\mathrm{I}} = (2\pi)^3 \delta^3\Big(\sum_{a=1}^3 \vecs[k,a]\Big) \Big[\prod_{a=1}^3 \frac{1}{(2k_a^{\,3})}\Big] \frac{H^4}{4\epsilon} \times\\ 
&\bigg[ \bigg\{ i \big[\text{cosh}(\alpha) - i \text{sinh}(\alpha)\big]^3 \Big( \text{cosh}^3(\alpha)\, \mathcal{I}_1 + i \, \text{cosh}^2(\alpha) \text{sinh}(\alpha) \, \mathcal{I}_2\\
&\hspace{27mm}- \text{cosh}(\alpha) \text{sinh}^2(\alpha) \, \mathcal{I}_3 - i \, \text{sinh}^3(\alpha) \, \mathcal{I}_4 \Big) + c.c. \bigg\}\\
&+ \big\{\vecs[k,1] \rightarrow \vecs[k,2] \rightarrow \vecs[k,3] \rightarrow \vecs[k,1] \big\} + \big\{\vecs[k,1] \rightarrow \vecs[k,3] \rightarrow \vecs[k,2] \rightarrow \vecs[k,1] \big\}\\
&+ 2\, \frac{2\epsilon-\eta}{\epsilon} \, \text{cosh}^2(2\alpha) \Big(\sum_{a=1}^3 k_a^{\,3} \Big)\bigg],
\end{split}
\end{equation}
where $\mathcal{I}_1, \mathcal{I}_2, \mathcal{I}_3$ and $\mathcal{I}_4$ are the integrals
\begin{equation}
\begin{split}
\label{i1}
\mathcal{I}_1 = \int\limits_{\eta_0}^0 d\eta\, \text{e}^{iK\eta} (1-ik_1\eta)\bigg[\mathcal{F}- \frac{1}{\eta^2}(1-ik_2\eta) (1-ik_3\eta) \vecs[k,2]\cdot\vecs[k,3]\bigg],
\end{split}
\end{equation}
\begin{equation}
\begin{split}
\label{i2}
\mathcal{I}_2 &= \int\limits_{\eta_0}^0 d\eta\, \text{e}^{i(k_1+k_2-k_3)\eta} (1-ik_1\eta)\bigg[ \mathcal{F}- \frac{1}{\eta^2}(1-ik_2\eta) (1+ik_3\eta) \vecs[k,2]\cdot\vecs[k,3]\bigg]\\
&+\int\limits_{\eta_0}^0 d\eta\, \text{e}^{i(k_1-k_2+k_3)\eta} (1-ik_1\eta)\bigg[\mathcal{F}
- \frac{1}{\eta^2}(1+ik_2\eta) (1-ik_3\eta) \vecs[k,2]\cdot\vecs[k,3]\bigg]\\
&+\int\limits_{\eta_0}^0 d\eta\, \text{e}^{i(-k_1+k_2+k_3)\eta} (1+ik_1\eta)\bigg[\mathcal{F}
- \frac{1}{\eta^2}(1-ik_2\eta) (1-ik_3\eta) \vecs[k,2]\cdot\vecs[k,3]\bigg],
\end{split}
\end{equation}
\begin{equation}
\begin{split}
\label{i3}
\mathcal{I}_3 &= \int\limits_{\eta_0}^0 d\eta\, \text{e}^{i(k_1-k_2-k_3)\eta} (1-ik_1\eta)\bigg[ \mathcal{F}- \frac{1}{\eta^2}(1+ik_2\eta) (1+ik_3\eta) \vecs[k,2]\cdot\vecs[k,3]\bigg]\\
&+\int\limits_{\eta_0}^0 d\eta\, \text{e}^{i(-k_1+k_2-k_3)\eta} (1+ik_1\eta)\bigg[\mathcal{F}
- \frac{1}{\eta^2}(1-ik_2\eta) (1+ik_3\eta) \vecs[k,2]\cdot\vecs[k,3]\bigg]\\
&+\int\limits_{\eta_0}^0 d\eta\, \text{e}^{i(-k_1-k_2+k_3)\eta} (1+ik_1\eta)\bigg[\mathcal{F}
- \frac{1}{\eta^2}(1+ik_2\eta) (1-ik_3\eta) \vecs[k,2]\cdot\vecs[k,3]\bigg],
\end{split}
\end{equation}
and
\begin{equation}
\begin{split}
\label{i4}
\mathcal{I}_4 = \int\limits_{\eta_0}^0 d\eta\, \text{e}^{-iK\eta} (1+ik_1\eta)\bigg[\mathcal{F}- \frac{1}{\eta^2}(1+ik_2\eta) (1+ik_3\eta) \vecs[k,2]\cdot\vecs[k,3]\bigg],
\end{split}
\end{equation}
where the quantity $\mathcal{F}$ is a function of $\vecs[k,1], \vecs[k,2], \vecs[k,3]$, given by
\be
\label{def}
\mathcal{F} = k_2^{\,2} k_3^{\,2} - k_2^{\,2} \vecs[k,1]\cdot\vecs[k,3] - k_3^{\,2} \vecs[k,1]\cdot\vecs[k,2].
\ee

Now, to evaluate the three point function, we need to calculate the $\eta$ integrals in $\mathcal{I}_1$ to $\mathcal{I}_4$. These are given by\footnote{A $'$ on the delta function denotes a derivative with respect to its argument. For e.g. in eq.\eqref{i21}, we have
\begin{equation*}
\delta'(k_1+k_2-k_3) = \frac{\del \delta(k_1+k_2-k_3)}{\del (k_1+k_2-k_3)}.
\end{equation*}}
\begin{equation}
\begin{split}
\label{i11}
\mathcal{I}_1 + c.c. &= 2 \, (\vecs[k,2]\cdot\vecs[k,3]) \,\frac{\text{cos}(K\sigma)}{\sigma} ,\\
i \mathcal{I}_1 + c.c. &= \frac{2\mathcal{F}}{K} \bigg(1 + \frac{k_1}{K}\bigg) + 2 \, (\vecs[k,2]\cdot\vecs[k,3]) \bigg( \frac{k_1k_2k_3}{K^2} + \frac{k_1k_2+k_2k_3+k_3k_1}{K} - K \bigg) ,
\end{split}
\end{equation}
\begin{equation}
\begin{split}
\label{i21}
\mathcal{I}_2 + c.c. = &(\vecs[k,2]\cdot\vecs[k,3])  \Bigg[ \bigg\lbrace \frac{2}{\sigma} \, \text{cos}[(k_1+k_2-k_3)\sigma] \\
&\hspace{2mm}+ (k_1k_2-k_2k_3-k_3k_1) \, \delta(k_1+k_2-k_3) + k_1k_2k_3  \, \delta\,'(k_1+k_2-k_3) \bigg\rbrace \\
&\hspace{27mm}+\big\lbrace k_1\rightarrow k_2\rightarrow k_3\rightarrow k_1\big\rbrace + \big\lbrace k_1\rightarrow k_3\rightarrow k_2\rightarrow k_1\big\rbrace \Bigg]\\
&\hspace{2mm}+\mathcal{F} \Big[\delta(k_1+k_2-k_3)  + \delta(k_1-k_2+k_3) + \delta(-k_1+k_2+k_3)\Big]\\
&\hspace{2mm}-k_1\mathcal{F} \Big[\delta\,'(k_1+k_2-k_3)  + \delta\,'(k_1-k_2+k_3) - \delta\,'(-k_1+k_2+k_3)\Big],
\end{split}
\end{equation}
\begin{equation}
\begin{split}
\label{i22}
i \mathcal{I}_2 + c.c. = &- 2 K (\vecs[k,2]\cdot\vecs[k,3]) \\
&+ 2 \, (\vecs[k,2]\cdot\vecs[k,3]) \Bigg[ \bigg\lbrace \frac{k_1k_2-k_2k_3-k_3k_1}{k_1+k_2-k_3}  - \frac{k_1k_2k_3}{(k_1+k_2-k_3)^2}\bigg\rbrace\\
&\hspace{20mm}+ \big\lbrace k_1\rightarrow k_2\rightarrow k_3\rightarrow k_1\big\rbrace + \big\lbrace k_1\rightarrow k_3\rightarrow k_2\rightarrow k_1\big\rbrace \Bigg] \\
&+2\mathcal{F} \bigg[ \frac{1}{k_1+k_2-k_3} + \frac{1}{k_1-k_2+k_3} + \frac{1}{-k_1+k_2+k_3}  \bigg]\\
&+2k_1\mathcal{F} \bigg[ \frac{1}{(k_1+k_2-k_3)^2} + \frac{1}{(k_1-k_2+k_3)^2} - \frac{1}{(-k_1+k_2+k_3)^2}  \bigg] ,
\end{split}
\end{equation}
where $\sigma \rightarrow 0$ is a small $|\eta|$ cutoff. Also, note that in the shorthand notation used in eqs.\eqref{i21} and \eqref{i22}, the cyclic permutations in $k_1, k_2, k_3$ act only within the square brackets, leaving the external $\vecs[k,2]\cdot\vecs[k,3]$ factor unaffected. Finally, the results for the integrals $\mathcal{I}_3$ and $\mathcal{I}_4$ can be easily expressed as
\begin{equation}
\begin{split}
\label{i31}
\mathcal{I}_3 + c.c. &= \mathcal{I}_2 +c.c.\\
-\, i\mathcal{I}_3 + c.c. &= i\mathcal{I}_2 +c.c.
\end{split}
\end{equation}
and
\begin{equation}
\begin{split}
\label{i41}
\mathcal{I}_4 + c.c. &= \mathcal{I}_1 +c.c.\\
-\, i\mathcal{I}_4 + c.c. &= i\mathcal{I}_1 +c.c.
\end{split}
\end{equation}

Note that the sum of any two amongst $(k_1, k_2, k_3)$ is always greater than the third, as $\vecs[k,1], \vecs[k,2]$ and $\vecs[k,3]$ form the three sides of a triangle. Hence, the terms involving $\delta$-functions or derivatives of $\delta$-functions in the equations above never click. Substituting results from eqs.\eqref{i11}-\eqref{i41} into eq.\eqref{tpall1}, we get the three point function given in eq.\eqref{tpall2}.

\section{Estimating the Back-reaction for $\alpha$-vacua}
\label{back-reaction}
In this appendix, we compute the quantum stress-energy tensor in an $\alpha$-vacua state for the free scalar field theory in de Sitter space, eq.\eqref{mlact}. This simple example helps in illustrating the fact that the contribution of the quantum stress tensor to the total stress-energy in an $\alpha$-vacua state is divergent. 

We start by considering the expression for the stress tensor given in eq.\eqref{mlem}, and use the mode expansion eq.\eqref{bdmode1} for the free field in the Bunch-Davies vacuum. Using the prescription of normal ordering, we get
\be
\label{br1}
\langle 0| :\!T_{\mu\nu}\!:|0\rangle = 0,
\ee
ensuring that there is no back-reaction on the background geometry for the Bunch-Davies vacuum. Note that the normal ordering is being done here with respect to the creation and annihilation operators $a_{\vec[k]}, a^\dagger_{\vec[k]}$, used in the definition of the Bunch-Davies vacuum. The rationale for this is that the short distance modes in the Bunch-Davies vacuum are in essentially the same state as in the Minkowski vacuum in flat space, and their contribution to $T_{\mu\nu}$, which is the source term for gravity, should therefore vanish. 

Now, using the same normal ordering prescription, we can compute the quantum stress tensor in the $\alpha$-vacua state. This gives, for instance,
\begin{equation}
\label{br2}
\langle \alpha| :\!T_{00}\!:|\alpha\rangle = \frac{H^2}{4\pi^2} \int\limits_0^\infty dk \, k \bigg[ (1+ 2k^2 \eta^2) \, \text{sinh}^2\alpha + \frac{\text{sinh}(2\alpha)}{2} \big(\sin(2k\eta) - 2k\eta \cos(2k\eta)\big)\bigg],
\end{equation}
which clearly shows that the back-reaction is divergent for large values of $k$. For the $T_{0i}$ component of the stress-energy tensor, we find that
\begin{equation}
\label{br3}
\langle \alpha| :\!T_{0i}\!:|\alpha\rangle = i H^2 \eta \int \frac{d^3k}{(2\pi)^3} \, \frac{k_i}{k} \bigg[ -\sinh^2\alpha + \frac{\sinh(2\alpha)}{2} \big( k\eta \cos(2k\eta) - \sin(2k\eta)\big)\bigg],
\ee
which vanishes due to the integrand being an odd function of $k_i$. 
We also get 
\begin{equation}
\begin{split}
\label{br4}
\langle \alpha| &:\!T_{ij}\!:|\alpha\rangle = H^2 \int \frac{d^3k}{(2\pi)^3} \frac{k_i k_j}{k^3} \bigg[ (1+k^2\eta^2)\sinh^2\alpha \\
&\hspace{48mm}- \frac{\sinh(2\alpha)}{2} \big( 2k\eta\cos(2k\eta) + (k^2\eta^2 - 1) \sin(2k\eta) \big) \bigg]\\
&+\delta_{ij} \, \frac{H^2}{4\pi^2}  \int\limits_0^\infty dk\,k \bigg[-\sinh^2\alpha + \frac{\sinh(2\alpha)}{2} \big( 2k\eta\cos(2k\eta) + (2k^2\eta^2 - 1) \sin(2k\eta) \big)\bigg],
\end{split}
\end{equation}
which is divergent. 

During inflation, linearizing about the homogeneous background, the scalar perturbations behave like a massless scalar field in approximately de Sitter space. The calculation above therefore shows that in the $\alpha$-vacua states, the back-reaction of the quantum stress tensor will diverge in the inflationary setting also, as mentioned in section \ref{mcccheck}.

\section{Calculating the EAdS on-shell Action}
\label{bulkdet}
In this appendix, we provide some details for calculating the EAdS on-shell action to determine the unknown coefficient $\langle OOO \rangle$.

\subsection{The Longitudinal Graviton Contribution}
\label{longitudinal}
We first compute the contribution to the EAdS on-shell action coming from the longitudinal part of the exchanged graviton, which we have denoted by $\mathcal{R}$; see eq.\eqref{remainder}. For the purpose of calculations, it will be convenient to express $\mathcal{R}_1, \mathcal{R}_2, \mathcal{R}_3$ given in eq.\eqref{r123pos} in momentum space. We get
\begin{equation}
\begin{split}
\label{r123mom}
&\mathcal{R}_1 = \int \frac{dz}{z^2} \,\frac{d^3k}{(2\pi)^3}\,  T_{zj}(z,-\vec[k]) \, \frac{1}{k^2} \, T_{zj}(z,\vec[k]),\\
&\mathcal{R}_2 = - \frac{i}{2} \int \frac{dz}{z} \,\frac{d^3k}{(2\pi)^3} \,  k_j T_{zj}(z,-\vec[k]) \, \frac{1}{k^2} \, T_{zz}(z,\vec[k]),\\
&\mathcal{R}_3 = - \frac{1}{4} \int \frac{dz}{z^2} \, \frac{d^3k}{(2\pi)^3}\,  k_j T_{zj}(z,-\vec[k]) \, \frac{1}{k^4} \, k_i T_{zi}(z,\vec[k]).
\end{split}
\end{equation}
Now, to compute the remainder terms, we need to find the quantities $T_{zj}(z,\vec[k])$ and $T_{zz}(z,\vec[k])$, with the energy-momentum tensor given in eq.\eqref{emads}. Note that to get the coefficient $\langle OOO \rangle$, out of the two factors of $T_{\mu\nu}$ in each of the remainder terms, one must give a contribution proportional to $\delta\phi \delta\phi$, and the other must give a contribution proportional to $\bar{\phi}\delta\phi$; see figure \ref{stuch}. Consider first the case when the energy-momentum tensor contributes a factor of $\delta\phi \delta\phi$. For this case
\be
\label{tzj2dp}
T_{zj}(z,\vec[x]) = \del_z\delta\phi \, \del_j \delta\phi,
\ee
which on substituting the expression for $\delta\phi$ from eq.\eqref{modedp} and converting to momentum space gives\footnote{For $\delta\phi$ given in eq.\eqref{modedp}, we have
\begin{equation*}
\begin{split}
&\del_z \delta\phi_{\vec[k]}(z,\vec[x]) = - k^2 z \, \phi_0(\vec[k]) \text{e}^{-kz} \text{e}^{i\vec[k]\cdot\vec[x]},\\
&\del_j \delta\phi_{\vec[k]}(z,\vec[x]) = i k_j \, \phi_0(\vec[k]) (1+kz) \, \text{e}^{-kz} \text{e}^{i\vec[k]\cdot\vec[x]}.
\end{split}
\end{equation*}
}
\be
\label{tzj2dpm}
T_{zj}(z,\vec[k]) = -iz \,(2\pi)^3 \delta^3(\vec[p]+\vec[q]-\vec[k]) \, \phi_{0}(\vec[p]) \phi_{0}(\vec[q]) \,p^{2} q_j\, (1+qz)\, \text{e}^{-(p+q)z},
\ee
where $\vec[p], \vec[q]$ are the momentum labels carried by the two external $\delta\phi$ legs. Similarly, we also have
\be
\label{tzz2dp}
T_{zz}(z,\vec[x]) = \half \Big( (\del_z\delta\phi)^2 - (\del_i \delta\phi)^2 \Big),
\ee
which in momentum space takes the form
\begin{equation}
\begin{split}
\label{tzz2dpm}
T_{zz}(z,\vec[k]) = \half\,(2\pi)^3 &\delta^3(\vec[p]+\vec[q]-\vec[k]) \, \phi_{0}(\vec[p]) \phi_{0}(\vec[q]) \times\\
&\big[p^2 q^2 z^2+ (\vec[p]\cdot\vec[q]) (1+pz)(1+qz)\big] \, \text{e}^{-(p+q)z}.
\end{split}
\end{equation}

Consider now the case when the energy-momentum tensor contributes one factor of $\bar\phi$ and one factor of $\delta\phi$. For this case, we have
\be
\label{tzj1dp}
T_{zj}(z,\vec[x]) = \del_z\bar\phi \, \del_j \delta\phi,
\ee
which in momentum space is given by
\be
\label{tzj1dpm}
T_{zj}(z,\vec[k]) = i \,(2\pi)^3 \delta^3(\vec[p]-\vec[k]) \,\del_z\bar\phi \, \phi_{0}(\vec[p]) \, p_j (1+pz)\, \text{e}^{-pz},
\ee
where $\vec[p]$ is the momentum carried by the external leg $\delta\phi$. We also have\footnote{We use the notation $V'(\bar\phi) \equiv {dV(\bar\phi)}/{d\bar\phi}$. Also note that we have an additional factor of $\text{R}_{\text{AdS}}^2$ multiplying the $V'(\bar\phi)$ term in eq. \eqref{tzz1dp} as compared to eq.\eqref{emads}. It is because in writing the EAdS on-shell action eq.\eqref{3pons}, we extracted out an overall factor of $\text{R}_{\text{AdS}}^2$.}
\be
\label{tzz1dp}
T_{zz}(z,\vec[x]) = \del_z\bar\phi \, \del_z \delta\phi - \frac{\text{R}_{\text{AdS}}^2}{z^2} \, V'(\bar\phi)\, \delta\phi,
\ee
implying that
\be
\label{tzz1dpm}
T_{zz}(z,\vec[k]) = - (2\pi)^3 \delta^3(\vec[p]-\vec[k]) \bigg[p^2 z \del_z\bar\phi + \frac{\text{R}_{\text{AdS}}^2}{z^2} \, V'(\bar\phi) (1+pz)\bigg] \phi_{0}(\vec[p]) \, \text{e}^{-pz}.
\ee

We can now jump into the computation of the remainder terms. Consider for instance the s-channel process shown in figure \ref{stuch}. The contribution from this process to $\mathcal{R}_1$, defined in eq.\eqref{r123mom}, is
\begin{equation}
\begin{split}
\label{r1sdet}
\mathcal{R}_1^s = 2 \int \frac{dz}{z^2} &\frac{d^3k}{(2\pi)^3} \bigg[-iz \,(2\pi)^3 \delta^3(\vecs[k,1]+\vecs[k,2]+\vec[k]) \, \phi_{0}(\vecs[k,1]) \phi_{0}(\vecs[k,2]) \times\\
&\hspace{11mm}\text{e}^{-(k_1+k_2)z} \Big\lbrace k_1^{\,2} k_{2j} (1+k_2z) + k_2^{\,2} k_{1j} (1+k_1z) \Big\rbrace\bigg]\\
& \times\frac{1}{k^2} \bigg[ i \,(2\pi)^3 \delta^3(\vecs[k,3]-\vec[k]) \,\del_z\bar\phi \, \phi_{0}(\vecs[k,3]) \, k_{3j} (1+k_3z)\, \text{e}^{-k_3z} \bigg],
\end{split}
\end{equation}
which on simplification yields
\begin{equation}
\begin{split}
\label{r1ssim}
\mathcal{R}_1^s = - 2 \, &\frac{\dot{\bar\phi}}{H} \, (2\pi)^3\delta^3\Big(\sum_{a=1}^3 \vecs[k,a]\Big) \Big( \prod_{a=1}^3 \phi_0(\vecs[k,a])\Big) \frac{1}{k_3^{\,2}} \times\\
&\int_{0}^{\infty} \frac{dz}{z^2} \text{e}^{-Kz} \bigg[ \Big\lbrace k_1^{\,2} (\vecs[k,2]\cdot\vecs[k,3]) +  k_2^{\,2} (\vecs[k,1]\cdot\vecs[k,3]) \Big\rbrace \\
&\hspace{20mm}+ z \Big\lbrace k_1^{\,2} (\vecs[k,2]\cdot\vecs[k,3])(k_2 + k_3) +  k_2^{\,2} (\vecs[k,1]\cdot\vecs[k,3]) (k_1 + k_3) \Big\rbrace\\
&\hspace{20mm}+ z^2 \Big\lbrace k_1^{\,2} k_2 k_3 (\vecs[k,2]\cdot\vecs[k,3]) +  k_2^{\,2} k_1 k_3 (\vecs[k,1]\cdot\vecs[k,3])\Big\rbrace \bigg],
\end{split}
\end{equation}
where we have used the fact that under the analytic continuation given in eq.\eqref{anaz}, $z\del_z\bar\phi$ goes over to $-\dot{\bar\phi}/H$, and since we are working to the leading order in slow-roll, we can consider this factor to be a constant and pull it out of the $z$-integral. Eq.\eqref{r1ssim} gives
\begin{equation}
\begin{split}
\label{r1sinp}
\mathcal{R}_1^s =& - 2 \, \frac{\dot{\bar\phi}}{H} \, (2\pi)^3\delta^3\Big(\sum_{a=1}^3 \vecs[k,a]\Big) \Big( \prod_{a=1}^3 \phi_0(\vecs[k,a])\Big) \frac{1}{k_3^{\,2}} \times\\
&\Bigg[ \frac{1}{K} \Big\lbrace k_1^{\,2} k_2 k_3 (\vecs[k,2]\cdot\vecs[k,3]) +  k_2^{\,2} k_1 k_3 (\vecs[k,1]\cdot\vecs[k,3])\Big\rbrace\\
&+ \Big\lbrace k_1^{\,2} (\vecs[k,2]\cdot\vecs[k,3])(k_2 + k_3) +  k_2^{\,2} (\vecs[k,1]\cdot\vecs[k,3]) (k_1 + k_3) \Big\rbrace \bigg(\int_0^\infty \frac{dz}{z} \, \text{e}^{-Kz}\bigg)\\
&+ \Big\lbrace k_1^{\,2} (\vecs[k,2]\cdot\vecs[k,3]) +  k_2^{\,2} (\vecs[k,1]\cdot\vecs[k,3]) \Big\rbrace  \bigg(\int_0^\infty \frac{dz}{z^2} \, \text{e}^{-Kz}\bigg)\Bigg].
\end{split}
\end{equation}
Now, the remaining two integrals in eq.\eqref{r1sinp} are divergent. Let us replace the lower limit of the remaining integrals by $\rho$, where $\rho \rightarrow 0$. Then
\be
\label{iads1}
\int_\rho^\infty \frac{dz}{z} \, \text{e}^{-Kz} = \Gamma\big[0,K\rho\big],
\ee
and
\be
\label{iads2}
\int_\rho^\infty \frac{dz}{z^2} \, \text{e}^{-Kz} = \frac{1}{\rho} - K - K\,\Gamma\big[0,K\rho\big].
\ee
Using the results of eqs.\eqref{iads1} and \eqref{iads2} in eq.\eqref{r1sinp}, we get
\begin{equation}
\begin{split}
\label{r1sfin}
\mathcal{R}_1^s = -& 2 \, \frac{\dot{\bar\phi}}{H} \, (2\pi)^3\delta^3\Big(\sum_{a=1}^3 \vecs[k,a]\Big) \Big( \prod_{a=1}^3 \phi_0(\vecs[k,a])\Big) \frac{1}{k_3^{\,2}} \times\\
&\Bigg[ \frac{1}{K} \Big\lbrace k_1^{\,2} k_2 k_3 (\vecs[k,2]\cdot\vecs[k,3]) +  k_2^{\,2} k_1 k_3 (\vecs[k,1]\cdot\vecs[k,3])\Big\rbrace\\
&- \bigg(K + K\,\Gamma\big[0,K\rho\big] - \frac{1}{\rho} \bigg) \Big\lbrace k_1^{\,2} (\vecs[k,2]\cdot\vecs[k,3]) +  k_2^{\,2} (\vecs[k,1]\cdot\vecs[k,3]) \Big\rbrace\\
&+ \Gamma\big[0,K\rho\big] \Big\lbrace k_1^{\,2} (\vecs[k,2]\cdot\vecs[k,3])(k_2 + k_3) +  k_2^{\,2} (\vecs[k,1]\cdot\vecs[k,3]) (k_1 + k_3) \Big\rbrace \Bigg].
\end{split}
\end{equation}
This is our final answer for $\mathcal{R}_1^s$. From figure \ref{stuch}, it is clear that the t and u channel contributions to $\mathcal{R}_1$ can be easily obtained from $\mathcal{R}_1^s$ by performing momentum interchanges. In particular,
\be
\label{r1t}
\mathcal{R}_1^t = \mathcal{R}_1^s(\vecs[k,2] \leftrightarrow \vecs[k,3]) ,
\ee
and
\be
\label{r1u}
\mathcal{R}_1^u = \mathcal{R}_1^s(\vecs[k,1] \leftrightarrow \vecs[k,3]).
\ee
The complete expression for $\mathcal{R}_1$ is then
\be
\label{rcomp}
\mathcal{R}_1 = \mathcal{R}_1^s + \mathcal{R}_1^t + \mathcal{R}_1^u .
\ee

Following exactly the same procedure as above, we can now compute $\mathcal{R}_2$, eq.\eqref{r123mom}. We get\footnote{In calculating $\mathcal{R}_2$, we have used the slow-roll approximation $V'(\bar\phi) \approx - 3 H \dot{\bar\phi}$.}
\begin{equation}
\begin{split}
\label{r2sfin}
\mathcal{R}_2^{s}& = - \half \, \frac{\dot{\bar\phi}}{H} \, (2\pi)^3\delta^3\Big(\sum_{a=1}^3 \vecs[k,a]\Big) \Big( \prod_{a=1}^3 \phi_0(\vecs[k,a])\Big) \frac{1}{k_3^{\,2}}\times\\
&\Bigg[ \frac{1}{K^2} \Big\lbrace k_1^{\,2} k_2 k_3^{\,2} (\vecs[k,2]\cdot\vecs[k,3]) + k_1 k_2^{\,2} k_3^{\,2} (\vecs[k,1]\cdot\vecs[k,3])- k_1 k_2 k_3^{\,3} (\vecs[k,1]\cdot\vecs[k,2])- k_1^{\,2} k_2^{\,2} k_3^{\,3} \Big\rbrace\\
&+ \frac{1}{K} \Big\lbrace k_1^{\,2} k_3^{\,2} (\vecs[k,2]\cdot\vecs[k,3]) + k_2^{\,2} k_3^{\,2} (\vecs[k,1]\cdot\vecs[k,3])- k_1 k_2 k_3^{\,2} (\vecs[k,1]\cdot\vecs[k,2])- k_1^{\,2} k_2^{\,2} k_3^{\,2}\\
&\hspace{10mm}-k_3^{\,3}(\vecs[k,1]\cdot\vecs[k,2]) (k_1+k_2) - 3 k_1^{\,2} k_2 k_3 (\vecs[k,2]\cdot\vecs[k,3]) - 3 k_1 k_2^{\,2} k_3 (\vecs[k,1]\cdot\vecs[k,3]) \Big\rbrace\\
&+ k_3^{\,2}\bigg( K - \frac{1}{\rho}\bigg) (\vecs[k,1]\cdot\vecs[k,2]) + 3\bigg(K + K\,\Gamma\big[0,K\rho\big] - \frac{1}{\rho} \bigg) \Big\lbrace k_1^{\,2} (\vecs[k,2]\cdot\vecs[k,3]) +  k_2^{\,2} (\vecs[k,1]\cdot\vecs[k,3]) \Big\rbrace\\
&- 3  \Gamma\big[0,K\rho\big] \Big\lbrace k_1^{\,2} (\vecs[k,2]\cdot\vecs[k,3])(k_2 + k_3) +  k_2^{\,2} (\vecs[k,1]\cdot\vecs[k,3]) (k_1 + k_3) \Big\rbrace\Bigg].
\end{split}
\end{equation}
As before, we have
\be
\label{r2tu}
\mathcal{R}_2^t = \mathcal{R}_2^s(\vecs[k,2] \leftrightarrow \vecs[k,3]),\,\, \mathcal{R}_2^u = \mathcal{R}_2^s(\vecs[k,1] \leftrightarrow \vecs[k,3]),
\ee
and the complete expression is
\be
\label{r2}
\mathcal{R}_2 = \mathcal{R}_2^s + \mathcal{R}_2^t + \mathcal{R}_2^u .
\ee

Finally, evaluating $\mathcal{R}_3$, we find that
\be
\label{r3}
\mathcal{R}_3 = -\, \frac{1}{4}\, \mathcal{R}_1 .
\ee

Combining the expressions for $\mathcal{R}_1, \mathcal{R}_2$ and $\mathcal{R}_3$, we get an expression for $\mathcal{R}$, eq.\eqref{remainder}, given by
\begin{equation}
\begin{split}
\label{rpfin}
\mathcal{R} = \half \, \frac{\dot{\bar\phi}}{H}& \, (2\pi)^3\delta^3\Big(\sum_{a=1}^3 \vecs[k,a]\Big) \Big( \prod_{a=1}^3 \phi_0(\vecs[k,a])\Big) \times \\
&\Bigg[- \half \sum_{a=1}^3 k_a^{\,3} + \half \sum_{a \neq b} k_a k_b^{\,2} + \frac{4}{K}\sum_{a<b} k_a^{\,2} k_b^{\,2} - \frac{1}{2\rho} \sum_{a=1}^3 k_a^{\,2} \Bigg].
\end{split}
\end{equation}
The cut-off dependent term in eq.\eqref{rpfin} has to be removed by the addition of a suitable local counter-term. The final expression for the remainder term is then given by eq.\eqref{rfin}.

\subsection{The Transverse Graviton Contribution}
\label{transverse}
We now proceed to calculate the contribution to the EAdS on-shell action coming from the transverse part of the exchanged graviton. This contribution is denoted by $\mathcal{W}$, and is defined in eq.\eqref{tgcont}. Out of the two $T_{ij}$ terms in $\mathcal{W}$, one must contribute a piece proportional to $\delta\phi \delta\phi$, and the other must contribute a piece proportional to $\del_z\bar\phi \, \delta\phi$. Using the expression eq.\eqref{emads} for the energy-momentum tensor, we get
\be
\label{tijd}
T_{ij}(z,\vec[x]) = -\bigg( \del_z\bar\phi \, \del_z\delta\phi + \frac{\text{R}_{\text{AdS}}^2}{z^2} \, V'(\bar\phi) \, \delta\phi\bigg) \delta_{ij}.
\ee
The key point to note in eq.\eqref{tijd} is that $T_{ij}$ is proportional to $\delta_{ij}$. Thus, while calculating $\mathcal{W}$, the contraction of $T_{ij}$ with the transverse graviton propagator $\tilde{\mathcal{G}}_{ij,kl}$ gives the trace $\tilde{\mathcal{G}}_{ii,kl}$, which vanishes. Therefore, there is no contribution from the transverse graviton exchange to the EAdS on-shell action, i.e.
\be
\label{wv}
\mathcal{W} = 0.
\ee
Therefore, the only contribution to the EAdS on-shell action comes from the longitudinal part of the exchanged graviton, eq.\eqref{rfin}.

\bibliography{refs}{}

\providecommand{\href}[2]{#2}\begingroup\raggedright\begin{thebibliography}{10}

\bibitem{Antoniadis:1996dj}
I.~Antoniadis, P.~O. Mazur and E.~Mottola, \emph{{Conformal invariance and
  cosmic background radiation}},
  \href{http://dx.doi.org/10.1103/PhysRevLett.79.14}{\emph{Phys.Rev.Lett.} {\bf
  79} (1997) 14--17}, [\href{http://arxiv.org/abs/astro-ph/9611208}{{\tt
  astro-ph/9611208}}].

\bibitem{Larsen:2002et}
F.~Larsen, J.~P. van~der Schaar and R.~G. Leigh, \emph{{De Sitter holography
  and the cosmic microwave background}},
  \href{http://dx.doi.org/10.1088/1126-6708/2002/04/047}{\emph{JHEP} {\bf 0204}
  (2002) 047}, [\href{http://arxiv.org/abs/hep-th/0202127}{{\tt
  hep-th/0202127}}].

\bibitem{Larsen:2003pf}
F.~Larsen and R.~McNees, \emph{{Inflation and de Sitter holography}},
  \href{http://dx.doi.org/10.1088/1126-6708/2003/07/051}{\emph{JHEP} {\bf 0307}
  (2003) 051}, [\href{http://arxiv.org/abs/hep-th/0307026}{{\tt
  hep-th/0307026}}].

\bibitem{McFadden:2010vh}
P.~McFadden and K.~Skenderis, \emph{{Holographic Non-Gaussianity}},
  \href{http://dx.doi.org/10.1088/1475-7516/2011/05/013}{\emph{JCAP} {\bf 1105}
  (2011) 013}, [\href{http://arxiv.org/abs/1011.0452}{{\tt 1011.0452}}].

\bibitem{Antoniadis:2011ib}
I.~Antoniadis, P.~O. Mazur and E.~Mottola, \emph{{Conformal Invariance, Dark
  Energy, and CMB Non-Gaussianity}},
  \href{http://dx.doi.org/10.1088/1475-7516/2012/09/024}{\emph{JCAP} {\bf 1209}
  (2012) 024}, [\href{http://arxiv.org/abs/1103.4164}{{\tt 1103.4164}}].

\bibitem{McFadden:2011kk}
P.~McFadden and K.~Skenderis, \emph{{Cosmological 3-point correlators from
  holography}},
  \href{http://dx.doi.org/10.1088/1475-7516/2011/06/030}{\emph{JCAP} {\bf 1106}
  (2011) 030}, [\href{http://arxiv.org/abs/1104.3894}{{\tt 1104.3894}}].

\bibitem{Creminelli:2011mw}
P.~Creminelli, \emph{{Conformal invariance of scalar perturbations in
  inflation}},
  \href{http://dx.doi.org/10.1103/PhysRevD.85.041302}{\emph{Phys.Rev.} {\bf
  D85} (2012) 041302}, [\href{http://arxiv.org/abs/1108.0874}{{\tt
  1108.0874}}].

\bibitem{Bzowski:2011ab}
A.~Bzowski, P.~McFadden and K.~Skenderis, \emph{{Holographic predictions for
  cosmological 3-point functions}},
  \href{http://dx.doi.org/10.1007/JHEP03(2012)091}{\emph{JHEP} {\bf 1203}
  (2012) 091}, [\href{http://arxiv.org/abs/1112.1967}{{\tt 1112.1967}}].

\bibitem{Kehagias:2012pd}
A.~Kehagias and A.~Riotto, \emph{{Operator Product Expansion of Inflationary
  Correlators and Conformal Symmetry of de Sitter}},
  \href{http://dx.doi.org/10.1016/j.nuclphysb.2012.07.004}{\emph{Nucl.Phys.}
  {\bf B864} (2012) 492--529}, [\href{http://arxiv.org/abs/1205.1523}{{\tt
  1205.1523}}].

\bibitem{Kehagias:2012td}
A.~Kehagias and A.~Riotto, \emph{{The Four-point Correlator in Multifield
  Inflation, the Operator Product Expansion and the Symmetries of de Sitter}},
  \href{http://dx.doi.org/10.1016/j.nuclphysb.2012.11.025}{\emph{Nucl.Phys.}
  {\bf B868} (2013) 577--595}, [\href{http://arxiv.org/abs/1210.1918}{{\tt
  1210.1918}}].

\bibitem{Schalm:2012pi}
K.~Schalm, G.~Shiu and T.~van~der Aalst, \emph{{Consistency condition for
  inflation from (broken) conformal symmetry}},
  \href{http://dx.doi.org/10.1088/1475-7516/2013/03/005}{\emph{JCAP} {\bf 1303}
  (2013) 005}, [\href{http://arxiv.org/abs/1211.2157}{{\tt 1211.2157}}].

\bibitem{Bzowski:2012ih}
A.~Bzowski, P.~McFadden and K.~Skenderis, \emph{{Holography for inflation using
  conformal perturbation theory}},
  \href{http://dx.doi.org/10.1007/JHEP04(2013)047}{\emph{JHEP} {\bf 1304}
  (2013) 047}, [\href{http://arxiv.org/abs/1211.4550}{{\tt 1211.4550}}].

\bibitem{McFadden:2014nta}
P.~McFadden, \emph{{Soft limits in holographic cosmology}},
  \href{http://dx.doi.org/10.1007/JHEP02(2015)053}{\emph{JHEP} {\bf 02} (2015)
  053}, [\href{http://arxiv.org/abs/1412.1874}{{\tt 1412.1874}}].

\bibitem{Kehagias:2015jha}
A.~Kehagias and A.~Riotto, \emph{{High Energy Physics Signatures from Inflation
  and Conformal Symmetry of de Sitter}},
  \href{http://dx.doi.org/10.1002/prop.201500025}{\emph{Fortsch. Phys.} {\bf
  63} (2015) 531--542}, [\href{http://arxiv.org/abs/1501.03515}{{\tt
  1501.03515}}].

\bibitem{Mata:2012bx}
I.~Mata, S.~Raju and S.~P. Trivedi, \emph{{CMB from CFT}},
  \href{http://dx.doi.org/10.1007/JHEP07(2013)015}{\emph{JHEP} {\bf 1307}
  (2013) 015}, [\href{http://arxiv.org/abs/1211.5482}{{\tt 1211.5482}}].

\bibitem{Ghosh:2014kba}
A.~Ghosh, N.~Kundu, S.~Raju and S.~P. Trivedi, \emph{{Conformal Invariance and
  the Four Point Scalar Correlator in Slow-Roll Inflation}},
  \href{http://dx.doi.org/10.1007/JHEP07(2014)011}{\emph{JHEP} {\bf 1407}
  (2014) 011}, [\href{http://arxiv.org/abs/1401.1426}{{\tt 1401.1426}}].

\bibitem{Kundu:2014gxa}
N.~Kundu, A.~Shukla and S.~P. Trivedi, \emph{{Constraints from Conformal
  Symmetry on the Three Point Scalar Correlator in Inflation}},
  \href{http://dx.doi.org/10.1007/JHEP04(2015)061}{\emph{JHEP} {\bf 1504}
  (2015) 061}, [\href{http://arxiv.org/abs/1410.2606}{{\tt 1410.2606}}].

\bibitem{Maldacena:2002vr}
J.~M. Maldacena, \emph{{Non-Gaussian features of primordial fluctuations in
  single field inflationary models}},
  \href{http://dx.doi.org/10.1088/1126-6708/2003/05/013}{\emph{JHEP} {\bf 0305}
  (2003) 013}, [\href{http://arxiv.org/abs/astro-ph/0210603}{{\tt
  astro-ph/0210603}}].

\bibitem{Maldacena:2011nz}
J.~M. Maldacena and G.~L. Pimentel, \emph{{On graviton non-Gaussianities during
  inflation}}, \href{http://dx.doi.org/10.1007/JHEP09(2011)045}{\emph{JHEP}
  {\bf 1109} (2011) 045}, [\href{http://arxiv.org/abs/1104.2846}{{\tt
  1104.2846}}].

\bibitem{Kundu:2015xta}
N.~Kundu, A.~Shukla and S.~P. Trivedi, \emph{{Ward Identities for Scale and
  Special Conformal Transformations in Inflation}},
  \href{http://dx.doi.org/10.1007/JHEP01(2016)046}{\emph{JHEP} {\bf 01} (2016)
  046}, [\href{http://arxiv.org/abs/1507.06017}{{\tt 1507.06017}}].

\bibitem{Weinberg:2003sw}
S.~Weinberg, \emph{{Adiabatic modes in cosmology}},
  \href{http://dx.doi.org/10.1103/PhysRevD.67.123504}{\emph{Phys.Rev.} {\bf
  D67} (2003) 123504}, [\href{http://arxiv.org/abs/astro-ph/0302326}{{\tt
  astro-ph/0302326}}].

\bibitem{Creminelli:2004yq}
P.~Creminelli and M.~Zaldarriaga, \emph{{Single field consistency relation for
  the 3-point function}},
  \href{http://dx.doi.org/10.1088/1475-7516/2004/10/006}{\emph{JCAP} {\bf 0410}
  (2004) 006}, [\href{http://arxiv.org/abs/astro-ph/0407059}{{\tt
  astro-ph/0407059}}].

\bibitem{Cheung:2007sv}
C.~Cheung, A.~L. Fitzpatrick, J.~Kaplan and L.~Senatore, \emph{{On the
  consistency relation of the 3-point function in single field inflation}},
  \href{http://dx.doi.org/10.1088/1475-7516/2008/02/021}{\emph{JCAP} {\bf 0802}
  (2008) 021}, [\href{http://arxiv.org/abs/0709.0295}{{\tt 0709.0295}}].

\bibitem{Weinberg:2008nf}
S.~Weinberg, \emph{{Non-Gaussian Correlations Outside the Horizon}},
  \href{http://dx.doi.org/10.1103/PhysRevD.78.123521}{\emph{Phys.Rev.} {\bf
  D78} (2008) 123521}, [\href{http://arxiv.org/abs/0808.2909}{{\tt
  0808.2909}}].

\bibitem{Creminelli:2011sq}
P.~Creminelli, C.~Pitrou and F.~Vernizzi, \emph{{The CMB bispectrum in the
  squeezed limit}},
  \href{http://dx.doi.org/10.1088/1475-7516/2011/11/025}{\emph{JCAP} {\bf 1111}
  (2011) 025}, [\href{http://arxiv.org/abs/1109.1822}{{\tt 1109.1822}}].

\bibitem{Bartolo:2011wb}
N.~Bartolo, S.~Matarrese and A.~Riotto, \emph{{Non-Gaussianity in the Cosmic
  Microwave Background Anisotropies at Recombination in the Squeezed limit}},
  \href{http://dx.doi.org/10.1088/1475-7516/2012/02/017}{\emph{JCAP} {\bf 1202}
  (2012) 017}, [\href{http://arxiv.org/abs/1109.2043}{{\tt 1109.2043}}].

\bibitem{Creminelli:2012ed}
P.~Creminelli, J.~Norena and M.~Simonovic, \emph{{Conformal consistency
  relations for single-field inflation}},
  \href{http://dx.doi.org/10.1088/1475-7516/2012/07/052}{\emph{JCAP} {\bf 1207}
  (2012) 052}, [\href{http://arxiv.org/abs/1203.4595}{{\tt 1203.4595}}].

\bibitem{Hinterbichler:2012nm}
K.~Hinterbichler, L.~Hui and J.~Khoury, \emph{{Conformal Symmetries of
  Adiabatic Modes in Cosmology}},
  \href{http://dx.doi.org/10.1088/1475-7516/2012/08/017}{\emph{JCAP} {\bf 1208}
  (2012) 017}, [\href{http://arxiv.org/abs/1203.6351}{{\tt 1203.6351}}].

\bibitem{Senatore:2012wy}
L.~Senatore and M.~Zaldarriaga, \emph{{A Note on the Consistency Condition of
  Primordial Fluctuations}},
  \href{http://dx.doi.org/10.1088/1475-7516/2012/08/001}{\emph{JCAP} {\bf 1208}
  (2012) 001}, [\href{http://arxiv.org/abs/1203.6884}{{\tt 1203.6884}}].

\bibitem{Assassi:2012zq}
V.~Assassi, D.~Baumann and D.~Green, \emph{{On Soft Limits of Inflationary
  Correlation Functions}},
  \href{http://dx.doi.org/10.1088/1475-7516/2012/11/047}{\emph{JCAP} {\bf 1211}
  (2012) 047}, [\href{http://arxiv.org/abs/1204.4207}{{\tt 1204.4207}}].

\bibitem{Creminelli:2012qr}
P.~Creminelli, A.~Joyce, J.~Khoury and M.~Simonovic, \emph{{Consistency
  Relations for the Conformal Mechanism}},
  \href{http://dx.doi.org/10.1088/1475-7516/2013/04/020}{\emph{JCAP} {\bf 1304}
  (2013) 020}, [\href{http://arxiv.org/abs/1212.3329}{{\tt 1212.3329}}].

\bibitem{Goldberger:2013rsa}
W.~D. Goldberger, L.~Hui and A.~Nicolis, \emph{{One-particle-irreducible
  consistency relations for cosmological perturbations}},
  \href{http://dx.doi.org/10.1103/PhysRevD.87.103520}{\emph{Phys.Rev.} {\bf
  D87} (2013) 103520}, [\href{http://arxiv.org/abs/1303.1193}{{\tt
  1303.1193}}].

\bibitem{Hinterbichler:2013dpa}
K.~Hinterbichler, L.~Hui and J.~Khoury, \emph{{An Infinite Set of Ward
  Identities for Adiabatic Modes in Cosmology}},
  \href{http://arxiv.org/abs/1304.5527}{{\tt 1304.5527}}.

\bibitem{Creminelli:2013cga}
P.~Creminelli, A.~Perko, L.~Senatore, M.~Simonovic and G.~Trevisan, \emph{{The
  Physical Squeezed Limit: Consistency Relations at Order $q^2$}},
  \href{http://dx.doi.org/10.1088/1475-7516/2013/11/015}{\emph{JCAP} {\bf 1311}
  (2013) 015}, [\href{http://arxiv.org/abs/1307.0503}{{\tt 1307.0503}}].

\bibitem{Pimentel:2013gza}
G.~L. Pimentel, \emph{{Inflationary Consistency Conditions from a
  Wavefunctional Perspective}},
  \href{http://dx.doi.org/10.1007/JHEP02(2014)124}{\emph{JHEP} {\bf 1402}
  (2014) 124}, [\href{http://arxiv.org/abs/1309.1793}{{\tt 1309.1793}}].

\bibitem{Berezhiani:2013ewa}
L.~Berezhiani and J.~Khoury, \emph{{Slavnov-Taylor Identities for Primordial
  Perturbations}},  \href{http://arxiv.org/abs/1309.4461}{{\tt 1309.4461}}.

\bibitem{Sreenath:2014nka}
V.~Sreenath and L.~Sriramkumar, \emph{{Examining the consistency relations
  describing the three-point functions involving tensors}},
  \href{http://dx.doi.org/10.1088/1475-7516/2014/10/021}{\emph{JCAP} {\bf 1410}
  (2014) 021}, [\href{http://arxiv.org/abs/1406.1609}{{\tt 1406.1609}}].

\bibitem{Mirbabayi:2014zpa}
M.~Mirbabayi and M.~Zaldarriaga, \emph{{Double Soft Limits of Cosmological
  Correlations}},  \href{http://arxiv.org/abs/1409.6317}{{\tt 1409.6317}}.

\bibitem{Joyce:2014aqa}
A.~Joyce, J.~Khoury and M.~Simonovic, \emph{{Multiple Soft Limits of
  Cosmological Correlation Functions}},
  \href{http://arxiv.org/abs/1409.6318}{{\tt 1409.6318}}.

\bibitem{Sreenath:2014nca}
V.~Sreenath, D.~K. Hazra and L.~Sriramkumar, \emph{{On the scalar consistency
  relation away from slow roll}},  \href{http://arxiv.org/abs/1410.0252}{{\tt
  1410.0252}}.

\bibitem{Binosi:2015obq}
D.~Binosi and A.~Quadri, \emph{{The Cosmological Slavnov-Taylor Identity from
  BRST Symmetry in Single-Field Inflation}},
  \href{http://dx.doi.org/10.1088/1475-7516/2016/03/045}{\emph{JCAP} {\bf 1603}
  (2016) 045}, [\href{http://arxiv.org/abs/1511.09309}{{\tt 1511.09309}}].

\bibitem{Chowdhury:2016yrh}
D.~Chowdhury, V.~Sreenath and L.~Sriramkumar, \emph{{The scalar-scalar-tensor
  inflationary three-point function in the axion monodromy model}},
  \href{http://arxiv.org/abs/1605.05292}{{\tt 1605.05292}}.

\bibitem{Chen:2006nt}
X.~Chen, M.-x. Huang, S.~Kachru and G.~Shiu, \emph{{Observational signatures
  and non-Gaussianities of general single field inflation}},
  \href{http://dx.doi.org/10.1088/1475-7516/2007/01/002}{\emph{JCAP} {\bf 0701}
  (2007) 002}, [\href{http://arxiv.org/abs/hep-th/0605045}{{\tt
  hep-th/0605045}}].

\bibitem{Chen:2009bc}
X.~Chen, B.~Hu, M.-x. Huang, G.~Shiu and Y.~Wang, \emph{{Large Primordial
  Trispectra in General Single Field Inflation}},
  \href{http://dx.doi.org/10.1088/1475-7516/2009/08/008}{\emph{JCAP} {\bf 0908}
  (2009) 008}, [\href{http://arxiv.org/abs/0905.3494}{{\tt 0905.3494}}].

\bibitem{PhysRevD.32.3136}
B.~Allen, \emph{Vacuum states in de sitter space},
  \href{http://dx.doi.org/10.1103/PhysRevD.32.3136}{\emph{Phys. Rev. D} {\bf
  32} (Dec, 1985) 3136--3149}.

\bibitem{PhysRevD.35.3771}
B.~Allen and A.~Folacci, \emph{Massless minimally coupled scalar field in de
  sitter space}, \href{http://dx.doi.org/10.1103/PhysRevD.35.3771}{\emph{Phys.
  Rev. D} {\bf 35} (Jun, 1987) 3771--3778}.

\bibitem{Arkani-Hamed:2015bza}
N.~Arkani-Hamed and J.~Maldacena, \emph{{Cosmological Collider Physics}},
  \href{http://arxiv.org/abs/1503.08043}{{\tt 1503.08043}}.

\bibitem{ArmendarizPicon:1999rj}
C.~Armendariz-Picon, T.~Damour and V.~F. Mukhanov, \emph{{k - inflation}},
  \href{http://dx.doi.org/10.1016/S0370-2693(99)00603-6}{\emph{Phys. Lett.}
  {\bf B458} (1999) 209--218}, [\href{http://arxiv.org/abs/hep-th/9904075}{{\tt
  hep-th/9904075}}].

\bibitem{Garriga:1999vw}
J.~Garriga and V.~F. Mukhanov, \emph{{Perturbations in k-inflation}},
  \href{http://dx.doi.org/10.1016/S0370-2693(99)00602-4}{\emph{Phys. Lett.}
  {\bf B458} (1999) 219--225}, [\href{http://arxiv.org/abs/hep-th/9904176}{{\tt
  hep-th/9904176}}].

\bibitem{Silverstein:2003hf}
E.~Silverstein and D.~Tong, \emph{{Scalar speed limits and cosmology:
  Acceleration from D-cceleration}},
  \href{http://dx.doi.org/10.1103/PhysRevD.70.103505}{\emph{Phys.Rev.} {\bf
  D70} (2004) 103505}, [\href{http://arxiv.org/abs/hep-th/0310221}{{\tt
  hep-th/0310221}}].

\bibitem{Alishahiha:2004eh}
M.~Alishahiha, E.~Silverstein and D.~Tong, \emph{{DBI in the sky}},
  \href{http://dx.doi.org/10.1103/PhysRevD.70.123505}{\emph{Phys.Rev.} {\bf
  D70} (2004) 123505}, [\href{http://arxiv.org/abs/hep-th/0404084}{{\tt
  hep-th/0404084}}].

\bibitem{PhysRevD.31.754}
E.~Mottola, \emph{Particle creation in de sitter space},
  \href{http://dx.doi.org/10.1103/PhysRevD.31.754}{\emph{Phys. Rev. D} {\bf 31}
  (Feb, 1985) 754--766}.

\bibitem{Bousso:2001mw}
R.~Bousso, A.~Maloney and A.~Strominger, \emph{{Conformal vacua and entropy in
  de Sitter space}},
  \href{http://dx.doi.org/10.1103/PhysRevD.65.104039}{\emph{Phys. Rev.} {\bf
  D65} (2002) 104039}, [\href{http://arxiv.org/abs/hep-th/0112218}{{\tt
  hep-th/0112218}}].

\bibitem{Kaloper:2002cs}
N.~Kaloper, M.~Kleban, A.~Lawrence, S.~Shenker and L.~Susskind, \emph{{Initial
  conditions for inflation}},
  \href{http://dx.doi.org/10.1088/1126-6708/2002/11/037}{\emph{JHEP} {\bf 11}
  (2002) 037}, [\href{http://arxiv.org/abs/hep-th/0209231}{{\tt
  hep-th/0209231}}].

\bibitem{Goldstein:2003ut}
K.~Goldstein and D.~A. Lowe, \emph{{A Note on alpha vacua and interacting field
  theory in de Sitter space}},
  \href{http://dx.doi.org/10.1016/j.nuclphysb.2003.07.014}{\emph{Nucl. Phys.}
  {\bf B669} (2003) 325--340}, [\href{http://arxiv.org/abs/hep-th/0302050}{{\tt
  hep-th/0302050}}].

\bibitem{Kundu:2011sg}
S.~Kundu, \emph{{Inflation with General Initial Conditions for Scalar
  Perturbations}},
  \href{http://dx.doi.org/10.1088/1475-7516/2012/02/005}{\emph{JCAP} {\bf 1202}
  (2012) 005}, [\href{http://arxiv.org/abs/1110.4688}{{\tt 1110.4688}}].

\bibitem{Kundu:2013gha}
S.~Kundu, \emph{{Non-Gaussianity Consistency Relations, Initial States and
  Back-reaction}},
  \href{http://dx.doi.org/10.1088/1475-7516/2014/04/016}{\emph{JCAP} {\bf 1404}
  (2014) 016}, [\href{http://arxiv.org/abs/1311.1575}{{\tt 1311.1575}}].

\bibitem{Mukhanov:2007zz}
V.~Mukhanov and S.~Winitzki, \emph{{Introduction to quantum effects in
  gravity}}.
\newblock Cambridge University Press, 2007.

\bibitem{Coriano:2013jba}
C.~Coriano, L.~Delle~Rose, E.~Mottola and M.~Serino, \emph{{Solving the
  Conformal Constraints for Scalar Operators in Momentum Space and the
  Evaluation of Feynman's Master Integrals}},
  \href{http://dx.doi.org/10.1007/JHEP07(2013)011}{\emph{JHEP} {\bf 07} (2013)
  011}, [\href{http://arxiv.org/abs/1304.6944}{{\tt 1304.6944}}].

\bibitem{Bzowski:2013sza}
A.~Bzowski, P.~McFadden and K.~Skenderis, \emph{{Implications of conformal
  invariance in momentum space}},
  \href{http://dx.doi.org/10.1007/JHEP03(2014)111}{\emph{JHEP} {\bf 03} (2014)
  111}, [\href{http://arxiv.org/abs/1304.7760}{{\tt 1304.7760}}].

\bibitem{Holman:2007na}
R.~Holman and A.~J. Tolley, \emph{{Enhanced Non-Gaussianity from Excited
  Initial States}},
  \href{http://dx.doi.org/10.1088/1475-7516/2008/05/001}{\emph{JCAP} {\bf 0805}
  (2008) 001}, [\href{http://arxiv.org/abs/0710.1302}{{\tt 0710.1302}}].

\bibitem{Ganc:2011dy}
J.~Ganc, \emph{{Calculating the local-type fNL for slow-roll inflation with a
  non-vacuum initial state}},
  \href{http://dx.doi.org/10.1103/PhysRevD.84.063514}{\emph{Phys. Rev.} {\bf
  D84} (2011) 063514}, [\href{http://arxiv.org/abs/1104.0244}{{\tt
  1104.0244}}].

\bibitem{Aravind:2013lra}
A.~Aravind, D.~Lorshbough and S.~Paban, \emph{{Non-Gaussianity from Excited
  Initial Inflationary States}},
  \href{http://dx.doi.org/10.1007/JHEP07(2013)076}{\emph{JHEP} {\bf 07} (2013)
  076}, [\href{http://arxiv.org/abs/1303.1440}{{\tt 1303.1440}}].

\bibitem{Ashoorioon:2010xg}
A.~Ashoorioon and G.~Shiu, \emph{{A Note on Calm Excited States of Inflation}},
  \href{http://dx.doi.org/10.1088/1475-7516/2011/03/025}{\emph{JCAP} {\bf 1103}
  (2011) 025}, [\href{http://arxiv.org/abs/1012.3392}{{\tt 1012.3392}}].

\bibitem{Ashoorioon:2013eia}
A.~Ashoorioon, K.~Dimopoulos, M.~M. Sheikh-Jabbari and G.~Shiu,
  \emph{{Reconciliation of High Energy Scale Models of Inflation with Planck}},
  \href{http://dx.doi.org/10.1088/1475-7516/2014/02/025}{\emph{JCAP} {\bf 1402}
  (2014) 025}, [\href{http://arxiv.org/abs/1306.4914}{{\tt 1306.4914}}].

\bibitem{Lee:2016vti}
H.~Lee, D.~Baumann and G.~L. Pimentel, \emph{{Non-Gaussianity as a Particle
  Detector}},  \href{http://arxiv.org/abs/1607.03735}{{\tt 1607.03735}}.

\bibitem{Raju:2010by}
S.~Raju, \emph{{BCFW for Witten Diagrams}},
  \href{http://dx.doi.org/10.1103/PhysRevLett.106.091601}{\emph{Phys. Rev.
  Lett.} {\bf 106} (2011) 091601}, [\href{http://arxiv.org/abs/1011.0780}{{\tt
  1011.0780}}].

\bibitem{Raju:2011mp}
S.~Raju, \emph{{Recursion Relations for AdS/CFT Correlators}},
  \href{http://dx.doi.org/10.1103/PhysRevD.83.126002}{\emph{Phys. Rev.} {\bf
  D83} (2011) 126002}, [\href{http://arxiv.org/abs/1102.4724}{{\tt
  1102.4724}}].

\end{thebibliography}\endgroup
\bibliographystyle{JHEP}

\end{document}